\documentclass[prb,preprint,footinbib,superscriptaddress]{revtex4-1}%
\usepackage[utf8]{inputenc}
\usepackage{amssymb}
\usepackage{amsfonts}
\usepackage{amsmath}
\DeclareMathOperator{\diag}{diag}
\DeclareMathOperator{\sign}{sign}
\DeclareMathOperator{\conv}{conv}
\DeclareMathOperator{\lcm}{lcm}
\usepackage{graphicx}

\usepackage{xcolor}
\usepackage[caption=false]{subfig}%
\providecommand{\U}[1]{\protect\rule{.1in}{.1in}}
\newtheorem{theorem}{Theorem}

\newtheorem{lemma}[theorem]{Lemma}

\begin{document}
\title{Geometry of the order-disorder surface of the mean-field square lattice Ising model with up to third-neighbor interactions}
\author{Rodolfo Subert}
\affiliation{Institute AMOLF, Science Park 104, 1098XG Amsterdam, the Netherlands}
\author{Bela M. Mulder}
\affiliation{Institute AMOLF, Science Park 104, 1098XG Amsterdam, the Netherlands}
\affiliation{Institute for Theoretical Physics, Utrecht University, the Netherlands}
\date{\today}

\begin{abstract}
We revisit the field-free Ising model on a square lattice with up to third-neighbour (nnnn) interactions, also known as the $J_{1}$--$J_{2}$--$J_{3}$ model, in the mean-field approximation. Using a systematic enumeration procedure, we show that the region of phase space in which the high-temperature disordered phase is stable against all modes representing periodic magnetisation patterns up to a given size is a convex polytope that can be obtained by solving a standard vertex enumeration problem. Each face of this polytope corresponds to a set of coupling constants for which a single set of modes, equivalent up to a symmetry of the lattice, bifurcates from the disordered solution. While the structure of this polytope is simple in the halfspace $J_{3}>0$, where the nnnn-interaction is ferromagnetic, it becomes increasingly complex in the halfspace $J_{3}<0$, where the antiferromagnetic nnnn-interaction induces strong frustration. We characterize a few salient properties of these `disorder polytopes' in terms of the geometry of the space of contributing modes. We then consider the limit $N\rightarrow\infty$ giving a closed form description of the order-disorder surface in the thermodynamic limit, which shows that for $J_3 <0$ the emergent ordered phases will have a `devil's surface'-like mode structure. Finally, using Monte Carlo simulations, we show that for small periodic systems the mean-field analysis correctly predicts the dominant modes of the ordered phases that develop for coupling constants associated with the centroid of the faces of the disorder polytope.
\end{abstract}
\maketitle

\section{Introduction}

A few years back Jacobs et al.\ \cite{Jacobs2015}, inspired by the advances in creating nanoparticles that interact highly specifically by leveraging the extreme selectivity of base-pairing interaction in DNA, introduced the notion of self-assembling systems with `addressable complexity', i.e.\ the creation of regular structures in which one has full control over the spatial arrangement of different particle types. Stylized prototypes of such systems are multicomponent lattice gases with isotropic interactions in which one is free to choose the strength, sign, selectivity and range(s) of the interparticle interactions. 

Arguably the simplest system of this type is the equal mole fraction binary lattice gas, which can be mapped onto the field free (= equal chemical potential) Ising model. If only nearest neighbour (nn) interactions with coupling constant $J_1$ are taken into account the results depend strongly on the underlying lattice structure. On the triangular lattice, when $J_1 >0$ we obtain a homogeneous ferromagnetic low-temperature phase (F), corresponding to a complete demixing of the particles, while for $J_1 <0$  no long range order develops and the system is caught in a finite-entropy ground state \cite{Wannier1950Antiferromagnetism.Net}. The square lattice, however, is bipartite and hence not frustrated by a $J_1 <0$ coupling, and exhibits a regular anti-ferromagnetic (AF) checkerboard phase at low temperatures. 

Thus, if one wishes to observe more complex ordering patterns on the square lattice, longer-ranged interactions are required, and specifically those that introduce frustration, effectively preempting the period-2 repeat of the AF state. Hence, starting in the 70's of the previous century, a long line of authors has studied the so-called frustrated Ising model obtained by introducing anti-ferromagnetic ($J_2 <0$) next-nearest-neighbour (nnn) interactions on the square lattice \cite{Nightingale1977Non-universalitySystems,Swendsen1979Monte2,Oitmaa1981TheInteractions,Binder1980PhaseInteractions,Landau1980PhaseInteractions,Landau1985PhaseCouplings,Moran-Lopez1993First-orderInteractions}, with more recent work appearing in the past decade or so \cite{dosAnjos2008PhaseLattice,Kalz2008PhaseInteractions,Kalz2011AnalysisLattice}. As this type of interaction penalizes equal spins across the diagonal of each square unit cell, it frustrates the nn-interactions independently of their sign.

However, increasing the range of interactions even further allows the degree of frustration also to be increased. Indeed, very general arguments suggest that in order to obtain the maximum complexity periodic patterns on a given lattice structure all symmetries implied by the point-group of the lattice must be suppressed by the interactions \cite{Tindemans2010b}. For the square lattice, this implies that also next-next-nearest-neighbor couplings (nnnn) need to be taken into account, as shown in Fig.\ \ref{fig:local}. Clearly, an anti-ferromagnetic nnnn interaction ($J_3 <0$) adds yet another level of frustration as it potentially frustrates \emph{both} the nn- and nnn- bonds independently of the sign of their interaction. In fact this latter extension was already studied actively a couple of decades back purely for its theoretical interest \cite{Kanamori1983ExactLattice,Brandt1983Ground-stateInteractions,Landau1985PhaseCouplings}.  Strikingly, interest in this nnnn-model, also known as the $J_1-J_2-J_3$ model, was revived in the past decade with a few theoretical studies appearing \cite{Kassan-Ogly2015IsingInteractions,Liu2016RoleFrustration}, as well as a significant paper showing that a model with up to third-neighbor coupling is actually relevant to understanding the magnetic origin of high-$T_{c}$ superconductivity in a class of iron chalcogenides \cite{Glasbrenner2015EffectChalcogenides}. 

Reviewing these works, however, reveals that we are far from having a complete picture of the phase behavior of these systems. Most of the effort was devoted to understanding the structure of the ground states, using either the method of inequalities introduced by Kanamori \cite{Kanamori1966MagnetizationSystem} or direct enumeration. These analyses are, however, all limited by implicit or explicit assumptions on the size of the repeating patterns considered. Characteristically, Landau and Binder \cite{Landau1985PhaseCouplings} remark \textquotedblleft\textit{Since the phase diagram is expected to be very complicated ("devil's staircase" of phases), no attempt to include these phases has been made}\textquotedblright. Where the behavior at finite temperature is concerned, the main tool has been Monte Carlo simulations, but again the attention was mostly devoted to the nature of the transitions towards certain specific states, or to the behavior in response to external fields.

Driven by the question to what extent one can `design' specific magnetisation patterns on the square lattice, our aim here is to provide a fresh perspective on the phase behavior of the field-free nnnn-model in a way that systematically allows the consideration of phases of increasing complexity. We do this in the framework of mean-field theory, which allows us to exactly formulate the criteria if and when the high-temperature disordered phase becomes unstable to magnetization modes belonging to periodicities with increasing unit cell size $N$. This analysis reveals that the region in phase space where the disordered phase is stable is a convex polytope whose complexity increases as we increase $N$. Each of the faces of this polytope defines the values of the coupling constants for which a specific equivalence class of magnetization modes is spontaneously excited. We probe the structure of this polytope as a function of the unit cell size of the periodicities included, which provides a fingerprint of the complexity of the predicted phase space.  On the basis of this analysis, we are able to analytically pass to the limit $N\rightarrow\infty$ to give a closed form description of the order-disorder surface in the thermodynamic limit. This shows that in the strongly-frustrated region of phase space $J_3 <0$, the mean-field theory predicts a `devil's surface'-like structure for the modes developing from the disordered phase, in which in an arbitrarily small neighborhood of any set of coupling parameters one can find phases of arbitrary spatial complexity becoming stable. 

While the mean field results are quantitatively at best a severe approximation to the true phase boundaries, its predictions regarding the possible symmetry breaking patterns, however, are potentially more robust. We explore this latter premise by performing MC simulations with the appropriate finite periodic boundary conditions along rays in phase space, corresponding to decreasing temperature at fixed coupling constants, that pass through the centers of the predicted mode instability faces. These show that the mean-field analysis consistently correctly predicts the dominant mode first appearing in the ordered region in the cases considered.

The structure of the paper is as follows: In Section \ref{sec:model} we set up the model. The mean field treatment is discussed in Section \ref{sec:mft}. The bifurcation analysis is presented in Section \ref{sec:bifurcation}, which introduces our main object of interest, the disorder polytope. In Section \ref{sec:geometry} we first discuss the phenomenology of the disorder polytope (Section \ref{sec:phenomenology}), then discuss some of its specific features (Section \ref{sec:specific-features}), and finally take the limit $N\rightarrow\infty$ (Section \ref{sec:limit}) leading to our major result, the prediction of the full order-disorder surface. Finally, in Section \ref{sec:simulations} we show using Monte Carlo simulations that for finite $N$, implemented through periodic boundary conditions, the mean-field analysis correctly predicts the bifurcating modes. 

\section{Model}
\label{sec:model}
\begin{figure}[ptb]
\centering
\includegraphics{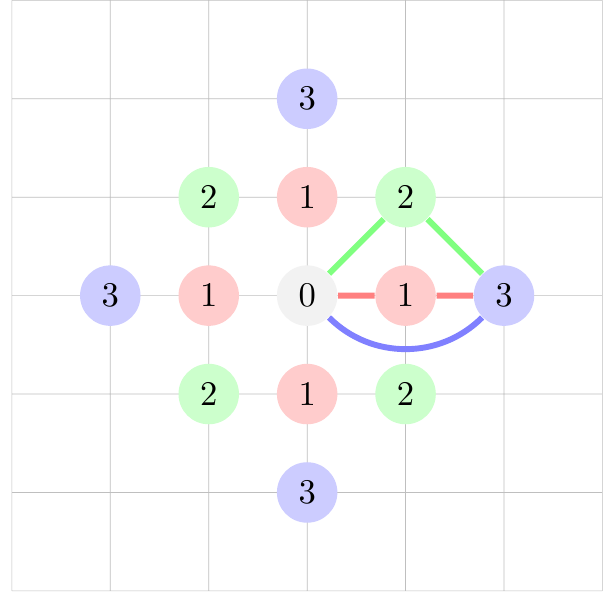} \caption{The interaction
neighborhoods of the origin site ($0$: grey site) in the range-3 Ising model
on the square lattice. $\mathrm{N}_{1}$: red sites, $\mathrm{N}_{2}$: green
sites, $\mathrm{N}_{3}$: blue sites. An anti-ferromagnetic nnnn-bond (blue line) between two sites, frustrates the spin arrangement both along the shortest nn-paths (red lines) and nnn-paths (green lines) that connect them.}%
\label{fig:local}%
\end{figure}

We consider the 2-dimensional square lattice $\mathrm{L}=\left\{
\mathrm{z}=\left(  z^{1},z^{2}\right)  |z^{1},z^{2}\in\mathbb{Z}\right\}  .$
Throughout, we will lower case roman letters to denote sites of the lattice,
and capital roman letters to denote sets of sites. We also make use of the
fact that the square lattice forms a group under vector addition, which is
generated by the basis vectors $\mathrm{e}_{1}=\left(  1,0\right)  $ and
$\mathrm{e}_{2}=\left(  0,1\right)  $ and can be equipped with an inner
product $\left\langle \mathrm{z},\mathrm{z}^{\prime}\right\rangle
=z^{1}z^{1\prime}+z^{2}z^{2\prime}.$ The sites of the lattice are occupied by
Ising spins $\sigma_{\mathrm{z}}\in\left\{  -1,1\right\}  .$ To denote a spin
configuration on a set of sites $\mathrm{C,}$ we use the notation
$\sigma_{\mathrm{C}}.$ We define the \emph{range} $r\left(  \mathrm{z,z}%
^{\prime}\right)  $ between two distinct sites as the index of the Euclidean
distance $\left\vert \mathrm{z-z}^{\prime}\right\vert $ in the ordered list of
distances between sites of the lattice, with $r=1$ denoting nearest neighbours
($\left\vert \mathrm{z-z}^{\prime}\right\vert =1$), $r=2$ next nearest
neighbours ($\left\vert \mathrm{z-z}^{\prime}\right\vert =\sqrt{2}$), $r=3$
next next nearest neighbours ($\left\vert \mathrm{z-z}^{\prime}\right\vert
=2$) and so on. We focus on the field-free range 3 Ising model, defined by the
Hamiltonian%
\begin{equation}
\mathcal{H}\left(  \sigma_{\mathrm{L}}\right)  =-J_{1}\sum_{r\left(
\mathrm{z,z}^{\prime}\right)  =1}\sigma_{\mathrm{z}}\sigma_{\mathrm{z}%
^{\prime}}-J_{2}\sum_{r\left(  \mathrm{z,z}^{\prime}\right)  =2}%
\sigma_{\mathrm{z}}\sigma_{\mathrm{z}^{\prime}}-J_{3}\sum_{r\left(
\mathrm{z,z}^{\prime}\right)  =3}\sigma_{\mathrm{z}}\sigma_{\mathrm{z}%
^{\prime}}, \label{eq:H}%
\end{equation}
where the minus sign in front of the \emph{coupling constants} $J_{1},J_{2}$
and $J_{3}$ is conventional. Further on, we will make regular use of the the
range $r$ neighborhoods of the origin
\begin{align}
\mathrm{N}_{1}  &  =\left\{  \mathrm{e}_{1},-\mathrm{e}_{1},\mathrm{e}%
_{2},-\mathrm{e}_{2}\right\}, \\
\mathrm{N}_{2}  &  =\left\{  \mathrm{e}_{1}+\mathrm{e}_{2},-\mathrm{e}%
_{1}-\mathrm{e}_{2},\mathrm{e}_{1}-\mathrm{e}_{2},-\mathrm{e}_{1}%
+\mathrm{e}_{2}\right\}, \\
\mathrm{N}_{3}  &  =\left\{  2\mathrm{e}_{1},-2\mathrm{e}_{1},2\mathrm{e}%
_{2},-2\mathrm{e}_{2}\right\},
\end{align}
which we show in Figure \ref{fig:local}. 

\section{Mean field theory}
\label{sec:mft}
Our approach to understanding the phase behaviour of the model
(\ref{eq:H}) is through mean-field theory (MFT). Although MFT is a drastic
approximation, and a fortiori so in lower dimensions, it nevertheless
generically is a good guide into the possible phases a system can display, as these are to a large extent determined by universal symmetry relations (see e.g.\ \cite{Boccara1976SymetriesDordre,Toledano1987TheTransitions}).  MFT is typically formulated as a set of self-consistent equations for the single
site spin probabilities%
\begin{equation}
P_{\mathrm{z}}\left(  \sigma_{\mathrm{z}}\right)  =\frac{e^{-\beta
V_{\mathrm{z}}\left(  \sigma_{\mathrm{z}}\right)  }}{\sum_{\sigma_{\mathrm{z}%
}}e^{-\beta V_{\mathrm{z}}\left(  \sigma_{\mathrm{z}}\right)  }},
\label{eq:MFT}%
\end{equation}
where $\beta=1/k_{B}T$ is the inverse temperature and the effective mean field
$V_{\mathrm{z}}\left(  \sigma_{\mathrm{z}}\right)  $ itself depends on the
spin probabilities on each site with the spin $\sigma_{\mathrm{z}}$ interacts%
\begin{multline}
V_{\mathrm{z}}\left(  \sigma_{\mathrm{z}}\right)  =-\sigma_{\mathrm{z}%
}\left\{  J_{1}\sum_{\mathrm{n}_{1}\mathrm{\in N}_{1}}\sum_{\sigma
_{\mathrm{z+n}_{1}}}\sigma_{\mathrm{z+n}_{1}}P_{\mathrm{z+n}_{1}}\left(
\sigma_{\mathrm{z+n}_{1}}\right)  +\right. \\
\left.  J_{2}\sum_{\mathrm{n}_{2}\mathrm{\in N}_{2}}\sum_{\sigma
_{\mathrm{z+n}_{2}}}\sigma_{\mathrm{z+n}_{2}}P_{_{\mathrm{z+n}_{2}}}\left(
\sigma_{\mathrm{z+n}_{2}}\right)  +J_{3}\sum_{\mathrm{n}_{3}\mathrm{\in N}%
_{3}}\sum_{\sigma_{\mathrm{z+n}_{2}}}\sigma_{\mathrm{z+n}_{3}}%
P_{_{\mathrm{z+n}_{3}}}\left(  \sigma_{\mathrm{z+n}_{3}}\right)  \right\}.
\end{multline}
The averages over the spin values in this expression can all be succinctly
summarized using the definition of the site magnetisation%
\begin{equation}
m\left(  \mathrm{z}\right)  =\sum_{\sigma_{\mathrm{z}}}\sigma_{\mathrm{z}%
}P_{\mathrm{z}}\left(  \sigma_{\mathrm{z}}\right),  \label{eq:mdef}%
\end{equation}
which allows us to reformulate (\ref{eq:MFT}) as%
\begin{equation}
m\left(  \mathrm{z}\right)  =\frac{\sum_{\sigma_{\mathrm{z}}}\sigma
_{\mathrm{z}}e^{-W_{\mathrm{z}}\left(  \sigma_{\mathrm{z}}\right)  }}%
{\sum_{\sigma_{\mathrm{z}}}e^{-W_{\mathrm{z}}\left(  \sigma_{\mathrm{z}%
}\right)  }}, \label{eq:MFTm}%
\end{equation}
with%
\begin{equation}\label{eq:W_z}
W_{\mathrm{z}}\left(  \sigma_{\mathrm{z}}\right)  =-\sigma_{\mathrm{z}%
}\left\{  K_{1}\sum_{\mathrm{n}_{1}\mathrm{\in N}_{1}}m\left(  \mathrm{z+n}%
_{1}\right)  +K_{2}\sum_{\mathrm{n}_{2}\mathrm{\in N}_{2}}m\left(
\mathrm{z+n}_{2}\right)  +K_{3}\sum_{\mathrm{n}_{3}\mathrm{\in N}_{3}}m\left(
\mathrm{z+n}_{3}\right)  \right\}, %
\end{equation}
where we have absorbed the common positive prefactor $\beta$ into the now
dimensionless coupling constants $K_{r}=\beta J_{r}$. 

In anticipation of the further developments below, it will turn out to be convenient to consider the triplets of possible values of the coupling constants $K_{1},K_{2}$ and $K_{3}$ as a linear vector space, whose elements we will denote by bold symbols, viz.
$\mathbf{K}=\left(  K_{1},K_{2},K_{3}\right)$. To further compactify notation, we also introduce summed neighborhood magnetizations
\begin{equation}
M_{r}\left(  \mathrm{z}\right)  =\sum_{\mathrm{n}_{r}\mathrm{\in N}_{r}%
}m\left(  \mathrm{z+n}_{r}\right)
\end{equation}
and define $\mathbf{M}\left(  \mathrm{z}\right)  =\left(  M_{1}\left(  \mathrm{z}%
\right)  ,M_{2}\left(  \mathrm{z}\right)  ,M_{3}\left(  \mathrm{z}\right)
\right)$, so that $W_{\mathrm{z}}(\sigma_{\mathrm{z}})=-\sigma_{\mathrm{z}} \mathbf{K}\cdot \mathbf{M}(\mathrm{z})$, where $\cdot$ is the Euclidean innerproduct.
Using these definitions, we can simplify Eq.~(\ref{eq:MFTm}) to take on the familiar form%
\begin{equation}
m\left(\mathrm{z}\right) =\tanh{\left(\mathbf{K}\cdot \mathbf{M}(\mathrm{z})\right)}  \label{eq:MFTSCm},%
\end{equation}
which constitutes an (infinite) set of coupled non linear
self-consistency equations for the magnetizations $\left\{m\left(
\mathrm{z}\right)  \right\} _{\mathrm{z}\in \mathrm{L}}$.

\section{Bifurcation analysis}
\label{sec:bifurcation}
We do not attempt to solve Eqs.~(\ref{eq:MFTSCm}) in all generality, but focus
on understanding the\ phases that develop from the high-temperature disordered
phase upon a temperature quench. First note that infinite temperature $\left(
\beta=0\right)  $ corresponds to the origin $\mathbf{K}=0$ of the
3-dimensional phase space of the model. It is easy to see that in this point
all spins are decoupled as the effective field vanishes, and we have $m\left(
\mathrm{z}\right)  =0.$ Moreover, by the same token, the disordered state with
$m\left(  \mathrm{z}\right)=0$ for which $\mathbf{M}(\mathrm{z})=0$ is in fact a solution for any value
of $\mathbf{K}$. We now inquire at which values of
$\mathbf{K}$ Eq.\ (\ref{eq:MFTSCm}) can support a non-zero
solution. To that end we expand Eq. (\ref{eq:MFTSCm}) to first order in the
magnetisations, yielding%
\begin{equation}\label{eq:bif}%
  m\left(  \mathrm{z}\right) = \mathbf{K}\cdot \mathbf{M}(\mathrm{z}).
\end{equation}
The values of the coupling constants $\mathbf{K}$ for which this set of equations,
admits a non-zero solution defines the set of \emph{order-disorder points}, in which an ordered solution to the self-consistency equation branches off from the disordered solution.

Since $\mathbf{M}(\mathrm{z})$ (c.f.\ Eq.~(\ref{eq:W_z})) involves the magnetisation of all sites in the interaction neighborhood of $\mathrm{z}$, even in the linear approximation defining the bifurcation equation, the magnetisations of all sites remain coupled. To proceed we therefore take the Fourier transform of (\ref{eq:bif}) with respect to lattice compatible wavevectors, which generically are of the form
\begin{equation}
\mathrm{q}=2\pi\left(  \frac{j_{1}}{n_{1}},\frac{j_{2}}{n_{2}}\right)
,\;j_{i}\in\mathbb{Z},n_{i}\in\mathbb{N}^{+},
\end{equation}
to obtain%
\begin{equation}
\hat{m}\left(\mathrm{q}\right)  = \mathbf{K\cdot F}\left(
\mathrm{q}\right)\,\hat{m}\left(\mathrm{q}\right), \label{eq:bifq}%
\end{equation}
where $\mathbf{F}\left(  \mathrm{q}\right)  \equiv\left(  F_{1}\left(
\mathrm{q}\right)  ,F_{2}\left(  \mathrm{q}\right)  ,F_{3}\left(
\mathrm{q}\right)  \right)$ is the set of Fourier transforms of the
indicator functions of the neighborhood clusters defined through%
\begin{equation}
F_{r}\left(  \mathrm{q}\right)  =\sum_{\mathrm{n}_{r}\mathrm{\in N}_{r}%
}e^{-i\left\langle \mathrm{n}_{r},\mathrm{q}\right\rangle }.
\end{equation}
For the range 3 model on the square lattice, the relevant lattice neighborhood
transforms are%
\begin{align}
F_{1}\left(  \mathrm{q}\right)   &  =2\cos q_{1}+2\cos q_{2}, \label{eq:F1}\\
F_{2}\left(  \mathrm{q}\right)   &  =2\cos\left(  q_{1}-q_{2}\right)
+2\cos\left(  q_{1}+q_{2}\right), \label{eq:F2}\\
F_{3}\left(  \mathrm{q}\right)   &  =2\cos2q_{1}+2\cos2q_{2}. \label{eq:F3}%
\end{align}
An important property of these functions is that they are invariant with
respect to the point symmetry group of the lattice -- here the dihedral group
$\mathfrak{D}_{4}$, the symmetry group of a square. Let $\mathrm{G}$ be the
real unitary 2D matrix representation of $\mathfrak{D}_{4},$ then for any
element $\mathrm{g}\in\mathrm{G}$%
\begin{equation}
F_{r}\left(  \mathrm{gq}\right)  =\sum_{\mathrm{n}_{r}\mathrm{\in N}_{r}%
}e^{-i\left\langle \mathrm{n}_{r},\mathrm{gq}\right\rangle }=\sum
_{g\mathrm{n}_{r}\mathrm{\in N}_{r}}e^{-i\left\langle \mathrm{gn}%
_{r},\mathrm{gq}\right\rangle }=\sum_{\mathrm{n}_{r}\mathrm{\in N}_{r}%
}e^{-i\left\langle \mathrm{n}_{r},\mathrm{q}\right\rangle }=F_{r}\left(
\mathrm{q}\right),  \label{eq:Finv}%
\end{equation}
where we have used the fact that $g$ simply permutes the sites of the lattice
neighborhoods $\mathrm{N}_{r}.$ This implies that instead of individual modes,
it suffices to consider the equivalence classes of modes defined by the orbits
$\mathrm{Gq=}\left\{  \mathrm{gq}|\mathrm{g}\in\mathrm{G}\right\}  $. In
passing, we also note that (\ref{eq:bifq}) is in fact readily generalised to
other lattices and models with longer-ranged pair interactions, as the lattice
structure enters only through the functions $\mathbf{F}\left(  \mathrm{q}%
\right)  ,$ and increasing the range of the pair interactions simply requires
increasing the dimensionality of the phase space spanned by the
coupling-constant vectors $\mathbf{K}$.

As Eq.\ (\ref{eq:bifq}) shows, close to a bifurcation, all magnetization modes
are decoupled. Also, it is clear that the loci in phase space at which the
state with zero magnetisation becomes unstable to the mode $\mathrm{q}$ lie on
the plane $L_{\mathrm{q}}=\left\{  \mathbf{K|K\cdot F}\left(  \mathrm{q}%
\right)  =1\right\}  $. Since at infinite temperature, where $\mathbf{K=0}$,
the system is surely disordered, we infer that the disordered phase is stable
against this mode in the half-space containing the origin bounded by
$L_{\mathrm{q}},$ i.e.
\begin{equation}
H_{\mathrm{q}}=\left\{  \mathbf{K|K\cdot F}\left(  \mathrm{q}\right)
<1\right\}  . \label{eq:half}%
\end{equation}
The problem we face, however, is that are in principle an infinite number of
modes to consider. In order to tackle this problem, we choose to
systematically enumerate the potential modes, ordering them by a natural
measure of the \textquotedblleft size\textquotedblright\ of the periodicity
they represent. Each periodically repeating pattern on the lattice
$\mathrm{L}$ is characterized by two basis vectors $\mathrm{p}_{1}=\left(
p_{1}^{1},p_{1}^{2}\right)  $ , $\mathrm{p}_{2}=\left(  p_{2}^{1},p_{2}%
^{2}\right)  \in\mathbb{Z}^{2}$ \ conveniently presented in matrix form%
\begin{equation}
\mathrm{P}=\left(
\begin{array}
[c]{cc}%
p_{1}^{1} & p_{1}^{2}\\
p_{2}^{1} & p_{2}^{2}%
\end{array}
\right)  ,
\end{equation}
where we choose the order of $\mathrm{p}_{1}$ and $\mathrm{p}_{2}$ such
$\det\mathrm{P}=N>0$. It is easy to see that $N$ is just the number of sites
in the unit cell $\mathcal{U}_{\mathrm{P}}$ of the periodic pattern. We call
it the \emph{index} of the periodicity, following the mathematical
nomenclature that associates it with the size of the quotient
group $\mathrm{L/P}$ when $\mathrm{P}$ is interpreted as a subgroup of
$\mathrm{L}$ \cite{Dummit2004AbstractAlgebra}. In Appendix \ref{app:periodic}
we review the construction of periodic patterns on L, their corresponding
discrete Brillouin zones  $\widehat{\mathcal{U}}_{_{\mathrm{P}}}$, and their
enumeration. An important result is that the structure of the set%
\begin{equation}
\widehat{\mathcal{U}}_{N}=%
{\displaystyle\bigcup\limits_{\left\{  \mathrm{P|}\left\vert
\widehat{\mathcal{U}}_{_{\mathrm{P}}}\right\vert =N\right\}  }}
\widehat{\mathcal{U}}_{_{\mathrm{P}}}=\left\{  \mathrm{q=}\frac{2\pi}%
{N}\left(  l_{1},l_{2}\right)  |0\leq l_{1},l_{2}<N\right\}  ,
\label{eq:UhatN}%
\end{equation}
which includes the wave vectors of all patterns of index $N$, is simply a
square array, and equal to the Brillouin zone of the square $N\times N$ periodicity $\mathrm{P}_{\square N}=\diag(N,N)$. For any lattice mode $\mathrm{q}$ we can define its \emph{complexity} as the smallest periodicity to which it belongs \footnote{Note that any mode $\mathfrak{q}$ compatible with periodicity $\mathrm{P}$ is trivially also compatible with periodicity $k\mathrm{P},\,k \ge 2$.}. If $\mathrm{q}=(2\pi n_1/d_1,2\pi n_2/d_2)$ with $n_i$ and $d_i$ relatively prime, then the complexity is simply given by $C(\mathrm{q})=\lcm(d_1,d_2)$.

In view of the invariance (\ref{eq:Finv}) of the
neighbourhood transforms $F_{r}\left(  \mathrm{q}\right)$, however, the proper
degrees of freedom for the mode analysis are the elements of the orbit space%
\begin{equation}
\widehat{\mathfrak{U}}_{N}\equiv\widehat{\mathcal{U}}_{N}/G=\left\{
\mathrm{Gq}|\mathrm{q}\in\widehat{\mathcal{U}}_{N}\right\}  ,
\end{equation}
which we will denote by $\mathfrak{q}$, throughout using a Gothic font to
indicate quantities related to orbits with respect to the point group
$G=\mathfrak{D}_{4}$. The set $\widehat{\mathfrak{U}}_{N}$ is commonly called the Irreducible Brillouin Zone, henceforth abbreviated as IBZ. Note that different $\mathrm{q}\in\widehat{\mathcal{U}}_{N}$ behave differently under the action of the point-symmetry group $G$, depending on their location within $\widehat{\mathcal{U}}_{N}$. Specifically, to each mode in $\mathfrak{q}\in \widehat{\mathfrak{U}}_{N}$ we can associate a \emph{multiplicity} $M\left(\mathfrak{q}\right)=|G\mathrm{q}|$, i.e. the length of the orbit under the action of $G$ to which is belongs, which will play an important role in the further analysis.  Details of the construction of the IBZ and the number of modes of complexity $N$ contained in it are discussed in Appendix \ref{app:Uhat}.

We now define our main object of interest, the region $D_{N}$ around the origin in phase space in which the
disordered solution is stable against all modes in $\widehat{\mathfrak{U}}_{N}$, which is formed by the intersection of all the pertinent half spaces of the type (\ref{eq:half})
\begin{equation}
D_{N}=%
{\displaystyle\bigcap\limits_{\mathfrak{q}\in\widehat{\mathfrak{U}}_{N}}}
H_{\mathfrak{q}}. \label{eq:Ddis_N}%
\end{equation}

Generically, the intersection of a finite number of half-spaces is a so-called convex \emph{polytope}, a bounded polyhedron \cite{Grunbaum2003ConvexPolytopes}. Our main goal here is to understand the
structure of these \emph{disorder polytopes} and their behavior as a function of $N$. The surface of the disorder polytopes is the locus in phase space where the disordered high-temperature solution becomes unstable, which we will call the \emph{order-disorder surface}. Note that not all modes in $\widehat{\mathfrak{U}}_{N}$ necessarily contribute a face to $D_{N}$: These \textquotedblleft faceless\textquotedblright\ modes are preempted by other modes whose instability surface lies closer to the origin. The problem of determining the structure of a polytope from the set of defining half-spaces is known as the \emph{vertex enumeration problem}. Intriguingly, the computational complexity of the vertex enumeration problem in its most general form is as yet undecided \cite{Reimers2014PolynomialBranch-width}. However, several well-developed
algorithms exist that are both polynomial in time and memory when the polytopes are known to be bounded \cite{Avis2015ComparativeCodes}. 

\begin{figure}
    \centering
    \includegraphics[width=\textwidth]{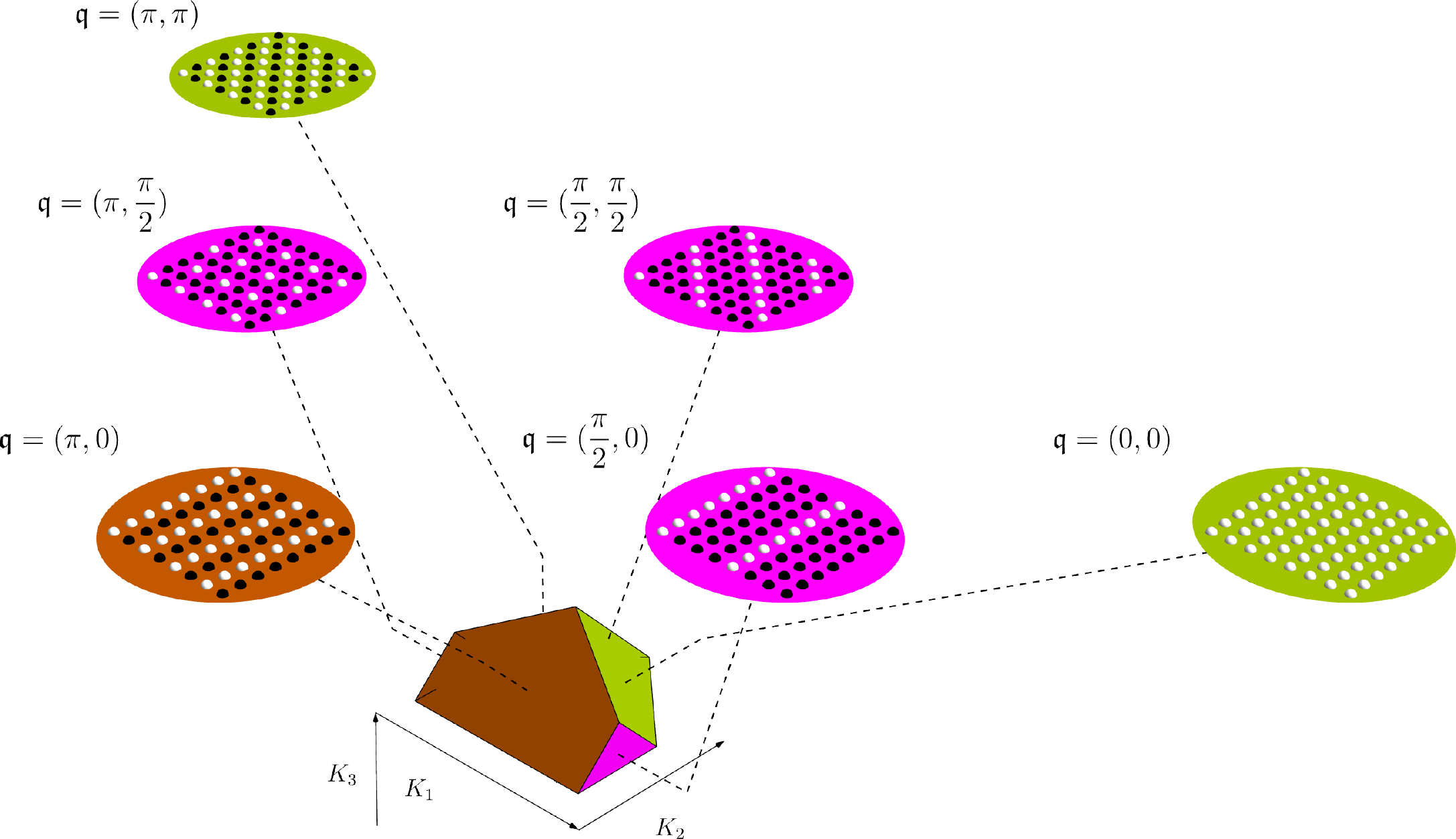}
    \caption{The order-disorder surface of the nnnn-Ising model for modes with periodic unit cell with size $N=4$ in the space of coupling dimensionless coupling constants $(K_1,K_2,K_3)$. Each of the faces of this polytope is labelled by the wavevector $\mathfrak{q}=2\pi\left(\frac{i}{6},\frac{j}{4}\right)$ of the mode in the Irreducible Brillouin Zone that becomes unstable at this face, and the corresponding periodic pattern of magnetisations is visualized.}
    \label{fig:overview}
\end{figure}

In Figure \ref{fig:overview} we illustrate the relationship between the faces of order-disorder surface, the boundary of the polytope $D_4$, the modes $\mathfrak{q}\in\widehat{\mathfrak{U}}_{N}$ in the IBZ which become unstable at these faces, and the periodic magnetisation patterns that these modes represent. 

Ultimately, we are of course interested in the limit $N\rightarrow\infty$, where all restrictions on the periodicity of the bifurcation modes is lifted, to obtain the full domain of stability of the disordered phase, i.e.\
\begin{equation}
D_{\infty}=\lim_{N\rightarrow\infty}D_{N}. \label{eq:D}%
\end{equation}
We will show how $D_{\infty}$ can be constructed and how the finite $N$ ``approximations'' approach this limit from below.
\section{The geometry of the disordered region}
\label{sec:geometry}
\subsection{Phenomenology}
\label{sec:phenomenology}
We first present an overview of the results on the disorder polytopes for finite $N$. These results were obtained using the vertex enumeration package \texttt{lrs} based on the algorithm developed by Avis and Fukuda \cite{Avis1992APolyhedra,Avis2018Mplrs:Code}, with bespoke post-processing to remove rationalization artifacts (for details see Appendix \ref{app:lrs}), and rendered with Mathematica. As we go along, we point out a number of features that are dealt with in more detail in Section \ref{sec:specific-features} below. 

\begin{figure}[htbp]
    \centering
    \subfloat[$D_4$]{\includegraphics[width=0.4\textwidth]{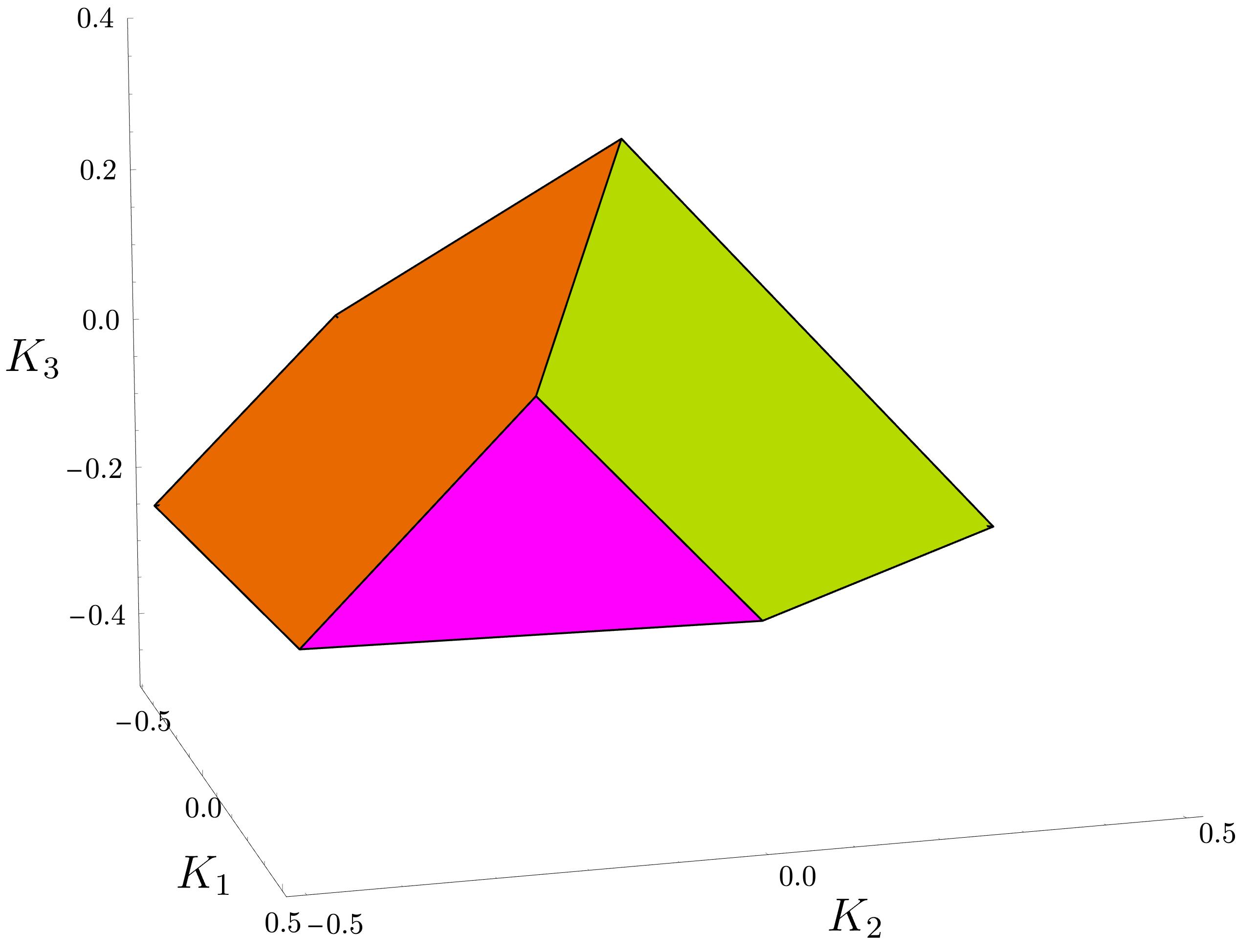}}
    \hfill
    \subfloat[$D_5$]{\includegraphics[width=0.4\textwidth]{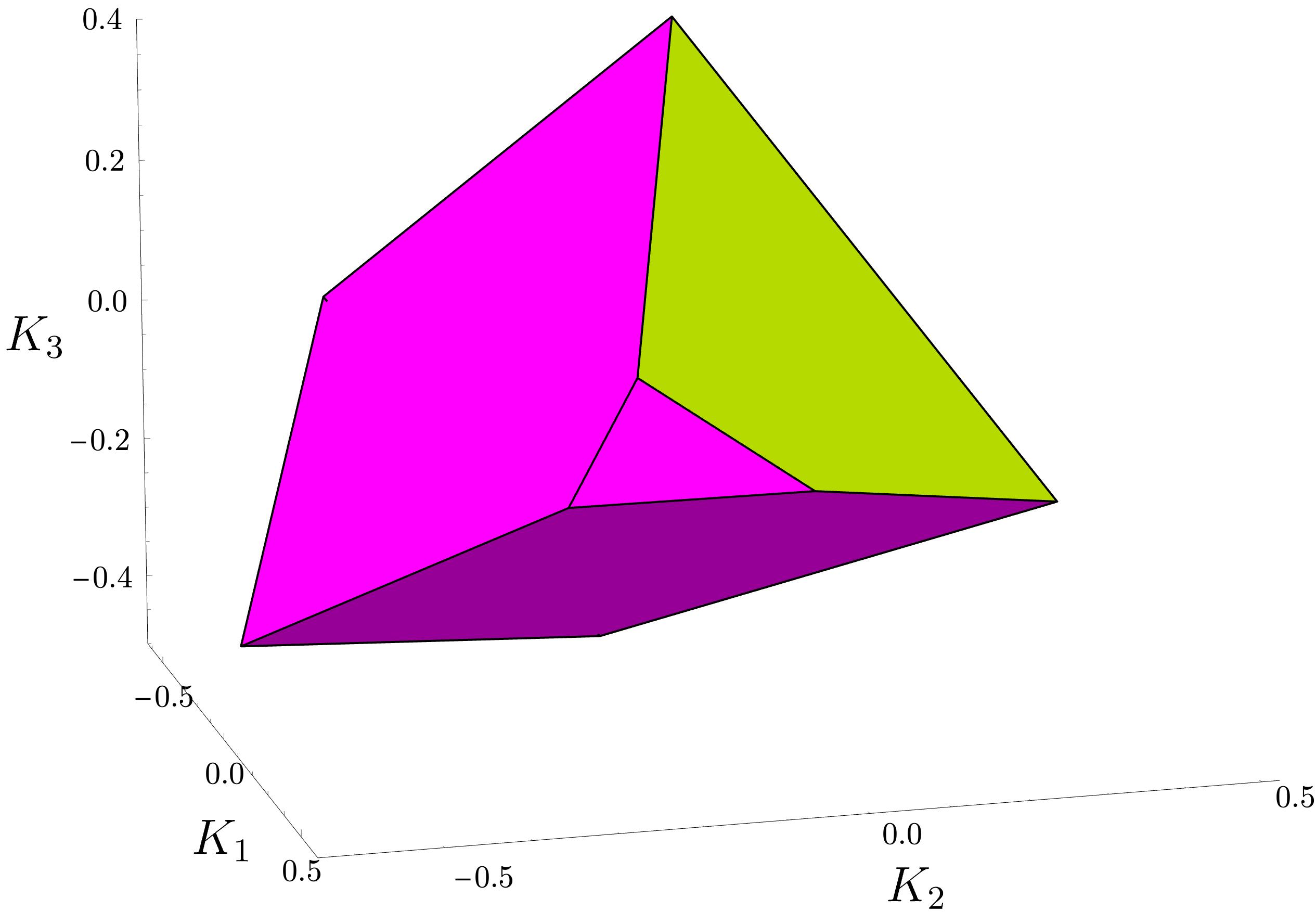}}
    \\
    \subfloat[$D_6$]{\includegraphics[width=0.4\textwidth]{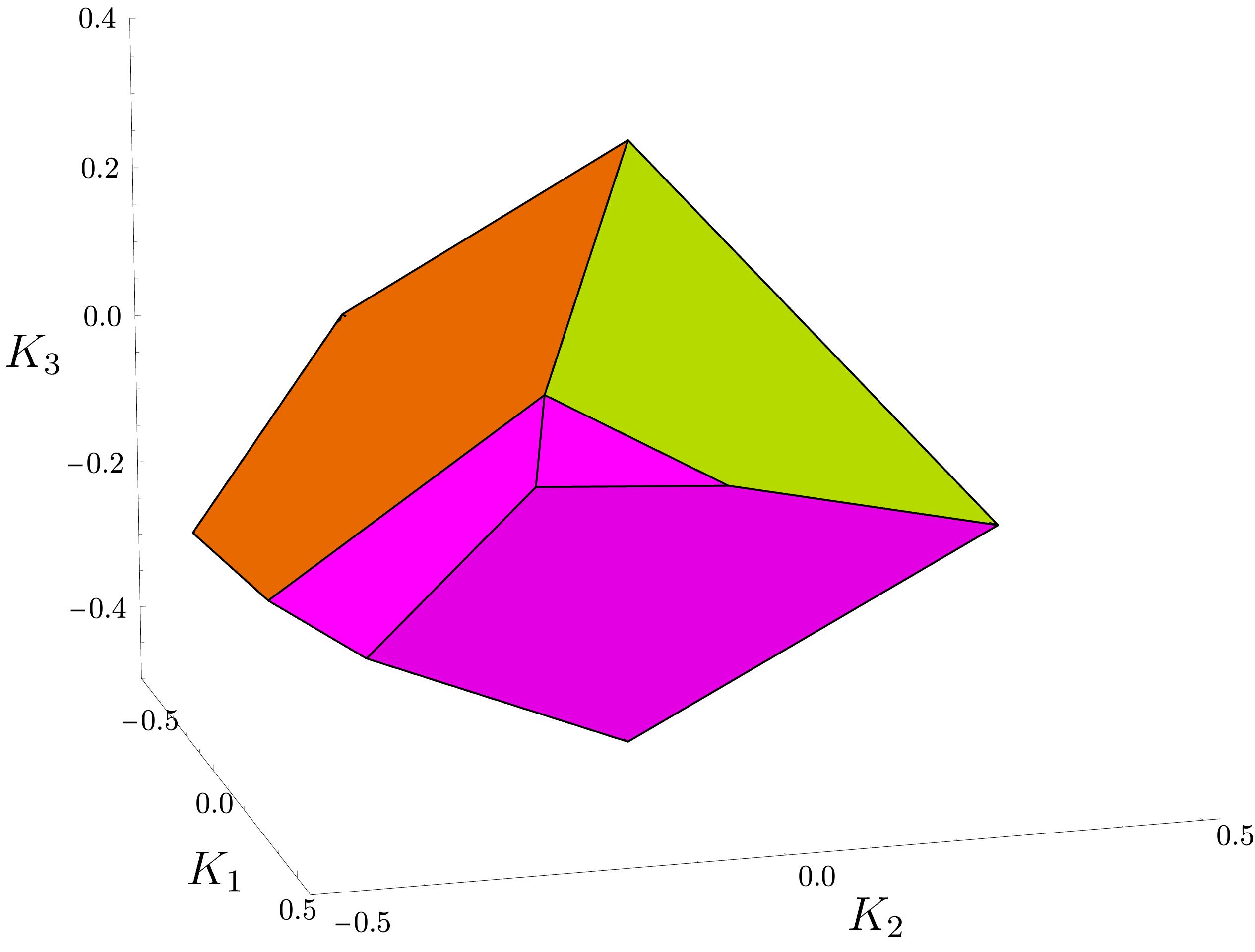}}
    \hfill
    \subfloat[$D_7$]{\includegraphics[width=0.4\textwidth]{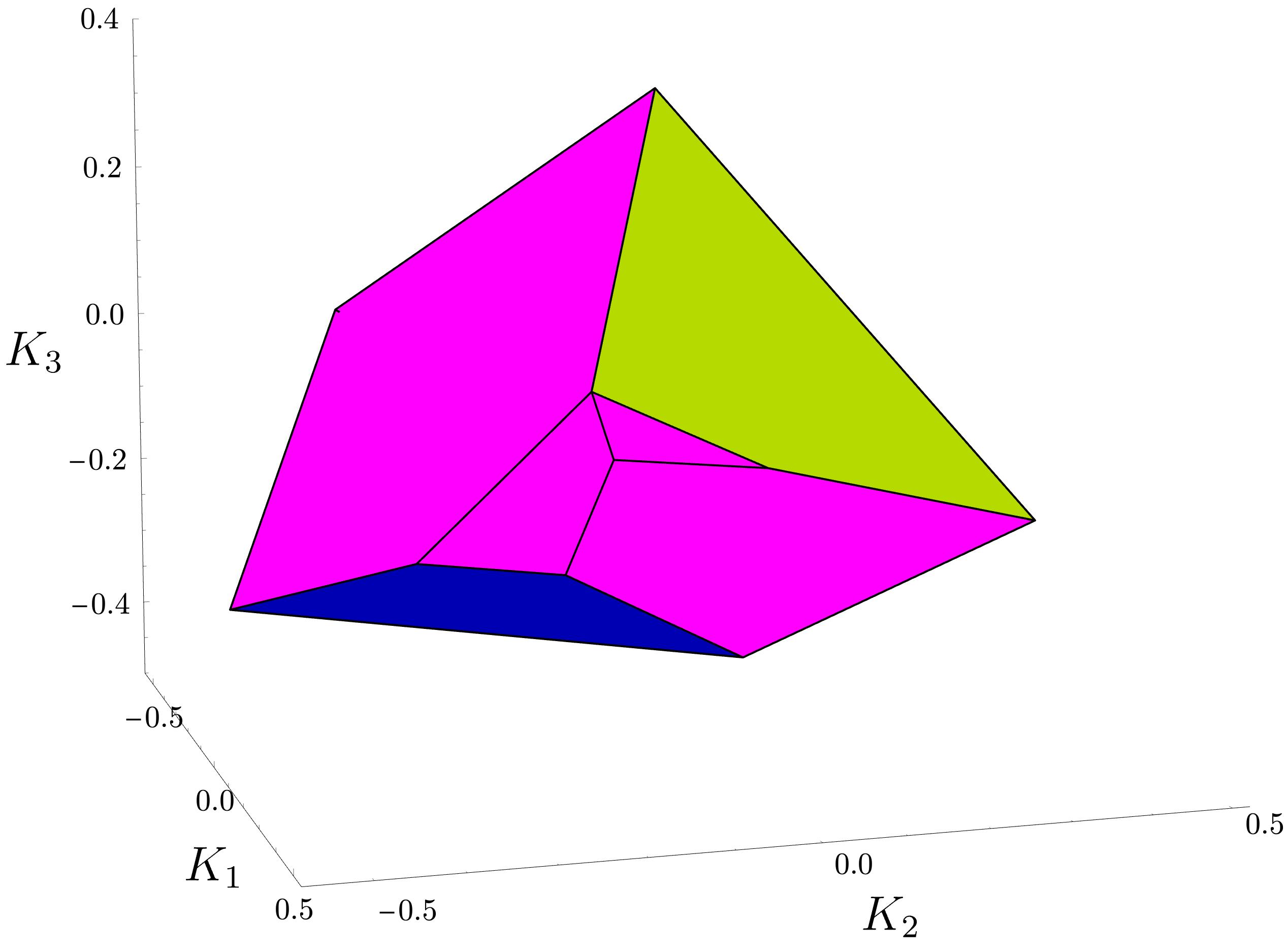}}
    \\
    \subfloat[$D_8$]{\includegraphics[width=0.4\textwidth]{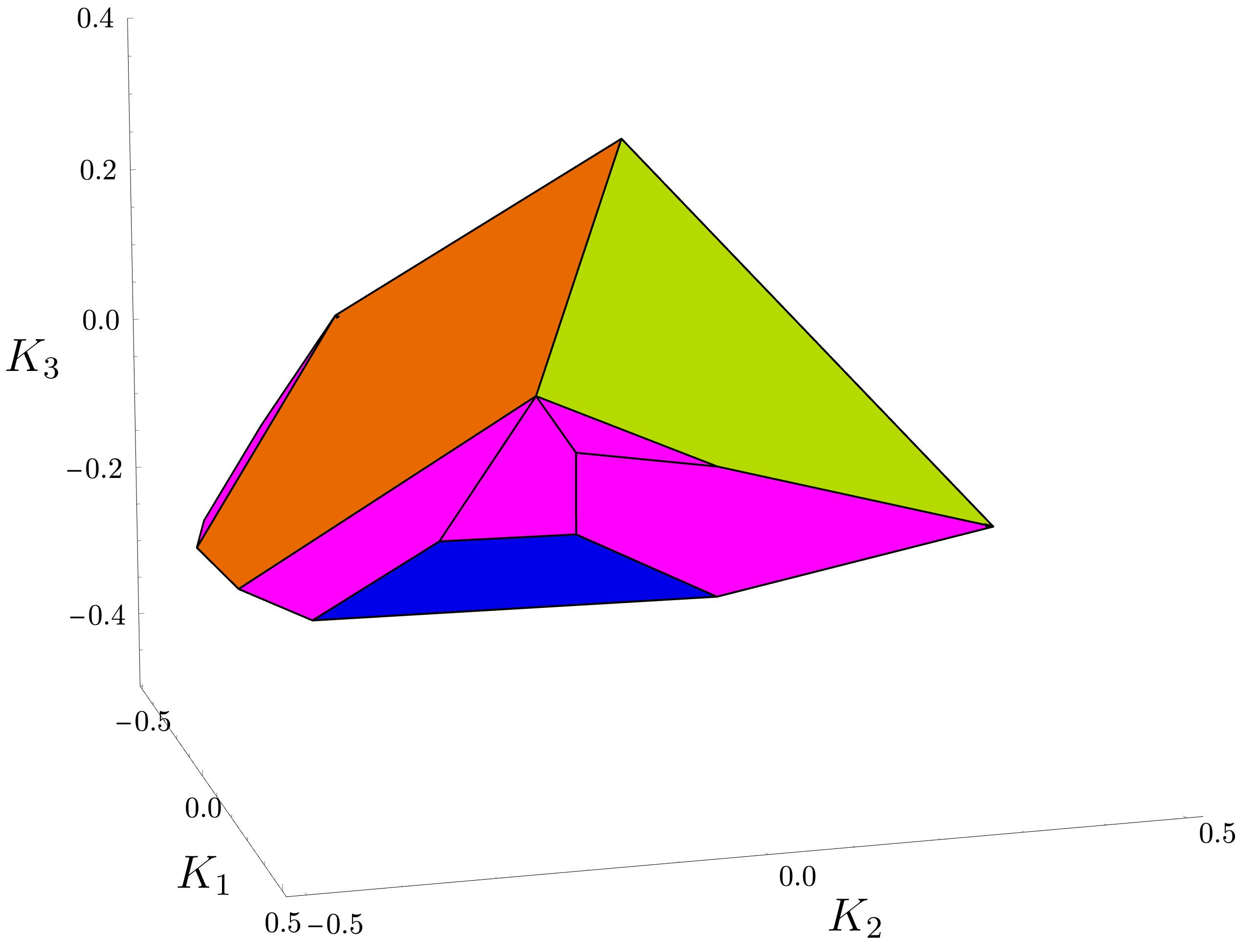}}
    \hfill
    \subfloat[$D_9$]{\includegraphics[width=0.4\textwidth]{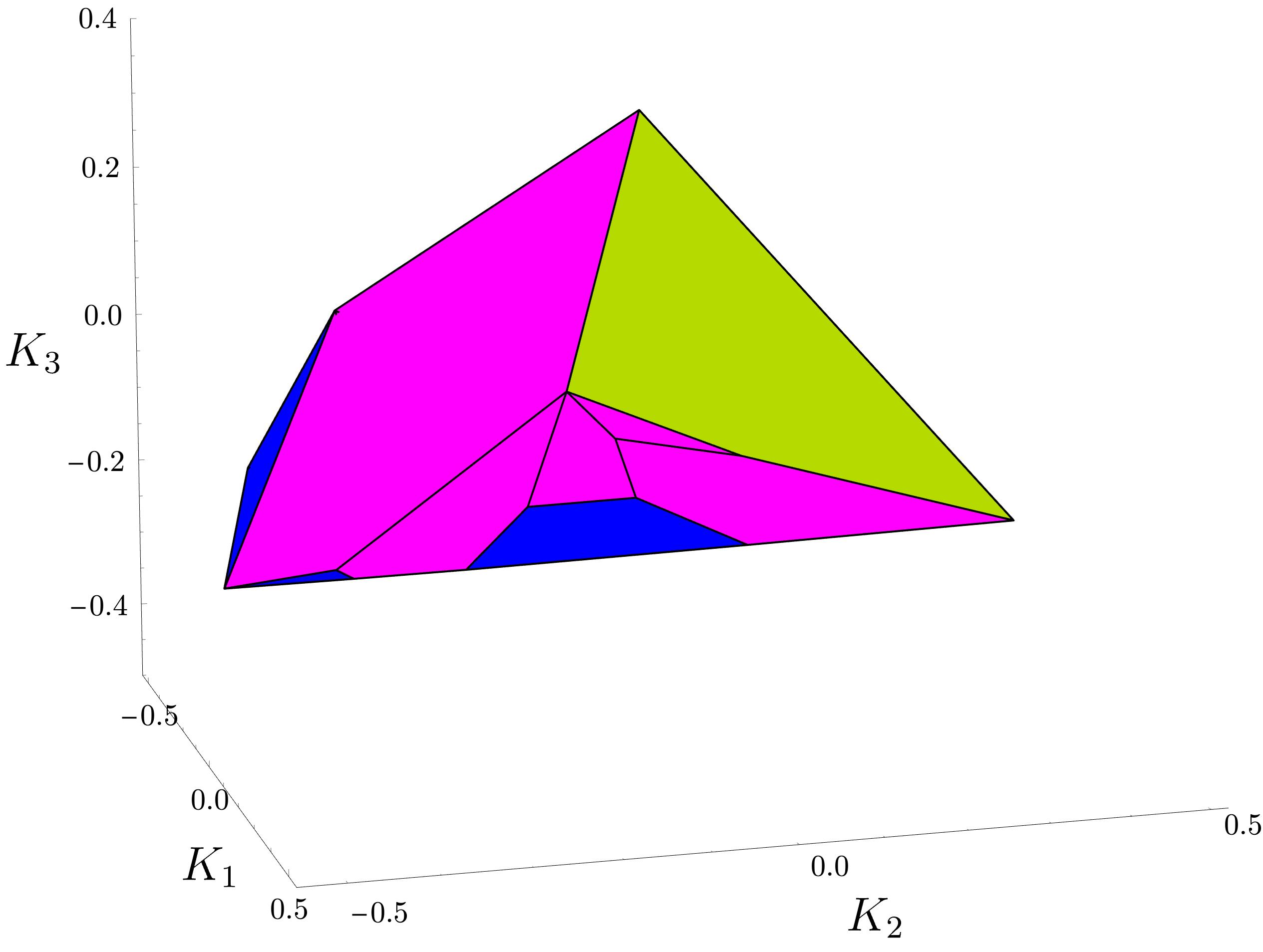}}
    \caption{The disorder polytopes $D_4$ through $D_9$. Faces are color coded for the multiplicity $M$ of the associated unstable mode: $M=1$: citrus, $M=2$: tawny, $M=4$: purple , $M=8$: blue.}
    \label{fig:D4-D9}
\end{figure}

We start off by noting that $D_1$, $D_2$ and $D_3$ are unbounded convex polyhedra, as they lack the requisite number of constraints to create a bounded domain, and we therefore do not display them.  In Figure \ref{fig:D4-D9} we show the disorder polytopes $D_4$ through $D_9$. Throughout, we will use a color code to indicate the multiplicity of the mode corresponding to each face of the poylytope: $M=1$: citrus, $M=2$: tawny, $M=4$: purple , $M=8$: blue.Two features immediately stand out. First, the polytopes with even $N$ appear symmetric upon changing the sign of $K_1$, whereas those with odd $N$ are clearly asymmetric in this respect. We discuss this symmetry in Section \ref{sec:odd-even}. Secondly, the top of the polytope in the halfspace $K_3 >0$ is bounded by just three faces, which moreover appear to be the same ones for all even $N$. The geometry of the top of the disorder polytope and the associated modes are examined more closely in Section \ref{sec:major-modes}.

We also notice that as $N$ increases the difference between the successive even and odd polytopes appears to decrease. As we will show explicitly later on in Section \ref{sec:limit} this difference indeed disappears in the limit $N\rightarrow\infty$. 

\begin{figure}[htbp]
    \centering
    \subfloat[$D_{10}$]{\includegraphics[width=0.4\textwidth]{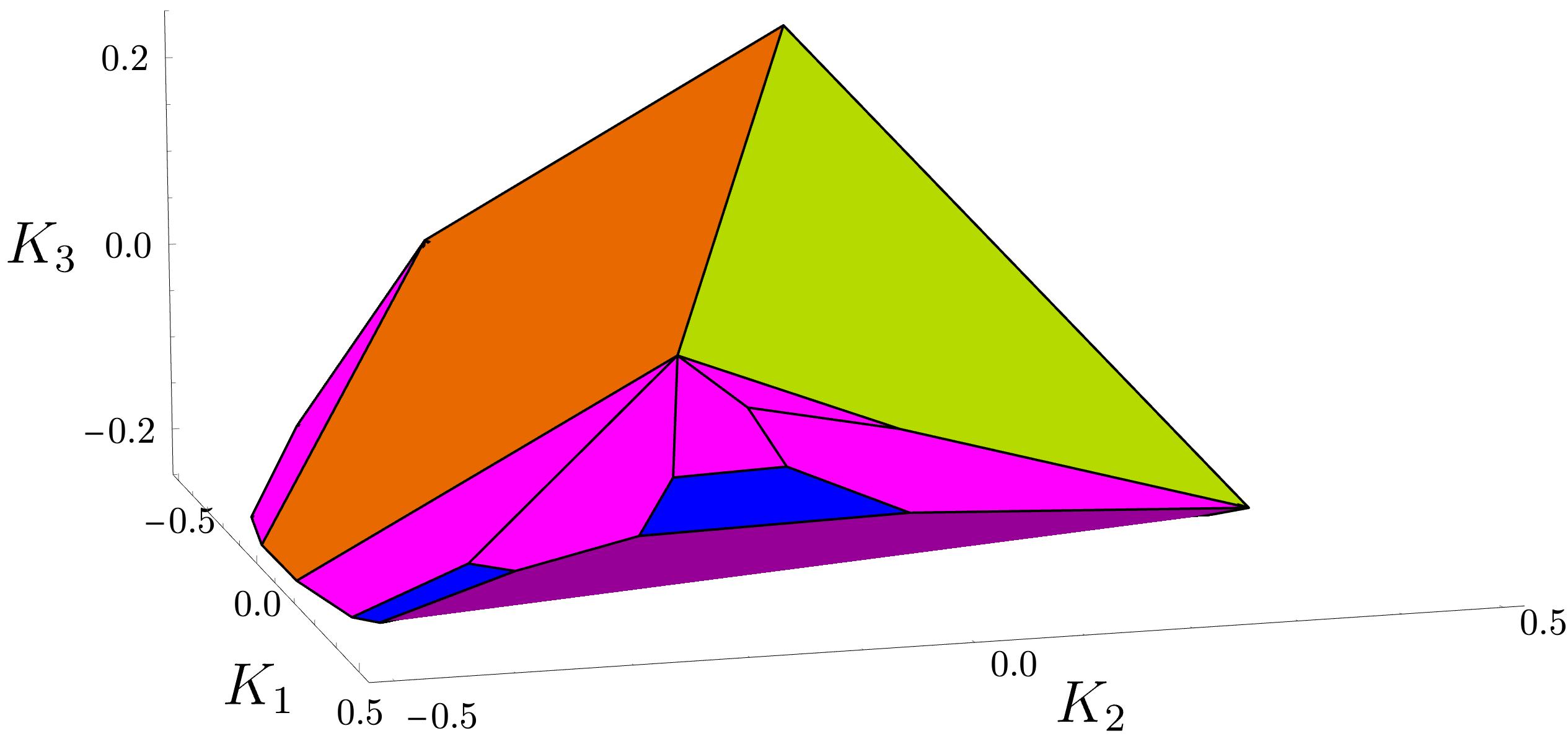}}
    \hfill
    \subfloat[$D_{12}$]{\includegraphics[width=0.4\textwidth]{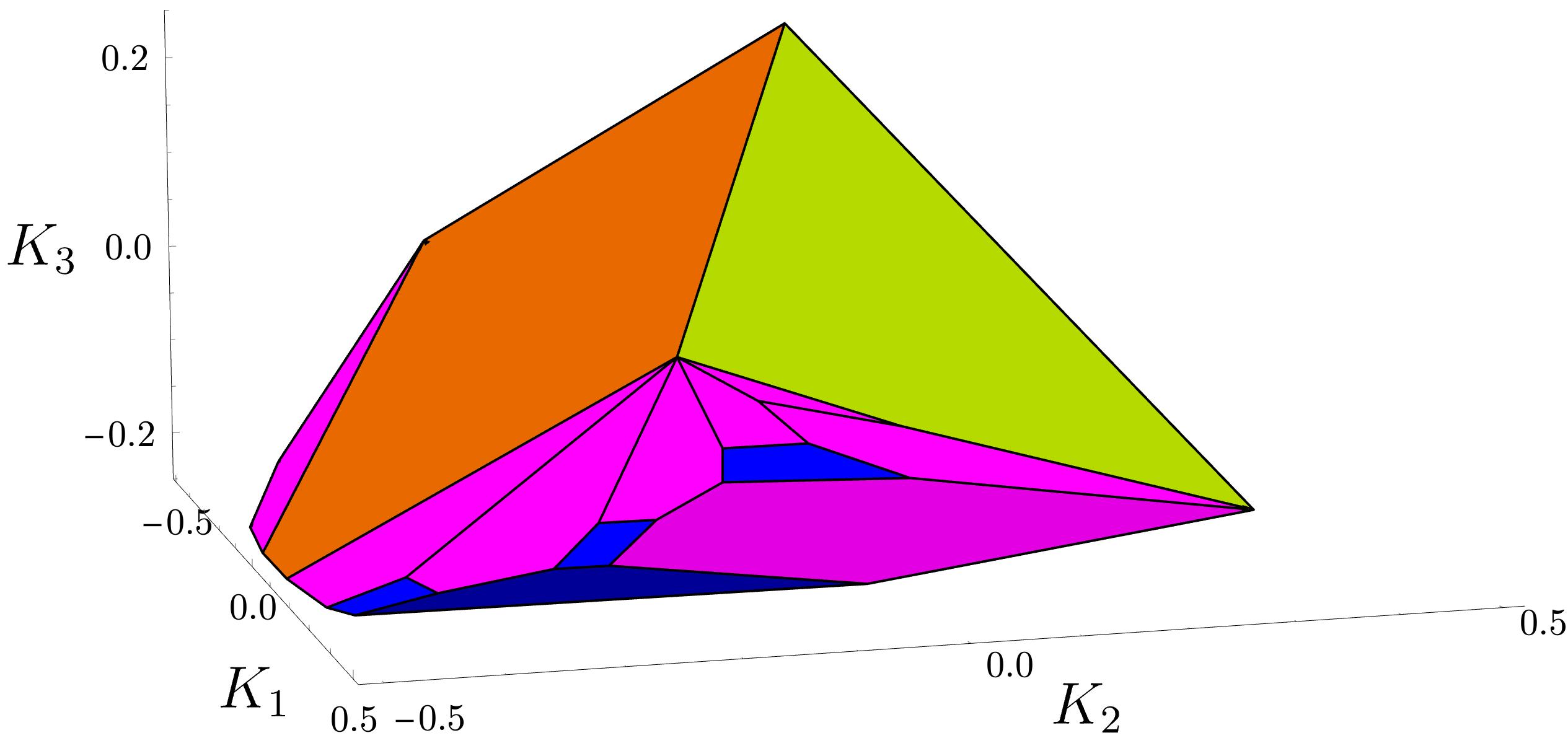}}
    \\
    \subfloat[$D_{14}$]{\includegraphics[width=0.4\textwidth]{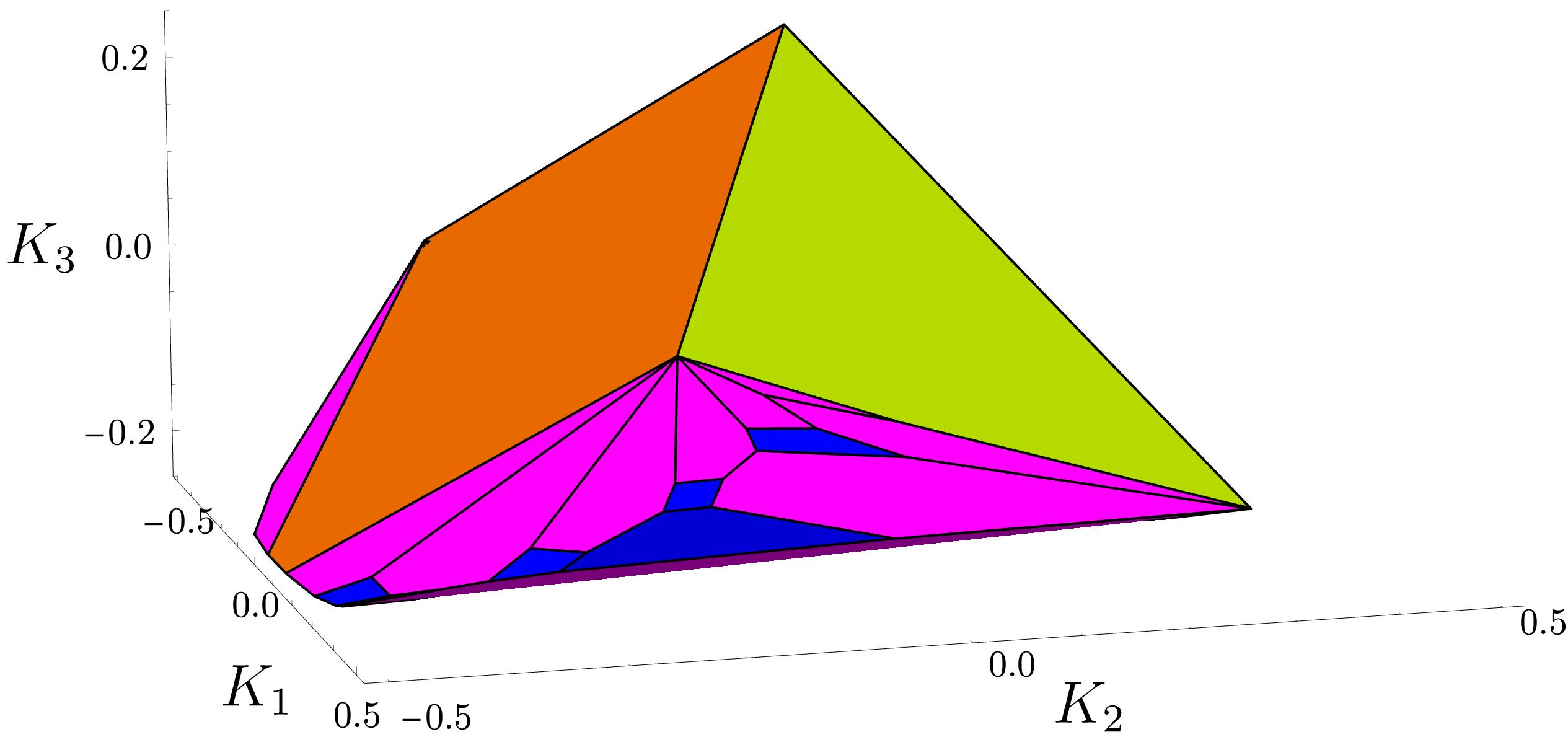}}
    \hfill
    \subfloat[$D_{16}$]{\includegraphics[width=0.4\textwidth]{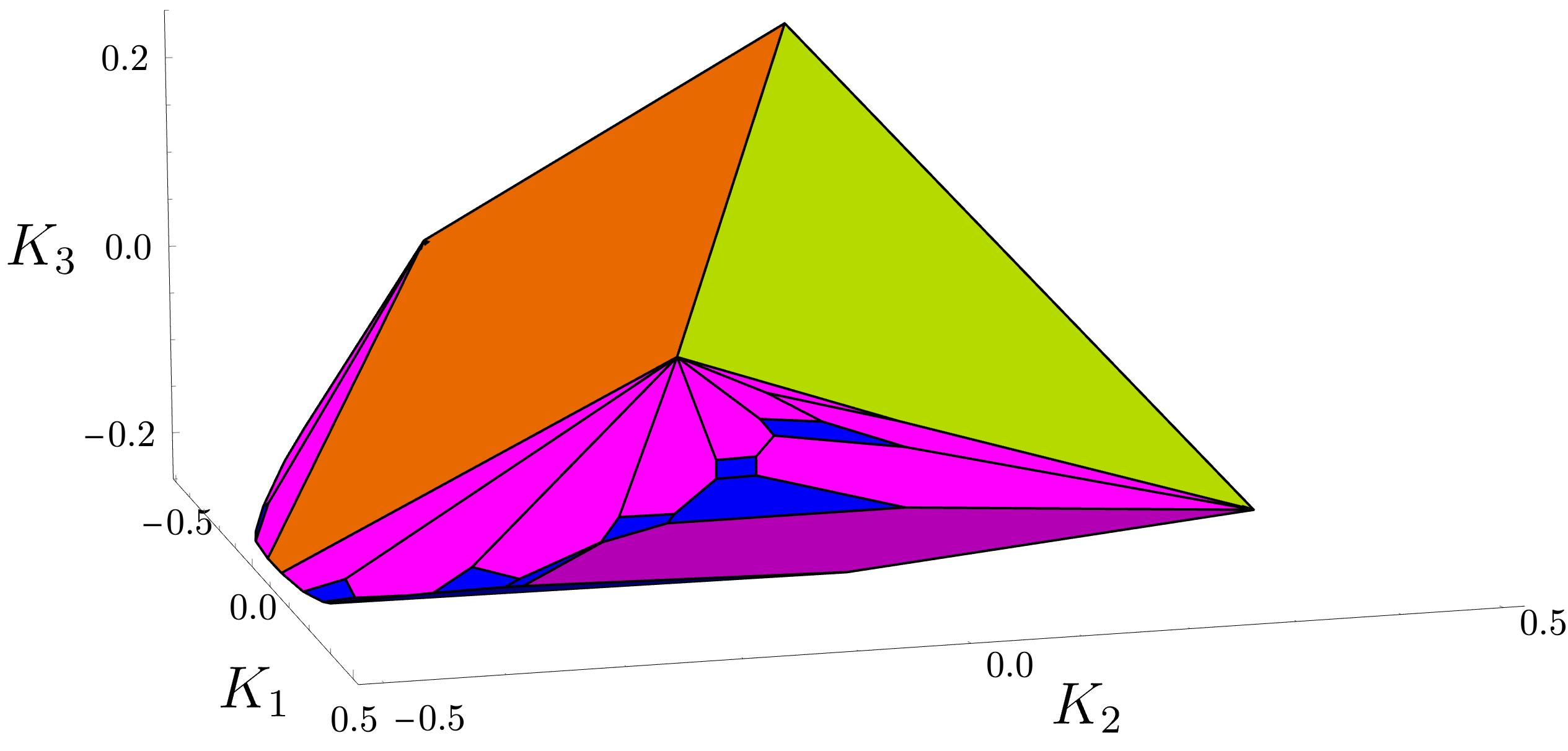}}
    \caption{The even disorder polytopes $D_{10}$ through $D_{16}$. Faces are color coded for the multiplicity $M$ of the associated unstable mode: $M=1$: citrus, $M=2$: tawny, $M=4$: purple , $M=8$: blue.}
    \label{fig:D10-D16}
\end{figure}

\begin{figure}[htbp]
    \centering
    \subfloat[$D_{16}$]{\includegraphics[width=0.5\textwidth]{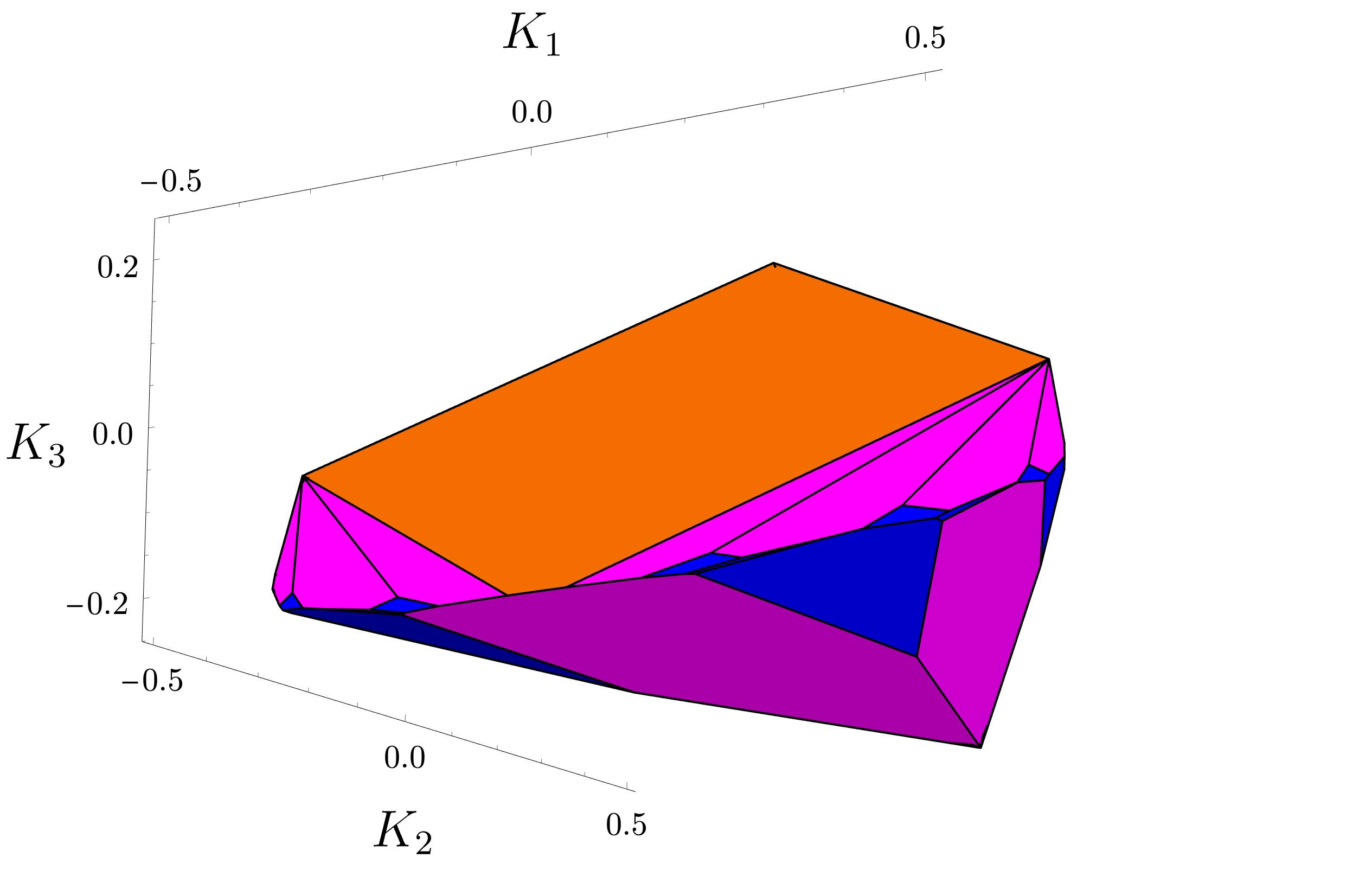}}
    \hfill
    \subfloat[$D_{32}$]{\includegraphics[width=0.5\textwidth]{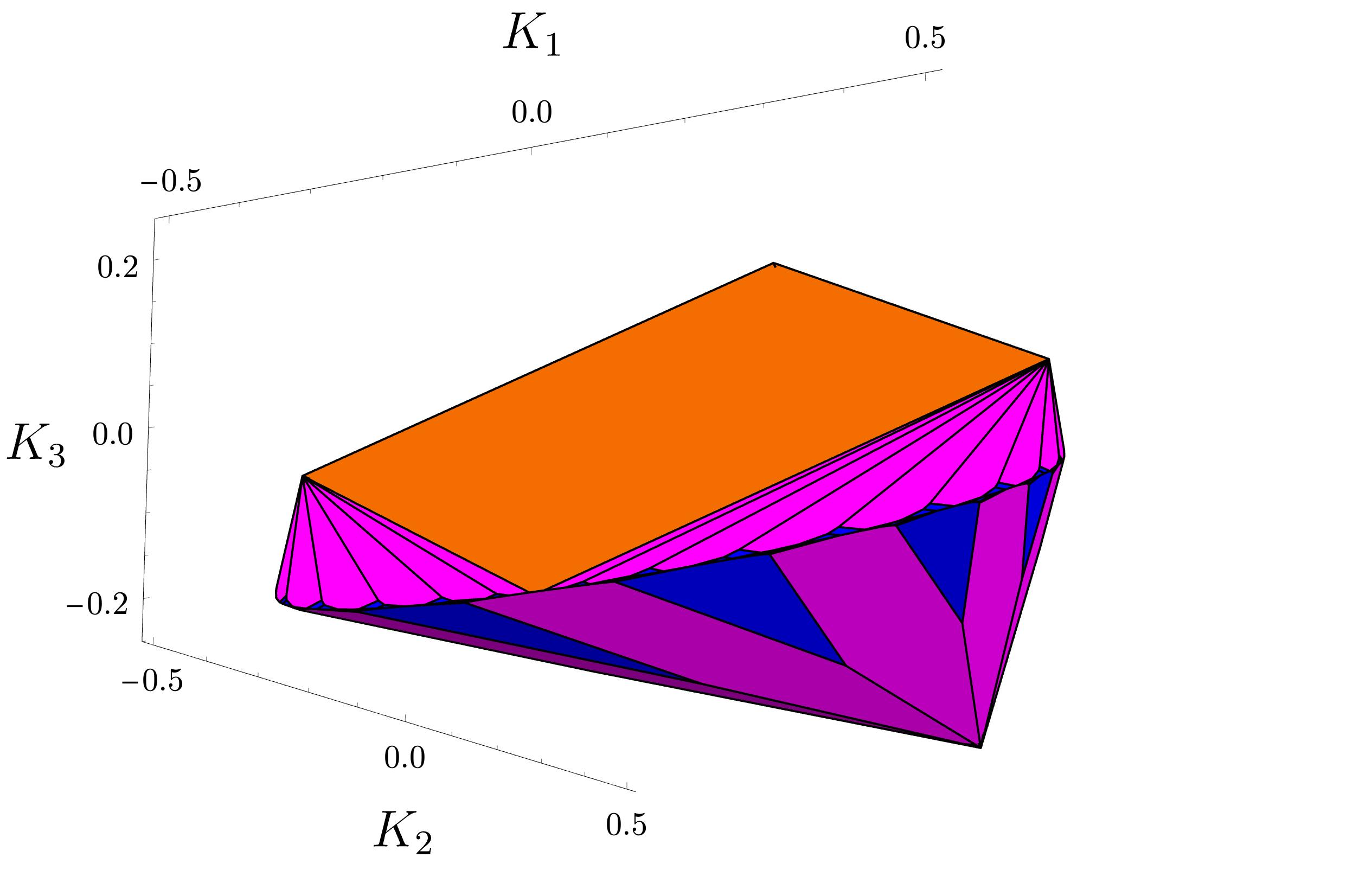}}
    \caption{``Bottom'' view of disorder polytopes $D_{16}$ and $D_{32}$, showing the  fans of striped, modulated-stripe and diagonal stripe $M=4$ modes (purple) emanating from the vertices $\mathbf{K}^S$, $\mathbf{K}^{MS}$ and $\mathbf{K}^{DS}$ respectively, as well as the $M=8$ modes (blue) that cluster around the so-called ridge. Note the decrease in area of the wedge-like $M=8$ modes that interdigitate the diagonal stripe fan as $N$ increases.}
    \label{fig:bottom}
\end{figure}

Next, in Figure \ref{fig:D10-D16} we show the even polytopes form $N=10$ to $N=16$. Again a number of features stand out. As $N$ increases, the complexity of the bottom of the polytope in the halfspace $K_3 <0$, where as we argued the system is strongly frustrated, increases. Moreover, we see a marked clustering of the faces corresponding to modes with multiplicity $M=4$ into \emph{fan}-like structures, while those belonging to modes with multiplicity $M=8$ seem to string out along a curve, which we will call the \emph{ridge}. These structures are brought into focus in Figure \ref{fig:bottom} where we show a view of $D_{16}$ and $D_{32}$ `from below' with a viewpoint on the negative $K_3$-axis. In Sections \ref{sec:fan-modes} and \ref{sec:ridge} we address the fans and ridge in more detail.

Finally, in Table \ref{tab:faceless} we list the number of faces of the disorder surface as a function $N$ compared to the maximal number of modes available, which indicates that for even $N$ a number of modes does not contribute a face to $D_N$. In Section \ref{sec:faceless} we characterise these \emph{`faceless'} modes.

\begin{table}[htbp]
    \centering
    \begin{tabular}{l|cccccccccc}
         N & 4 & 5 & 6 & 7 & 8 & 9 & 10 & 12 & 14 & 16 \\
         \hline
        \#faces & 6 &  6 & 10 & 10 & 14 & 15& 20 & 26 & 34 & 42 \\
        $|\hat{\mathfrak{U}}_N| $ & 6 & 6 & 10 & 10 & 15 & 15 & 21 & 28 & 36 & 45
    \end{tabular}
    \caption{Number of faces of the disorder polytopes as function $N$ compared to $|\hat{\mathfrak{U}}_N|$.}
    \label{tab:faceless}
\end{table}
%
%
\subsection{Specific features}
\label{sec:specific-features}
\subsubsection{Odd-even effects}
\label{sec:odd-even}
On the square lattice we can define a unique parity of each
site by defining $\left\Vert \mathrm{z}\right\Vert =\left(z_{1}+z_{2}\right) \mod 2$.  Considering Figure \ref{fig:local}, we see that the standard neighbourhood set $\mathrm{N}_{1}$ consists of sites with parity $1$, while both $\mathrm{N}_{2}$ and $\mathrm{N}_{3}$ only contain sites with parity $0$. This implies that for every solution $m_{\mathrm{z}}$ of the bifurcation equation Eq. (\ref{eq:bif}) with coupling constants $\mathbf{K}=\left(K_{1},K_{2},K_{3}\right)$ there is a solution $\bar{m}_{\mathrm{z}}=\left(  -\right)  ^{\left\Vert
\mathrm{z}\right\Vert }m_{\mathrm{z}}$ with coupling constants $\mathbf{\bar
{K}}=\left(-K_{1},K_{2},K_{3}\right)$. Fourier transforming $\bar
{m}_{\mathrm{z}}$, we find that $\mathrm{\bar{\mathrm{q}}=\mathrm{q}-(\pi},\pi)$. We also find that $F_{1}\left(\mathrm{\bar{q}}\right)=-F_{1}\left(\mathrm{q}\right)$, while $F_{2}\left(\mathrm{\bar{q}}\right)=F_{2}\left(\mathrm{q}\right)$ and $F_{3}\left(\mathrm{\bar{q}}\right)=F_{3}\left(\mathrm{q}\right)$, so that if $\mathbf{K\cdot F}\left(  \mathrm{q}\right)  =1$ then $\mathbf{\bar{K}\cdot F}\left(  \mathrm{\bar{q}}\right)=1$ and therefore also solves Eq.~(\ref{eq:bifq}). Referring to Figure \ref{fig:Uhat_infty}, we see that the mapping $\mathrm{q}\rightarrow\mathrm{q}-(\pi,\pi)$ corresponds to the reflection $r$ with respect to what we call the \emph{anti-diagonal}, the perpendicular bisector onto the hypotenuse of the the symmetry reduced Brillouin zone $\widehat{\mathfrak{U}}_{\infty}$.  We now ask under what conditions $\mathfrak{q}\in\widehat{\mathfrak{U}}_{N}\Rightarrow r\mathfrak{q}\in\widehat{\mathfrak{U}}_{N}$. As $\mathfrak{q}=\left(  2\pi\frac{i}{N},2\pi\frac{j}{N}\right)  ,0\leq j\leq i\leq\left\lfloor \frac{N}{2}\right\rfloor$, we have $r\mathfrak{q}=\left(  \frac{N-2j}{N}\pi,\frac{N-2i}{N}\pi\right)$, so that $r\mathfrak{q}\in\widehat{\mathfrak{U}}_{N}$ if and only if $N$ even, as is also illustrated in Figure \ref{fig:Uhat_odd_even}. Thus any facet of $D_{2N}$ associated with mode $\mathfrak{q}$ and normal vector $\mathbf{F}\left(  \mathfrak{q}\right)$ is paired with a facet with mode $\bar{\mathfrak{q}}$ and normal vector $\mathbf{F}\left(\mathfrak{q}\right)=\mathbf{F}\left(r\mathfrak{q}\right)$, and the whole polytope is mirror-symmetric with respect to the plane $K_{1}=0$ for all even $N$.

\subsubsection{The major modes for $K_3 >0$}
\label{sec:major-modes}
The three faces that bound the polytope in the half-space $K_3 >0$ are associated with the modes that are located at the extreme points of the IBZ  $\widehat{\mathfrak{U}}_{N}$. Defining
\begin{equation}
q^{m}_{N}=\left\lfloor \frac{N}{2}\right\rfloor \frac{2\pi}{N},%
\end{equation}
these are the modes $\mathfrak{q}_{0}=\left(0,0\right)$, $\mathfrak{q}_{1}=\left(
q^{m}_{N},0\right)$ and $\mathfrak{q}_{2}=\left(  q^{m}_{N},q^{m}_{N}\right)$. As $q^{m}_{2k}=\pi$ these facets are the same for all even $N$. In that case it is easy to see they represent the \emph{ferromagnetic}- ($\mathfrak{q}^{F}$), \emph{alternating striped}- ($\mathfrak{q}^{AS}$) and \emph{anti-ferromagnetic} ($\mathfrak{q}^{AF}$) ordering patterns respectively. A visualization of these modes can be found in Appendix \ref{app:visualization}. Also, as $q^{m}_{2k+1}=\pi\frac{2k}{2k+1}$, we see $\lim_{k\rightarrow \infty}q^{m}_{2k+1}=\pi$, so that as $N$ increases the odd top facets converge to the even ones. A direct computation of the location of the
top vertex $\mathbf{K}^{T}$ of the polytope, obtained by solving the conditions $\mathbf{K}\cdot\mathbf{F}(\mathfrak{q}^{F})=\mathbf{K}\cdot\mathbf{F}(\mathfrak{q}^{AS})=\mathbf{K}\cdot\mathbf{F}(\mathfrak{q}^{AF})=1$, then yields for even $N$ the vertex $\mathbf{K}^{T}  = \left( 0,0,\frac{1}{4}\right)$, while odd $N=2 k+1$ we have $\mathbf{K}^{T}_{odd}=\left(\frac{1}{4}\left(1+1/(2\cos\left(  \pi\frac{2k}{2k+1}\right)+1)\right),0,-1/(8\cos\left(\pi\frac{2k}{2k+1}\right)+4)\right)$. The latter, as expected, converges to $\mathbf{K}^{T}$ as $k\rightarrow\infty$.
\subsubsection{The fan modes}
\label{sec:fan-modes}
The three fans of faces shown most clearly in Figure \ref{fig:bottom} are associated with the multiplicity $M=4$ modes on the edges of the IBZ. We distinguish the modes of the form $\mathfrak{q}^{S}(i)=(2\pi i/N,0),\,i=1,\ldots,l(N)$ on the horizontal leg, which are associated with \emph{striped} ordering patterns, modes of the form $\mathfrak{q}^{MS}(i)=(\pi,2\pi i/N),\,i=1,l(N)$ on the vertical leg, which we associate with \emph{modulated-stripe} ordering patterns, and the modes on the hypotenuse of the form $\mathfrak{q}^{DS}(i)=(2\pi i/N,2 \pi i/N),\,k=1,\ldots,l(N)$, which we associate with \emph{diagonal stripe} ordering patterns, where $l(2k)=k-1$ and $l(2k+1)=k$. These modes are visualized in Appendix \ref{app:visualization}. 

We can show by explicit construction that the facets corresponding to any three successive fan modes share a common vertex, which moreover is independent of which triplet is considered. For the striped modes we find on solving $\mathbf{K}\cdot\mathbf{F}(\mathfrak{q}^{S}(i-1)=\mathbf{K}\cdot\mathbf{F}(\mathfrak{q}^{S}(i))=\mathbf{K}\cdot\mathbf{F}(\mathfrak{q}^{S}(i+1))=1$, the vertex $\mathbf{K}^{S}=(1/2,-1/4,0)$ for all $N$. The analogous calculation for the modulated-stripe modes yields for even $N$ the vertex $\mathbf{K}^{MS}=(-1/2,-1/4,0)$, consistent with the symmetry of $D_{2k}$ discussed above, while for odd $N=2 k+1$ we find $\mathbf{K}^{MS}_{odd}=\left\{\frac{1}{2} \sec \left(\frac{2 \pi  k}{2 k+1}\right),-\frac{1}{4} \sec ^2\left(\frac{2 \pi  k}{2 k+1}\right),0\right\}$, which converges to $\mathbf{K}^{MS}$ for $k\rightarrow\infty$. Finally, for the diagonal stripe modes we find $\mathbf{K}^{DS}=(0,1/2,-1/4)$ for all $N$.
Details on how these fans meet in the middle area of the bottom of the polytopes will be addressed in the following section. 

\subsubsection{The $M=8$ modes and the ridge}
\label{sec:ridge}
The modes with multiplicity $M=8$ have fewer remaining symmetries. A few examples are shown in Appendix \ref{app:visualization}. As Figure \ref{fig:bottom} suggests, the faces corresponding to these modes are directly connected to the striped- and modulated stripe fans and are clustered around an increasingly narrow quasi one-dimensional structure which we call the ridge. This structure can be characterised by considering the common vertex belonging to the faces corresponding to two successive modes along either of the legs of the IBZ and one of the interior $M=8$ modes nearest to this pair. Considering e.g.\ the pair striped modes $\left(\mathfrak{q}^{S}(i),\mathfrak{q}^{S}(i+1)\right)$ on the horizontal leg, the nearest interior mode is $\mathfrak{q}^{int}(i)=(2\pi i/N,2\pi/N)$, and we solve for $\mathbf{K}\cdot\mathbf{F}(\mathfrak{q}^{S}(i)=\mathbf{K}\cdot\mathbf{F}(\mathfrak{q}^{S}(i+1))=\mathbf{K}\cdot\mathbf{F}(\mathfrak{q}^{int}(i))=1$. For finite $N$, the resulting analytical expressions for the solution $\mathbf{K}^{\text{ridge}}(i)$ are rather unwieldy and we refrain from presenting them. However, by parameterizing $i=a N,\,a\in [0,1/2]$ we can take the limit $N\rightarrow\infty$ yielding
\begin{equation}
\label{eq:ridge}
    \mathbf{K}^{\text{Ridge}}(a)=\frac{1}{4 \cos (2 \pi  a)+\cos (4 \pi  a)+5}\left(4 \cos ^2(\pi  a),-1,-\frac{1}{2}\right).
\end{equation}
A similar analysis for the modulated stripe modes on the vertical leg, now parameterized by $i = (1/2-a)N,\,a\in [0,1/2]$ yields, as expected by the reflection symmetry in the anti-diagonal of the IBZ, the same result mirrored in the plane $K_1=0$. We also note that the ridge is a planar curve embedded in the plane $K_2=2 K_3$. For future reference we name the two end points of the ridge $\mathbf{K}^{R}_{\pm}=(\pm 2/5,-1/10,-1/20)$ and the lowest point on the curve $\mathbf{K}^{B}\equiv\mathbf{K}^{\text{Split}}(1/2)=(0,-1/2,-1/4)$. 

One also notices that the faces belonging to the diagonal stripe fan are ``split'' by wedge-shaped faces belonging to $M=8$ modes. The vertices at which this happens can be found by considering the common vertex between two subsequent diagonal stripe modes $\left(\mathfrak{q}^{DS}(i),\mathfrak{q}^{DS}(i+1)\right)$ with their common nearest interior mode $(2\pi (i+1)/N,2\pi i/N)$. Using a similar parameterization as above, i.e.\ $i=b N,\,b\in [0,1/2]$, and passing to the limit $N\rightarrow\infty$  we obtain the curve
\begin{equation}
     \mathbf{K}^{\text{Split}}(a)= \left(\frac{\cos (2 \pi  b)}{\cos (4 \pi  b)+2},0,-\frac{1}{4 (\cos (4 \pi  b)+2)}\right).
\end{equation}
However, by considering the angle between the pair of edges defined by the two pair of modes $\left(\mathfrak{q}^{DS}(i),(2\pi (i+1)/N,2\pi i/N)\right)$ and $\left((2\pi (i+1)/N,2\pi i/N),\mathfrak{q}^{DS}(i+1)\right)$, one can show that the surface area of these wedge-like $M=8$ faces vanishes in the limit $N\rightarrow\infty$.
\subsubsection{The faceless modes}
\label{sec:faceless}
The so-called faceless modes for even $N$ are all located on the \emph{anti-diagonal} that runs from the vertex $\mathfrak{q}^{AF}=(\pi,0)$ to the midpoint of the hypotenuse of the IBZ. These modes can generically be parameterized as $\mathfrak{q}^{AD}(\alpha) = (\pi-\alpha,\alpha),\,\alpha\in[0,\pi/2]$. It follows that $F(\mathfrak{q}^{AD}(\alpha))=\left(0,-2(1+\cos{2\alpha}),4\cos{2\alpha}\right)$. Considering the family of planes defined through $\mathbf{K}\cdot \mathbf{F}(\mathfrak{q}^{AD}(\alpha))=1$, we see that these share a common line of intersection given by $(K_1,-1/2,-1/4)$. Hence only the planes defined by the relevant endpoints, $\mathfrak{q}^{AS}=(\pi,0)$ and $\mathfrak{q}^{AD}=\left(\pi/2,\pi/2\right)$ for $N =4k$ or $\mathfrak{q}^{AD}=\left(2\pi (k+1)/(4k+2),2\pi k/(4k+2)\right)$ for $N= 4k+2$ (see Figure \ref{fig:Uhat_odd_even}) can contribute a face to $D_N$, and all the modes between these endpoints do not, which exactly explains the pattern observed in Table \ref{tab:faceless}. We note, however, that these modes will of course play a role for $\mathbf{K}$-values located on the common edge they share.
\subsection{The limit $N\rightarrow\infty$}
\label{sec:limit}
\subsubsection{The natural coordinate frame}
\label{sec:natural}
As $\mathbf{F}$ is a vector-valued mapping from the two dimensional domain $\widehat{\mathfrak{U}}_{\infty}$ to $\mathbb{R}^3$, it is clear that there must be a dependency between the elements of $\mathbf{F}\left(\mathfrak{q}\right)$. Indeed, we find that%
\begin{equation}
\begin{split}\label{eq:Fdepend}
F_{1}\left(\mathfrak{q}\right)^{2}=&\left(  2\cos q_{1}+2\cos q_{2}\right)
^{2}\\=&2\left(  2\cos\left(  q_{1}-q_{2}\right)  +2\cos\left(  q_{1}%
+q_{2}\right)  \right)  +2\cos2q_{1}+2\cos2q_{2}+4\\=&2F_{2}\left(
\mathfrak{q}\right)  +F_{3}\left(  \mathfrak{q}\right)  +4.
\end{split}
\end{equation}
This allow us to define a new coordinate frame with orthonormal basis vectors $\mathbf{\hat{n}}_{1}=\left(  1,0,0\right)
$, $\mathbf{\hat{n}}_{2}=\left(  0,1/\sqrt{5},-2/\sqrt{5}\right)$ and
$\mathbf{\hat{n}}_{3}=\left(  0,2/\sqrt{5},1/\sqrt{5}\right)$, which represents a clockwise rotation of the original frame by an angle $\chi = \arctan{2}$ around the $K_1$-axis. Defining the coordinates with respect to this frame through $\varphi_{j}=\mathbf{F}\left(\mathfrak{q}\right)\cdot\mathbf{\hat{n}}_{j}$ we find that $\varphi_{3}=\frac{1}{\sqrt{5}}\left(\varphi_{1}^{2}-4\right)$, so that we are left with the simple representation
\begin{equation}
\mathbf{F}\left(  \mathfrak{\varphi}_{1},\varphi_{2}\right)  =\varphi
_{1}\mathbf{\hat{n}}_{1}+\varphi_{2}\mathbf{\hat{n}}_{2}+\frac{1}{\sqrt{5}}\left(
\varphi_{1}^{2}-4\right)  \mathbf{\hat{n}}_{3}.\label{eq:F_phi}%
\end{equation}
The details of this transformation, as well as the shape of the IBZ in the new coordinates are presented in Appendix \ref{app:natural}.

\subsubsection{Surface reconstruction}
We now ask, given the relatively simple parametrization Eq.~(\ref{eq:F_phi}),
whether it is possible to reconstruct $D_{\infty}$ from the definition
$\mathbf{F}\left(  \mathfrak{\varphi}_{1},\varphi_{2}\right)  \cdot
\mathbf{K}\left(  \mathfrak{\varphi}_{1},\varphi_{2}\right)  =1$, the relation that characterizes the boundary points, cf.\ Eq.~(\ref{eq:bifq}). To that end, we introduce $\mathbf{\hat{u}}\left(  \mathfrak{\varphi}_{1},\varphi
_{2}\right)  =$ $\mathbf{F}\left(  \mathfrak{\varphi}_{1},\varphi_{2}\right)
/\left\vert \mathbf{F}\left(  \mathfrak{\varphi}_{1},\varphi_{2}\right)
\right\vert $ and note that this is the unit normal to the surface $\mathbf{K}%
\left(  \mathfrak{\varphi}_{1},\varphi_{2}\right)$. The defining equation
then reads%
\begin{equation}
\mathbf{\hat{u}}\left(  \mathfrak{\varphi}_{1},\varphi_{2}\right)
\cdot\mathbf{K}\left(  \mathfrak{\varphi}_{1},\varphi_{2}\right)  =\frac
{1}{\left\vert \mathbf{F}\left(  \mathfrak{\varphi}_{1},\varphi_{2}\right)
\right\vert }\equiv h\left(  \mathfrak{\varphi}_{1},\varphi_{2}\right)  ,
\end{equation}
which introduces the so-called \emph{support function} $h$.  It is a standard result of convexity theory (see e.g.\ \cite{Schneider2013ConvexBrunn-MinkowskiTheory}) that a convex body is fully determined by its support function. As the domain of our parameterization of the body is a compact set with only piecewise smooth boundary, we will need to preform the necessary inversion in the interior, the smooth boundary components, and the extreme points separately. 
\paragraph{Interior: the ridge}
\label{sec:ridge-infinity}
For notational brevity we omit the explicit dependence of all dependent variables on the coordinates $\varphi_j$, and denote the partial derivatives $\partial/\partial \varphi_j$ simply by $\partial_j$. The vectors $\partial_i\mathbf{K}$ are by definition tangent to the surface, so we have that%
\begin{equation}
\partial_{i}\left(\mathbf{\hat{u}}\cdot\mathbf{K}\right)  =\partial_{i}\mathbf{\hat{u}}\cdot\mathbf{K}  +\mathbf{\hat{u}}  \cdot\partial_{i}\mathbf{K} 
=\left(\partial_{i}\mathbf{\hat{u}}\right)  \cdot\mathbf{K}  =\partial_{i}h.
\end{equation}
Also, as $\mathbf{\hat{u}}\cdot\mathbf{\hat{u}}=1$, we have $ \partial_{i}\mathbf{\hat{u}} \cdot\mathbf{\hat{u}}=0,$ so that
$\partial_{i}\mathbf{\hat{u}}$ are also vectors in the tangent plane. This implies that
\begin{equation}
\mathbf{K}  =h \mathbf{\hat{u}}+\gamma_{1}\partial_{1}\mathbf{\hat{u}}
+\gamma_{2} \partial_{2}\mathbf{\hat{u}}.
\end{equation}
To obtain the unknown coefficient functions $\gamma_{i}$ we consider%
\begin{align}
\left(  \partial_{1}\mathbf{\hat{u}}\right)  \cdot\mathbf{K} &  =\gamma
_{1}\partial_{1}\mathbf{\hat{u}}\cdot\partial_{1}\mathbf{\hat{u}}+\gamma
_{2}\partial_{1}\mathbf{\hat{u}}\cdot\partial_{2}\mathbf{\hat{u}}=\partial
_{1}h,\\
\left(  \partial_{2}\mathbf{\hat{u}}\right)  \cdot\mathbf{K} &  =\gamma
_{1}\partial_{2}\mathbf{\hat{u}}\cdot\partial_{1}\mathbf{\hat{u}}+\gamma
_{2}\partial_{2}\mathbf{\hat{u}}\cdot\partial_{2}\mathbf{\hat{u}}=\partial
_{2}h,
\end{align}
which is readily solved by%
\begin{equation}
\left(
\begin{array}
[c]{c}%
\gamma_{1}\\
\gamma_{2}%
\end{array}
\right)  =\frac{1}{\Delta\left(  \mathbf{\hat{u}}\right)  }\left(
\begin{array}
[c]{cc}%
\partial_{2}\mathbf{\hat{u}}\cdot\partial_{2}\mathbf{\hat{u}} & -\partial
_{1}\mathbf{\hat{u}}\cdot\partial_{2}\mathbf{\hat{u}}\\
-\partial_{2}\mathbf{\hat{u}}\cdot\partial_{1}\mathbf{\hat{u}} & \partial
_{1}\mathbf{\hat{u}}\cdot\partial_{1}\mathbf{\hat{u}}%
\end{array}
\right)  \left(
\begin{array}
[c]{c}%
\partial_{1}h\\
\partial_{2}h
\end{array}
\right)  ,
\end{equation}
where the determinant is given by $\Delta\left(  \mathbf{\hat{u}}\right)
=\left(  \partial_{1}\mathbf{\hat{u}}\cdot\partial_{1}\mathbf{\hat{u}}\right)
\left(  \partial_{2}\mathbf{\hat{u}}\cdot\partial_{2}\mathbf{\hat{u}}\right)
-\left(  \partial_{1}\mathbf{\hat{u}}\cdot\partial_{2}\mathbf{\hat{u}}\right)
^{2}.$ 
The explicit calculation is performed using Mathematica and yields the curve
\begin{equation}\label{eq:Kridge}
    \mathbf{K}^{R}(\varphi_1) = \frac{2 \varphi_1}{4+\varphi_1^2} \hat{\mathbf{n}}_1-\frac{\sqrt{5}}{4+\varphi_1^2}\hat{\mathbf{n}}_3.
\end{equation}
 This result implies that for fixed $\varphi_1$ the mode instability surfaces with different values of $\varphi_2$ are all tangent to a single ridge-like structure. Substituting $\varphi_1=\sign(\varphi_1)\,4\cos^2{(a\pi)}$ and transforming back to the original frame then shows that this is in fact the ridge Eq.~(\ref{eq:ridge}) as introduced in Section \ref{sec:ridge}. This proves the perhaps surprising fact that, as we already hypothesized on the basis of the finite $N$ results, all the $M=8$ modes that make up the interior of the IBZ become unstable on a set of measure zero in phase space. 
\paragraph{The boundary: the fans}
\label{sec:fans-infinity}
Referring to Figure \ref{fig:phi_domain} and Eqs.~(\ref{eq:phi2max}) and (\ref{eq:phi2min}), we see that for each $\varphi_1$ there are two limiting tangent planes whose orientations are determined by $\mathbf{F}\left(\varphi_1,\varphi^{max}_2(\varphi_1)\right)$ and $\mathbf{F}\left(\varphi_1,\varphi^{min}_2(\varphi_1)\right)$ respectively. The former corresponds to a diagonal stripe mode, whereas the latter corresponds to striped ($\varphi_1>0$) and modulated stripe ($\varphi_1<0$) modes. Thus from each location on the ridge there are two straight lines with given orientation that end up in the already identified apices of the fans, the points $\mathbf{K}^{S}$, $\mathbf{K}^{MS}$ and $\mathbf{K}^{DS}$. Hence, in this limit the fans become sectors of a generalized cone with as base (a segment of) the ridge. These cone sectors are ruled surfaces, whose we can conveniently parametrize as
\begin{equation}
    \mathbf{K}^{X}(\varphi_1,l) = \mathbf{K}^{R}(\varphi_1)+l \left(  \mathbf{K}^{X}-\mathbf{K}^{R}(\varphi_1)\right),\,l\in[0,1],
\end{equation}
where $X$ labels the specific apical vertex of the cone sector. 
\paragraph{The extreme points: the major modes} The three extreme points of the IBZ simply yield the major modes already discussed in Section \ref{sec:major-modes} that dominate the phase diagram for $K_3>0$.
\subsubsection{The geometry of $D_{\infty}$}\label{sec:dinfty}
It is now straightforward to verify how the fans connect up with the major modes. With all these components in place we can now give the full description of $D_\infty$, by enumerating the components of its boundary $\partial D_{\infty}$.

\begin{figure}
    \centering
    \includegraphics[width=0.8\textwidth]{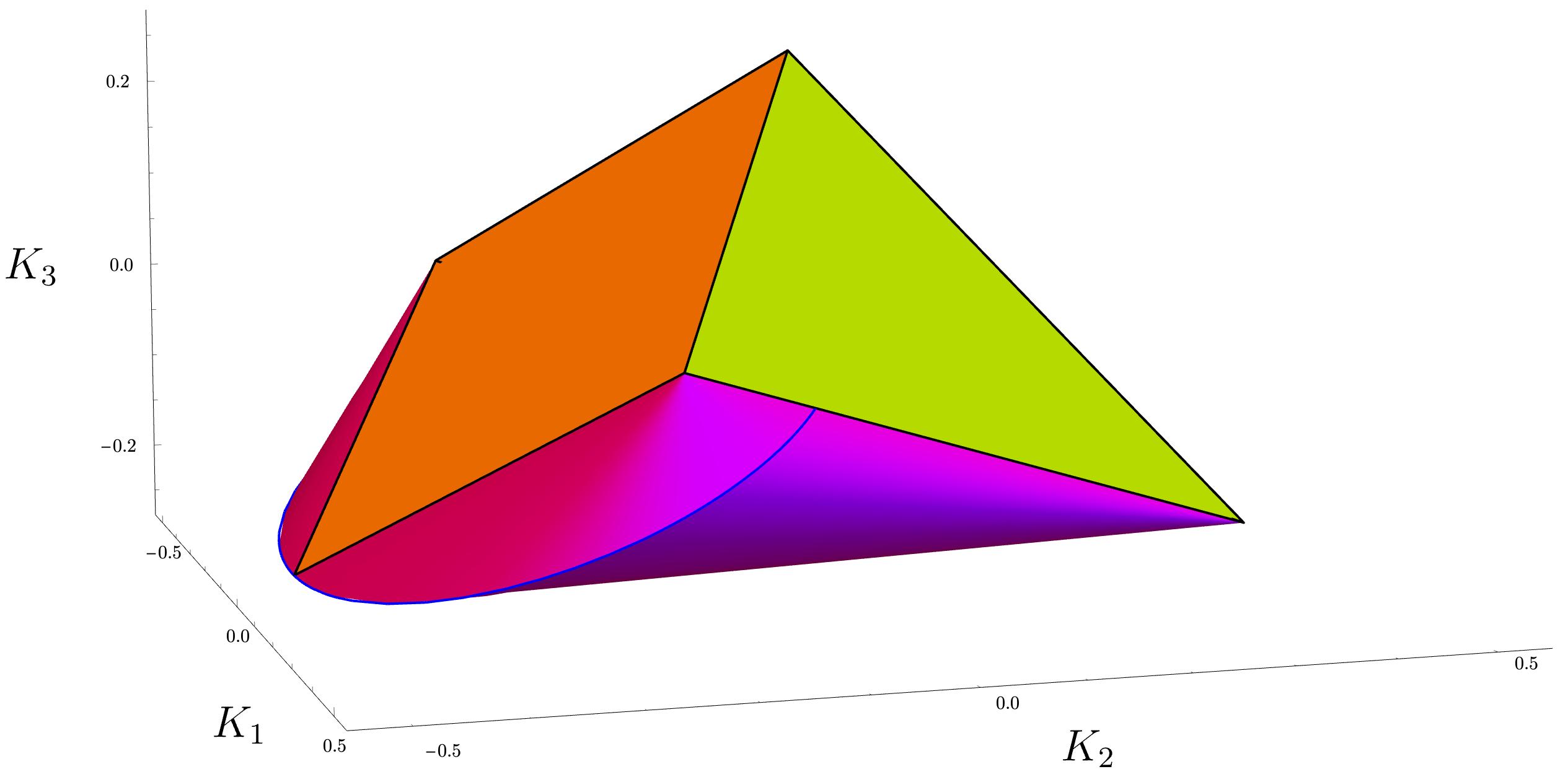}
    \caption{The disordered region $D_\infty$. Shading highlights the curved generalized cones with apices at the vertices $\mathbf{K}^{S}$, $\mathbf{K}^{MS}$ and $\mathbf{K}^{DS}$, which are the locus of the $M=4$ unstable modes. The ridge  $\mathbf{K}^{R}(\varphi_1)$, Eq.\ (\ref{eq:Kridge}), is the common boundary of these cones, and the locus of the $M=8$ unstable modes.}
    \label{fig:D_infinity}
\end{figure}

\begin{table}[htbp]
    \centering
    \begin{tabular}{|c|c|cc|c|lc|}
    \hline
         Type & Symbol & Mode(s) &  & M & Specification &\\
        \hline
                    & F  & $(0,0)$      &   & 1 & $\conv\left( \mathbf{K}^{T}, \mathbf{K}^{S},\mathbf{K}^{DS}\right)$ & \\
        Major modes & AF & $(\pi,\pi)$  &   & 1 & $\conv\left(\mathbf{K}^{T}, \mathbf{K}^{MS}, \mathbf{K}^{DS}\right)$ & \\
                    & AS & $(\pi,0)$    &   & 2 & $\conv\left( \mathbf{K}^{T}, \mathbf{K}^{S},\mathbf{K}^{MS},\mathbf{K}^{B}\right)$ & \\
        \hline
                    & S & $(a\pi,0)$    &   & 4 & $\mathbf{K}^{R}(\varphi_1)+l \left(                      \mathbf{K}^{S}-\mathbf{K}^{R}(\varphi_1)\right)$ & \\
        Fans        & MS& $(\pi,a\pi)$ & $a\in[0,1] $ & 4 & $\mathbf{K}^{R}(\varphi_1)+l \left(  \mathbf{K}^{MS}-\mathbf{K}^{R}(\varphi_1)\right)$ & $l\in[0,1]$ \\
                    & DS& $(a\pi,a\pi)$             &  &4 & $\mathbf{K}^{R}(\varphi_1)+l \left(  \mathbf{K}^{DS}-\mathbf{K}^{R}(\varphi_1)\right) $ & \\
        \hline
        Ridge        & R & all others    &  & 8 & $\mathbf{K}^{R}(\varphi_1) \quad  \varphi_1\in[-4,4]$ & \\
        \hline
        
    \end{tabular}
    \caption{The components of the surface of the order-disorder surface $\partial D_\infty$. Here $\conv(\mathbf{K}_1,\mathbf{K}_2,\ldots)$ denotes the convex hull of the set of points in the argument list.}
    \label{tab:D_infinity}
\end{table}
We visualize $D_\infty$ in Figure \ref{fig:D_infinity}. We now note that due to the fact that both the fans and the ridge are sets with non-zero curvature, the structure of the bifurcation modes in these regimes of phase space are inevitably of a `devil's surface' nature. Any variation of $\mathbf{K}$ in these regimes leads to a smooth non-constant variation of the critical modes $\mathfrak{q}$ that satisfy the bifurcation condition $\mathbf{K}\cdot F(\mathfrak{q})=1$.  As $2\pi\mathbb{Q}^2\cap\widehat{\mathfrak{U}}_{\infty}$ is dense in $\widehat{\mathfrak{U}}_{\infty}$, there are bifurcating modes of arbitrary complexity in the neighbourhood of any mode $\mathfrak{q}$ in this regime.

\section{Comparison with simulations}
\label{sec:simulations}
It is clearly infeasible to test the predicted devil's surface like complexity of the mode structure of the nascent phases at the order-disorder boundary by numerical means. However, our analysis of finite periodicities with fixed index $N$, which led to to the definition of the disorder polytopes $D_N$, showed that these are all realized on the common $N\times N$ square periodicity. The latter condition is readily realized by imposing periodic boundary conditions in a standard single spin-flip Metropolis simulation. To be able to limit ourselves to a finite number of simulations we make the following choice. For fixed $N$ we consider the set of bifurcating modes $\{\mathfrak{q}_f\}$, where $f$ indexes the set of faces of $D_N$. For each mode $\mathfrak{q}_f$ we determine a representative  coupling vector $\mathbf{K}^{*}_f$ as the centroid of the face it belongs to. We then perform a series of simulations along the ray in phase space $\beta\mathbf{K}^{*}_f,\,\beta\in[0,\infty)$. The scaled inverse temperature $\beta$ is thus chosen so that the predicted transition occurs at $\beta=1$, which allows for easy comparison with the simulations independent of the details of each face. 

In order to analyze the results of the simulation we need a suitable order parameter to signal the presence (or non-presence) of certain modes. As we will perform multiple replicates of the simulations at each inverse temperature, this order parameter has to be insensitive to any of the possible global symmetries that link different replicates. Defining the Fourier transform of the site magnetisation pattern by 
\begin{equation}\label{eq:antif}
     \hat{m}_{\mathrm{q}}=\frac{1}{N}\sum_{\mathrm{z}\in \mathcal{U}_{\mathrm{P}}} m_{\mathrm{z}}e^{-i\langle\mathrm{q},\mathrm{z}\rangle},
\end{equation}
we can define 
\begin{equation}\label{eq:op}
\mu_{\mathrm{q}} \equiv\frac
{1}{\left\vert \mathfrak{D}_4\right\vert }\sum_{g\in\mathfrak{D}_4}m_{g\mathrm{q}}^{\ast}m_{g\mathrm{q}}.%
\end{equation}
By virtue of being square in the magnetisations, this expression divides out the up-down symmetry of the Hamiltonian. By multiplying complex conjugates, the translation symmetries, which generate unitary phase factors, are divided out. Finally, the explicit ``averaging'' over the point group symmetries, divides out the remaining symmetries.

\begin{figure}[htbp]
    \centering
    \includegraphics[width=\textwidth]{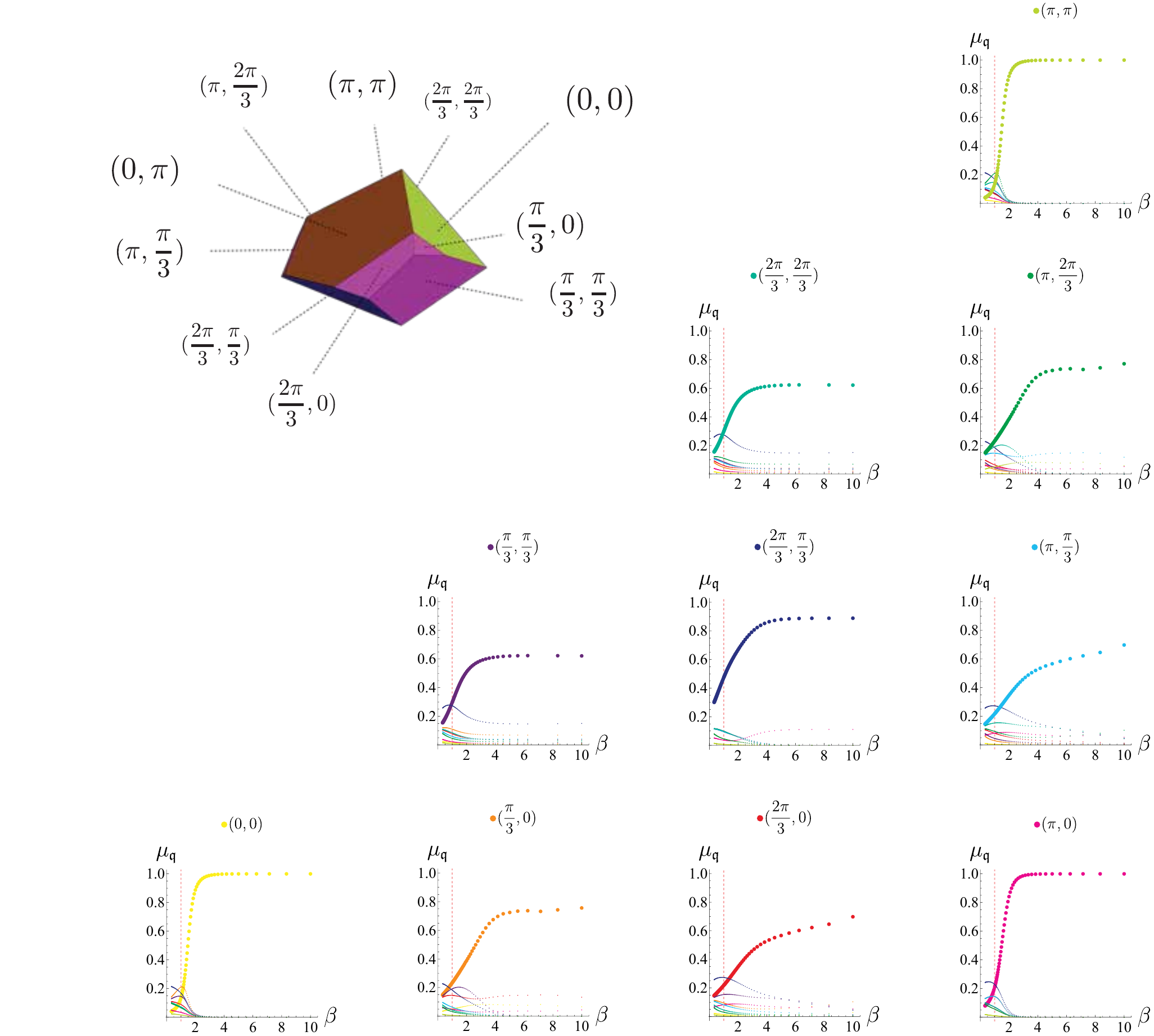}
    \\
    \caption{Order parameter $\mu_{\mathfrak{q}}$ as function of the reduced inverse temperature $\beta$ for all possible modes on a $6\times6$ periodic lattice, for coupling constants corresponding to the centroid of each of the $10$ possible faces of $D_6$, as visualized in the upper left corner. In all cases the predicted bifurcating mode is the sole dominant one. }
    \label{fig:sixmosaic}
\end{figure}

We performed simulations for $N=4,6,8,12$. As a proof of principle we show the order parameter values for the emergent modes beyond each of the $10$ faces of $D_6$ in Figure \ref{fig:sixmosaic}. The results for the other $N$ values were similar (data not shown). 

For ease of reference these plots are organized to mimic the geometry of the salient IBZ, $\mathfrak{\widehat{U}}_6$. In all cases the observed dominant mode is the one predicted by our mean-field analysis. Moreover, in all cases the other modes, which have fairly significant amplitudes due to inevitable finite-size effects in the disordered phase, appear to be suppressed in the ordered regime. Strikingly, the shape of the order parameter curves also obeys the predicted symmetry in the antidiagonal of the IBZ (see \ref{sec:odd-even}). Finally, and as expected, the mean-field analysis appears to underestimate the value of the inverse temperature at which the ordering transition occurs. In Appendix \ref{app:simulations} we provide a few more technical details about the simulations.  

\section{Conclusions}
Our analysis of the order-disorder transitions of the field-free nnnn Ising model on the square shows that the observation by Landau and Binder in their seminal paper on this topic almost four decades back \cite{Landau1985PhaseCouplings} that ``Using mean-field theory, we also find indications of interesting behavior for $T>0$'' was prescient. Our results indicate that in this approximation the strong frustration induced by an antiferromagnetic nnnn interactions produces fully developed complexity already at the level of the high-temperature order-disorder transition. Indeed, a large part of the order-disorder surface in the half-space $K_3 <0$ represents a `devil's surface', where bifurcating modes of arbitrary complexity are densely interspersed. 

Our results also bring to the fore a hitherto perhaps less appreciated role for the lattice symmetry group and its action on the space of lattice modes by showing that the multiplicity $M$ of these modes under the point group is a strong determinant whether and where in the phase space these modes become unstable. Strikingly, the $K_3>0$ part of the order-disorder surface is entirely determined by the three major modes at the extreme points of the IBZ, while the three one-parameter families of $M=4$ modes associated with the edges of the IBZ all located in the half-space $K_3 <0$ make up the remaining surface area. Thus a set of measure zero in the IBZ accounts for all the bifurcation modes except for a set of measure zero, the ridge,  to which all the $M=8$ modes, which represent the full measure of the IBZ, are compressed. It is our intuition that these results can possibly be interpreted within the setting of the so-called Equivariant Branching Lemma \cite{Golubitsky1988SingularitiesTheory,Golubitsky2003TheSpace}, a cornerstone of the theory of bifurcations with symmetry, which identifies a role for solutions with `maximal' residual symmetry with respect to the symmetry group being broken. 

All together these result provide a somewhat paradoxical answer to our original question of the designability of complex patterns in binary lattice gases. On the one hand, the antiferromagnetic nnnn interactions enable a vast array of patterns to become accessible directly from the disordered phase. On the other hand, the ultra-sensitive dependence on the precise values of the coupling constants, implied by the devil's surface for the $M=4$ modes and the collapse onto a set of zero measure of the $M=8$ modes, effectively precludes a requisite degree of control in selecting specific patterns. It is an interesting question whether it is possible to circumvent the latter defect, perhaps through multi-spin interactions, and construct a system with a more robust yet sufficiently rich phase behaviour. 

Obviously, the mean-field approach is a severe approximation, and one may well ask whether any of these features survive the inclusion of the inevitably strong correlations in a low-dimensional system such as the 2D square lattice. Here, we were able to provide limited evidence using Monte Carlo simulations that at least some of the predictions remain valid when we include these correlations up to cutoff imposed by periodic boundary conditions. Specifically, we correctly predict the dominant mode developing from the disordered phase along rays in phase space that pass through the center of the faces of the calculated disorder polytopes $D_N$. The `optimistic' view suggests that we can expect that results on the nature of symmetry-breaking events, which are to a large extent constrained by purely group-theoretical properties, may be more universal, and hence transcend the specific approximation chosen. 

There are several directions of further research suggested by our results. First, it would be interesting to study this system beyond the mean field approximation, perhaps using a variant of the Cluster Variation Method \cite{Pelizzola2005ClusterModels}. Obvious questions are: (i) does the order-disorder surface remain a convex polytope and (ii) if so, which of its features remain invariant. Next, one could explore the immediate generalisations of the bifurcation conditions Eq. (\ref{eq:bif}) to different lattices and/or longer-ranged interactions. The analysis framework we set up here can readily be extended in these directions, albeit that as we increase the interaction range we also increase the dimensionality of the disorder polytopes with concomitant increase of geometrical complexity. So far, we have also limited our analysis to the order-disorder surface. What happens beyond it is an open question. We have indications that, at least for finite $N$, the dimensionality of the solution spaces associated with the bifurcating modes is significantly smaller than $N$, which would possible make it tractable to at least numerically track these solutions to possible lower temperature transitions. We certainly expect that secondary transitions are likely to occur, as most of the bifurcating modes only partially break the symmetry of the underlying lattice. Although we did not dwell on this here, our simulations also point to the occurrence of such transitions.  

It would also be interesting to see what, if anything,  the present analysis reveals about the ground-state phase diagram. Here, the recently developed method of mapping the ground-state problem of arbitrary spin models into a Maximum Satisfiability problem \cite{Huang2016FindingMAX-SAT}, or tensor network approaches for frustrated systems \cite{Vanhecke2021SolvingNetworks} may prove useful. 

Finally, on a much more abstract level, there recently has been a series of papers that focus on the universality and complexity of classical spin models from the perspective of the theory of computation \cite{DeLasCuevas2016SimplePhysics,Kohler2019TranslationallyHamiltonians, Drexel2020DescribingMeasure}. It would be fascinating to explore what these insights could contribute to understanding the present system and frustrated systems in general. 

\begin{acknowledgments}
The authors would like thank David Avis and Charles `Skip' Jordan for their kind assistance in using \texttt{lrs}. This work is part of the Dutch Research Council (NWO) and was performed at the research institute AMOLF.
\end{acknowledgments}

\appendix

\section{Periodic patterns on $\mathbb{Z}^{2}\label{app:periodic}$}

A magnetization \emph{pattern} $m\left(  \mathrm{z}\right)  $ is periodic if
there exist two basis vectors $\mathrm{p}_{1}=\left(  p_{1}^{1},p_{1}%
^{2}\right)  $ , $\mathrm{p}_{2}=\left(  p_{2}^{1},p_{2}^{2}\right)
\in\mathbb{Z}^{2}$ such that%
\begin{equation}
\forall k_{1},k_{2}\in\mathbb{Z}:m\left(  \mathrm{z+}k_{1}\mathrm{p}_{1}%
+k_{2}\mathrm{p}_{2}\right)  =m\left(  \mathrm{z}\right).
\end{equation}
It is convenient to parametrize the periodicity through the matrix%
\begin{equation}
\mathrm{P}=\left(
\begin{array}
[c]{cc}%
p_{1}^{1} & p_{1}^{2}\\
p_{2}^{1} & p_{2}^{2}%
\end{array}
\right)  ,
\end{equation}
where, without loss of generality and by convention, we choose the order of
the basis vectors such that the \emph{index} of the periodicity $\det
(\mathrm{P})=N\equiv p_{1}^{1}p_{2}^{2}-p_{1}^{2}p_{2}^{1}>0.$ The unit cell associated
with this pattern is then defined as%
\begin{equation}
\mathcal{U}_{\mathrm{P}}=\left\{  \mathrm{z}\in\mathbb{Z}^{2}\Bigg\vert%
\begin{array}
[c]{c}%
0\leq\left\langle \mathrm{z},\mathrm{p}_{1}\right\rangle \left\langle
\mathrm{p}_{2},\mathrm{p}_{2}\right\rangle -\left\langle \mathrm{z}%
,\mathrm{p}_{2}\right\rangle \left\langle \mathrm{p}_{1},\mathrm{p}%
_{2}\right\rangle <N^{2},\\
0\leq\left\langle \mathrm{z},\mathrm{p}_{2}\right\rangle \left\langle
\mathrm{p}_{1},\mathrm{p}_{1}\right\rangle -\left\langle \mathrm{z}%
,\mathrm{p}_{1}\right\rangle \left\langle \mathrm{p}_{1},\mathrm{p}%
_{2}\right\rangle <N^{2}%
\end{array}
\right\}.
\end{equation}
Note that the number of lattice points in the unit cell is given by the index
$\left\vert \mathcal{U}\right\vert =N.$ The set $\widehat{\mathcal{U}%
}_{_{\mathrm{P}}}$ of wave vectors compatible with this periodicity must
satisfy%
\begin{align}
\left\langle \mathrm{q,p}_{1}\right\rangle  &  =2\pi k_{1},\label{eq:period1}\\
\left\langle \mathrm{q,p}_{2}\right\rangle  &  =2\pi k_{2}, \label{eq:period2}%
\end{align}
for some $\mathrm{k}=\left(  k_{1},k_{2}\right)  \in\mathbb{Z}^{2},$ so that
$\exp\left(  i\,\left\langle \mathrm{q,z+}l_{1}\mathrm{p}_{1}+l_{2}%
\mathrm{p}_{2}\right\rangle \right)  =\exp\left(  i\,\left\langle
\mathrm{q,z}\right\rangle \right)  $ for all $l_{1},l_{2}\in\mathbb{Z}.$
Writing (\ref{eq:period1}) and (\ref{eq:period2}) as%
\begin{equation}
\mathrm{Pq}=2\pi\mathrm{k},%
\end{equation}
we have%
\begin{equation}
\mathrm{q=}2\pi\mathrm{P}^{-1}\mathrm{k}.%
\end{equation}
We now introduce the dual basis $\mathrm{\hat{p}}_{1}=\left(  p_{1}^{1}%
,p_{2}^{1}\right)  $ and $\mathrm{\hat{p}}_{2}=\left(  p_{1}^{2},p_{2}%
^{2}\right)  ,$ and note that $\mathrm{P}^{-1}\mathrm{\hat{p}}_{1}=$
$\mathrm{e}_{1}$ and $\mathrm{P}^{-1}\mathrm{\hat{p}}_{2}=\mathrm{e}_{2}$.
\ Thus, $\mathrm{q}^{\prime}=2\pi\mathrm{P}^{-1}\left(  \mathrm{k}%
+l_{1}\mathrm{\hat{p}}_{1}+l_{2}\mathrm{\hat{p}}_{2}\right)  =\mathrm{q}+2\pi
l_{1}\mathrm{e}_{1}+2\pi l_{2}\mathrm{e}_{2},$ and hence $\mathrm{Pq}^{\prime
}=2\pi\mathrm{k+}2\pi l_{1}\mathrm{Pe}_{1}+2\pi l_{2}\mathrm{Pe}_{2}\equiv
2\pi\mathrm{k}^{\prime}.$ So, adding integer multiples of the dual basis
vectors to $\mathrm{k}$ does not yield additional information, and we can
restrict ourselves to solutions in the dual unit cell (discrete Brillouin
zone)
\begin{equation}
\widehat{\mathcal{U}}_{\mathrm{P}}=\left\{  \mathrm{q}=2\pi\mathrm{P}%
^{-1}\mathrm{k}\Bigg\vert\mathrm{k}\in\mathbb{Z}^{2},%
\begin{array}
[c]{c}%
0\leq\left\langle \mathrm{k},\mathrm{\hat{p}}_{1}\right\rangle \left\langle
\mathrm{\hat{p}}_{2},\mathrm{\hat{p}}_{2}\right\rangle -\left\langle
\mathrm{k},\mathrm{\hat{p}}_{2}\right\rangle \left\langle \mathrm{\hat{p}}%
_{1},\mathrm{\hat{p}}_{2}\right\rangle <N^{2}\\
0\leq\left\langle \mathrm{k},\mathrm{\hat{p}}_{2}\right\rangle \left\langle
\mathrm{\hat{p}}_{1},\mathrm{\hat{p}}_{1}\right\rangle -\left\langle
\mathrm{k},\mathrm{\hat{p}}_{1}\right\rangle \left\langle \mathrm{\hat{p}}%
_{1},\mathrm{\hat{p}}_{2}\right\rangle <N^{2}%
\end{array}
\right\}  ,
\end{equation}
where we have used that $\det\left(  \mathrm{P}^{T}\right)  =\det\left(
\mathrm{P}\right)  =N$ , which also shows that there are $\left\vert
\widehat{\mathcal{U}}_{_{\mathrm{P}}}\right\vert =\left\vert \mathcal{U}%
_{\mathrm{P}}\right\vert =N$ independent wave vectors that are compatible with
the periodicity.

In the following we would like to enumerate all possible periodicities,
classifying them according their index\ $N.$ This problem is equivalent to
enumerating all the subgroups of $\mathbb{Z}^{2}$ of index $N.$ This can be
performed employing a theorem due to Hermite \cite{Hermite1851}, which states
that for any matrix with integer entries $\mathrm{P}\in GL_{2}\left(
\mathbb{Z}\right)  $ and determinant $N$ there is a unimodular
(determinant-preserving up to sign) transformation $\mathrm{J}\in
GL_{2}\left(  \mathbb{Z}\right)  ,$ such that $\mathrm{P}^{\ast}=\mathrm{PJ}$,
where $\mathrm{P}^{\ast}$ is of the form%
\begin{equation}
\mathrm{P}^{\ast}=\left(
\begin{array}
[c]{cc}%
d_{1} & 0\\
0\leq s_{1}<d_{1} & d_{2}=\frac{N}{d_{1}}%
\end{array}
\right),
\end{equation}
the so-called lower-triangular Hermite normal form. These matrices thus fall
into equivalence classes, which are enumerated by considering that for any
divisor $d_{1}|N$ there are exactly $d_{1}$ inequivalent forms, and hence
$\left\vert \left\{  \mathrm{P|\det P=N}\right\}  \right\vert =\sum_{d_{1}%
|N}d_{1}\equiv\sigma_{1}\left(  N\right)  $. This implies that one can simply
choose as basis of our desired pattern the vectors $\mathrm{p}_{1}^{\ast
}=\left(  d_{1},0\right)  $ and $\mathrm{p}_{2}^{\ast}=\left(  0\leq
s_{1}<d_{1},d_{2}=N/d_{1}\right)  $. Note, however, that these vectors need
not be the set of minimal length basis vectors that generate the same periodic
sublattice. If necessary, these so-called Minkowski bases can be obtained from
the Hermite normal form basis, through an algorithm due to Lagrange
\cite{Nguyen2004Low-dimensionalRevisited}.

Next, we introduce
\begin{equation}
\widehat{\mathcal{U}}_{N}=%
{\displaystyle\bigcup\limits_{\left\{  \mathrm{P|}\left\vert
\widehat{\mathcal{U}}_{_{\mathrm{P}}}\right\vert =N\right\}  }}
\widehat{\mathcal{U}}_{_{\mathrm{P}}},
\end{equation}
i.e. the set of all lattice wave vectors compatible with periodic patterns
with index $N.$ We can now prove the following, as far as we can tell, non-trivial

\begin{lemma}
$\widehat{\mathcal{U}}_{N}=\left\{  \mathrm{q=}\frac{2\pi}{N}\left(
l_{1},l_{2}\right)  |0\leq l_{1},l_{2}<N\right\}  $ and hence $\left\vert
\widehat{\mathcal{U}}_{N}\right\vert =N^{2}.$
\end{lemma}

We prove this lemma in two steps. First, consider the periodicity
$\mathrm{\Pi}$, with basis vectors $\mathrm{\pi}_{1}=\left(  N,0\right)  $ and
$\mathrm{\pi}_{2}=\left(  0,N\right)  ,$ i.e. a square $N\times N$ unit cell.
We have $\mathcal{U}_{\mathrm{\Pi}}=\left\{  \mathrm{z}|0\leq z^{1}%
,z^{2}<N\right\}  .$ Also, $\widehat{\mathcal{U}}_{\mathrm{\Pi}}=\frac{2\pi
}{N}\mathcal{U}_{\mathrm{\Pi}}.$ Let the periodicity $\mathrm{P}$ with index
$N$ be given by the basis vectors $\mathrm{p}_{1}=$ $\left(  d,0\right)  $ and
$\mathrm{p}_{2}=$ $\left(  s,\bar{d}\right)  $ where $d\in\left[  N\right]  $
(the set of divisors of $N$) and the complementary divisor is defined as
$\bar{d}\equiv N/d.$ Now $\mathrm{\pi}_{1}=\bar{d}\mathrm{p}_{1}$ and
$\mathrm{\pi}_{2}=-s\mathrm{p}_{1}+d\mathrm{p}_{2}.$ It follows that any
$\mathrm{P}$ periodic pattern is also $\mathrm{\Pi}$ periodic, hence
$\widehat{\mathcal{U}}_{_{\mathrm{P}}}\subset\widehat{\mathcal{U}%
}_{\mathrm{\Pi}}$ and%
\begin{equation}
\widehat{\mathcal{U}}_{N}=%
{\displaystyle\bigcup\limits_{\left\{  \mathrm{P|}\left\vert
\widehat{\mathcal{U}}_{_{\mathrm{P}}}\right\vert =N\right\}  }}
\widehat{\mathcal{U}}_{_{\mathrm{P}}}\subset\widehat{\mathcal{U}}%
_{\mathrm{\Pi}}.%
\end{equation}
We then need to prove the reverse inclusion, by showing that for every
$\mathrm{q=}\frac{2\pi}{N}\left(  l_{1},l_{2}\right)  $ there is a
$\mathrm{P}$ with $\det\mathrm{P=N}$ such that $\mathrm{q}\in
\widehat{\mathcal{U}}_{_{\mathrm{P}}}.$ Recall that $\mathrm{P}^{-1}%
\mathrm{\hat{p}}_{1}=$ $\mathrm{e}_{1}$ and $\mathrm{P}^{-1}\mathrm{\hat{p}%
}_{2}=\mathrm{e}_{2},$ so that $\mathrm{k=}\frac{l_{1}}{N}\mathrm{\hat{p}}%
_{1}+\frac{l_{2}}{N}\mathrm{\hat{p}}_{1}$ would be a valid solution provided
$\mathrm{k}\in\mathbb{Z}^{2}.$ In the Hermite normal form representation
$\mathrm{\hat{p}}_{1}=\left(  d,s\right)  $ and $\mathrm{\hat{p}}_{2}=\left(
0,\bar{d}\right)  $. Thus the question reduces to whether $d\in\left[
N\right]  $ and $0\leq s<d$ can be chosen such that the congruences%
\begin{align}
l_{1}d\;  &  \equiv0\;\operatorname{mod}N,\label{eq:cond1}\\
l_{1}s+l_{2}\bar{d}  &  \equiv0\;\operatorname{mod}N \label{eq:cond2}%
\end{align}
hold$.$ We distinguish two cases:

\begin{itemize}
\item $l_{1}\nmid N$: In this case congruence (\ref{eq:cond1}) can only be
solved by taking $d=N$. This reduces the second one to $l_{1}s\equiv
N-l_{2}\;\operatorname{mod}N$. In this case $m\equiv\operatorname{GCD}\left(
l_{1},N\right)  =1$ and hence $m\mid N-l_{2}$ guaranteeing a solution
\cite{Schroeder1997NumberCommunication}.

\item $l_{1}\mid N$: In this case $l_{1}=d_{1}\in\left[  N\right]  /\left\{
N\right\}  $ so that (\ref{eq:cond1}) is solved by $d=\bar{d}_{1}$, and the
second condition reduces to $d_{1}s+l_{2}d_{1}\equiv0\;\operatorname{mod}N$,
which in turn reduces to $s+l_{2}\equiv0\;\operatorname{mod}\bar{d}_{1}.$ Let
$l_{2}=n\bar{d}_{1}+r_{2},$ then the latter congruence is trivially solved by
$s=\bar{d}_{1}-r_{2}.$
\end{itemize}
This shows that
\begin{equation}
\widehat{\mathcal{U}}_{\mathrm{\Pi}}\subset\widehat{\mathcal{U}}%
_{_{\mathrm{P}}}\subset\widehat{\mathcal{U}}_{N}%
\end{equation}
and, hence, $\widehat{\mathcal{U}}_{\mathrm{\Pi}}=\widehat{\mathcal{U}}_{N}.$

\section{The sets $\widehat{\mathfrak{U}}_{N}$}
\label{app:Uhat}
\begin{figure}[b]
\includegraphics{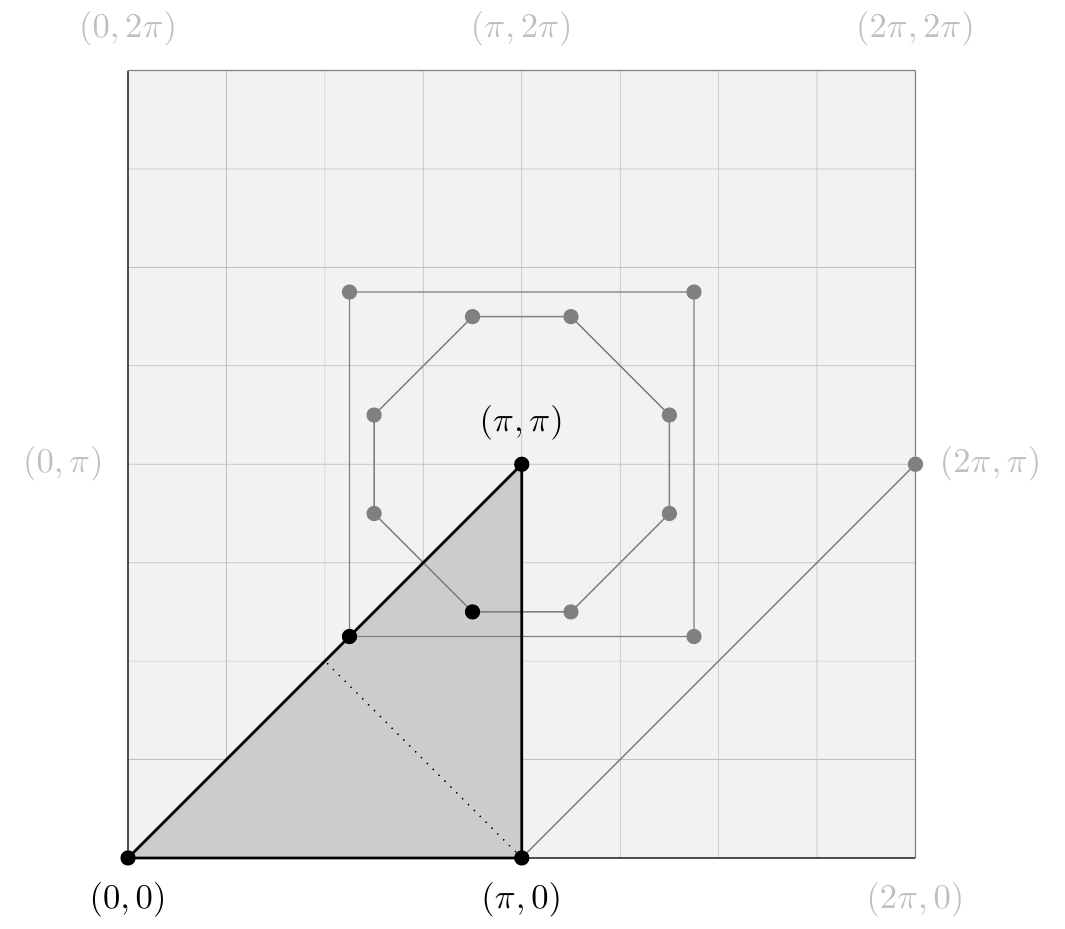}   
\caption{Geometry of the quotient space
$\widehat{\mathfrak{U}}_{\infty}=\widehat{\mathcal{U}}_{\infty}/G$, depicted as the shaded triangle. Multiplicities: $M=1$: The
points $(0,0)$ and $(\pi,\pi)$, $M=2$: The point $(\pi,0)$, $M=4$: All
boundary points, excluding the vertices, $M=8$: all interior points. Also shown
are three representative orbits in the embedding space
$\widehat{\mathcal{U}}_{\infty}=[0,2\pi)\times[0,2\pi)$. The dotted anti-diagonal is the unique line of mirror symmetry of $\widehat{\mathfrak{U}}_{\infty}$.}%
\label{fig:Uhat_infty}%
\end{figure}

We first note that (\ref{eq:UhatN}) shows that $\widehat{\mathcal{U}}%
_{N}\subset\widehat{\mathcal{U}}_{\infty}\equiv\lbrack0,2\pi)\times
\lbrack0,2\pi)\subset\mathbb{R}^{2}$ for all $N$. According to the fundamental
theorem of group actions, the possible orbit lengths, which we will call \emph{multiplicities}, of the group
$\mathfrak{D}_{4}$ is given by the four divisors $\left\{
1,2,4,8\right\}  $ of its order $\left\vert \mathfrak{D}_{4}\right\vert =8.$
The structure of the quotient space, also called an orbifold \cite{Caramello2019IntroductionOrbifolds},
$\widehat{\mathfrak{U}}_{\infty}=\widehat{\mathcal{U}}_{\infty}/G$, including
the multiplicities associated with isolated points or subsets, is illustrated
in Figure \ref{fig:Uhat_infty}. 

\begin{figure}[ptb]
\centering
\subfloat{\includegraphics{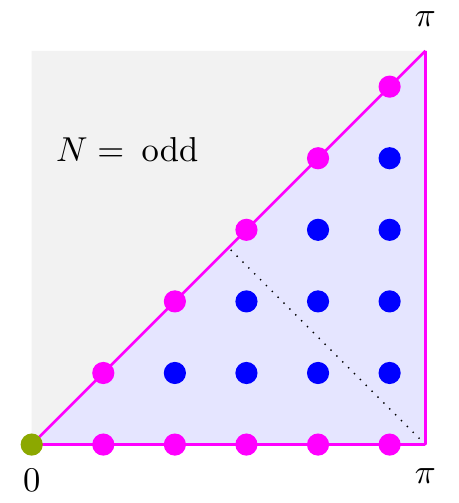}}
\subfloat{\includegraphics{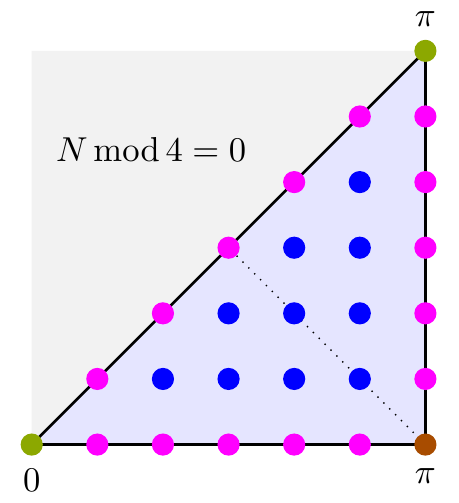}} 
\subfloat{\includegraphics{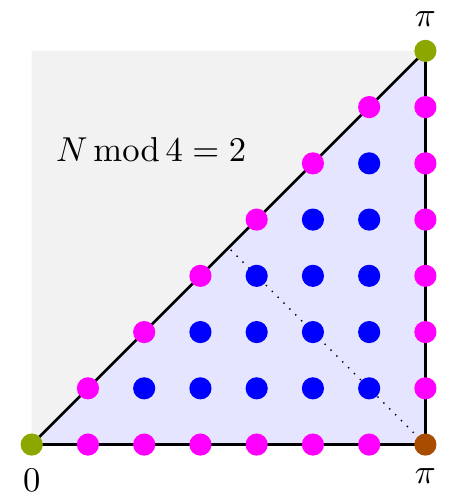}} 
\caption{Modes contributing to $\widehat{\mathfrak{U}}_{N}$ for $N=\,$ odd, $N\mod 4 = 0$ and $N\mod 4 = 2$. Multiplicities are color coded: $M=1$: citrus, $M=2$: tawny, $M=4$: purple , $M=8$: blue. $\widehat{\mathfrak{U}}_{N}$ is mirror symmetric with respect to the dotted anti-diagonal for even $N$.}%
\label{fig:Uhat_odd_even}%
\end{figure}

Now, let $s_{M}\left(  N\right)$ count the numbers of orbits with multiplicity $M$
in $\widehat{\mathcal{U}}_{N}$. The only two modes which have an orbit of
length $M=1$ are $(0,0)$ and $(\pi,\pi)$, hence $s_{1}(N)=1$ for $N$ odd \ and
$s_{1}(N)=2$ for $N$ even. $(0,\pi)$ is the only mode with orbit length $M=2$,
accordingly $s_{2}(N)=1$ for $N$ even and zero otherwise. For the remaining
two cases, it is convenient to define $N_{s}=\lfloor\frac{N-1}{2}\rfloor$ for
$N$ odd and $N_{s}=\lfloor\frac{N-2}{2}\rfloor$ for $N$ even. The number of
modes with orbit length $M=4$ is given by $s_{4}(N)=3N_{s}$ for $N$ even and
$s_{4}(N)=2N_{s}$ for $N$ odd. The orbits of length $M=8$ are, independently
of the parity of $N,$ given by $s_{8}(N)=\sum_{n=1}^{N_{s}}n=\frac{1}{2}%
N_{s}(N_{s}+1)$. \ Thus, the size of $\widehat{\mathfrak{U}}_{N}$ is simply
given by
\begin{equation}
|\widehat{\mathfrak{U}}_{N}|=s_{1}(N)+s_{2}(N)+s_{4}(N)+s_{8}(N)=\frac{1}%
{2}\left\lfloor {\frac{N+2}{2}}\right\rfloor (\left\lfloor {\frac{N+2}{2}%
}\right\rfloor +1).
\end{equation}
This is equivalent to the more intuitive formula
\begin{equation}
|\widehat{\mathfrak{U}}_{N}|=\sum_{n=1}^{\lfloor\frac{N}{2}+1\rfloor}n,
\end{equation}
which can be understood geometrically by referring to the figure
\ref{fig:Uhat_odd_even}. Note that this analysis also correctly reproduces the size of
$\widehat{\mathcal{U}}_{N}$
\begin{equation}
|\widehat{\mathcal{U}}_{N}|=s_{1}(N)+2s_{2}(N)+4s_{4}(N)+8s_{8}(N)=N^{2}.
\end{equation}

\section{The vertex enumeration algorithm}
\label{app:lrs}
The coefficients of the inequalities Eq.\ref{eq:half} defining the halfspaces that bound the disorder polytopes are generically irrational by virtue of the definition of $\mathbf{F}(\mathrm{q})$ (Eqs.~(\ref{eq:F1}-\ref{eq:F3})). However, the vertex enumeration algorithm \texttt{lrs} we employed intrinsically uses exact integer arithmetic. Thus we are forced to rationalize the components $\mathbf{F}(q)$ as input to the program. As a consequence of this approximation some artifacts are to be expected, primarily in the form of spurious vertices. We empirically observed, e.g., that the resulting
polytopes output by \texttt{lrs} are all \emph{simple}, i.e.\ all of their vertices have degree $3$, where \emph{degree} of a vertex is the number of edges, and hence also the number of faces, to which it belongs. Our analysis presented in Section \ref{sec:fan-modes}, however, shows that the vertices at the apex of the so-called fan modes $\mathbf{K}^{S}$, $\mathbf{K}^{MS}$ and $\mathbf{K}^{DS}$ are \emph{degenerate} as their degree in fact diverges in the limit $N\rightarrow \infty$. 

We have identified three sources of spurious vertices and developed appropriate corrective procedures for all of them. First, as already alluded to above, the vast majority of the spurious vertices appear around the apices of the fans, which we know analytically, and hence readily detected and removed. In fact, if a fan apex has degree $k$, we find that exactly $k-1$ spurious vertices are generated which are much closer among each other than the typical distance $d \sim 0.1$ to the nearest distinct vertex. The second category of spurious vertices is associated with the faceless modes (\ref{sec:faceless}) that are all tangent to a common edge. We empirically found that each faceless mode contributed exactly two spurious vertices in the neighbourhood of the common edge, which we subsequently removed. The final category of spurious vertices are ``accidental'' and only occur for the interior $M=8$ modes. They occur when components of $\mathbf{F}(q)$ for neighbouring $\mathrm{q}$'s happen to be identical, but because of finite precision arithmetic are mapped to different rational approximations. These then have to be dealt with through explicit analytical recalculation. In practice we never observed more than $4$ such spurious vertices for all $N$'s considered (up to $N=256$). Defining the degree of rationalization $D$ as the number of digits allowed for numerator and/or denominator in the approximants, we found that it is a useful rule of thumb to suspect all pairs vertices closer than $D^{-3}$ as being spurious. 

\section{Visualization of ordering patterns}
\label{app:visualization}
Here we visualize the ordering patterns corresponding to various (families of) bifurcating modes identified in the main text for periodicity index $N=12$. The amplitude of the modes is normalized by setting the value of the origin site to $1$ represented by the color red. The other amplitude values are then shown on a temperature-type scale, with dark blue corresponding to the minimum value $-1$.

\begin{figure}[htbp]
    \centering
    \subfloat{\includegraphics[width=0.3\textwidth]{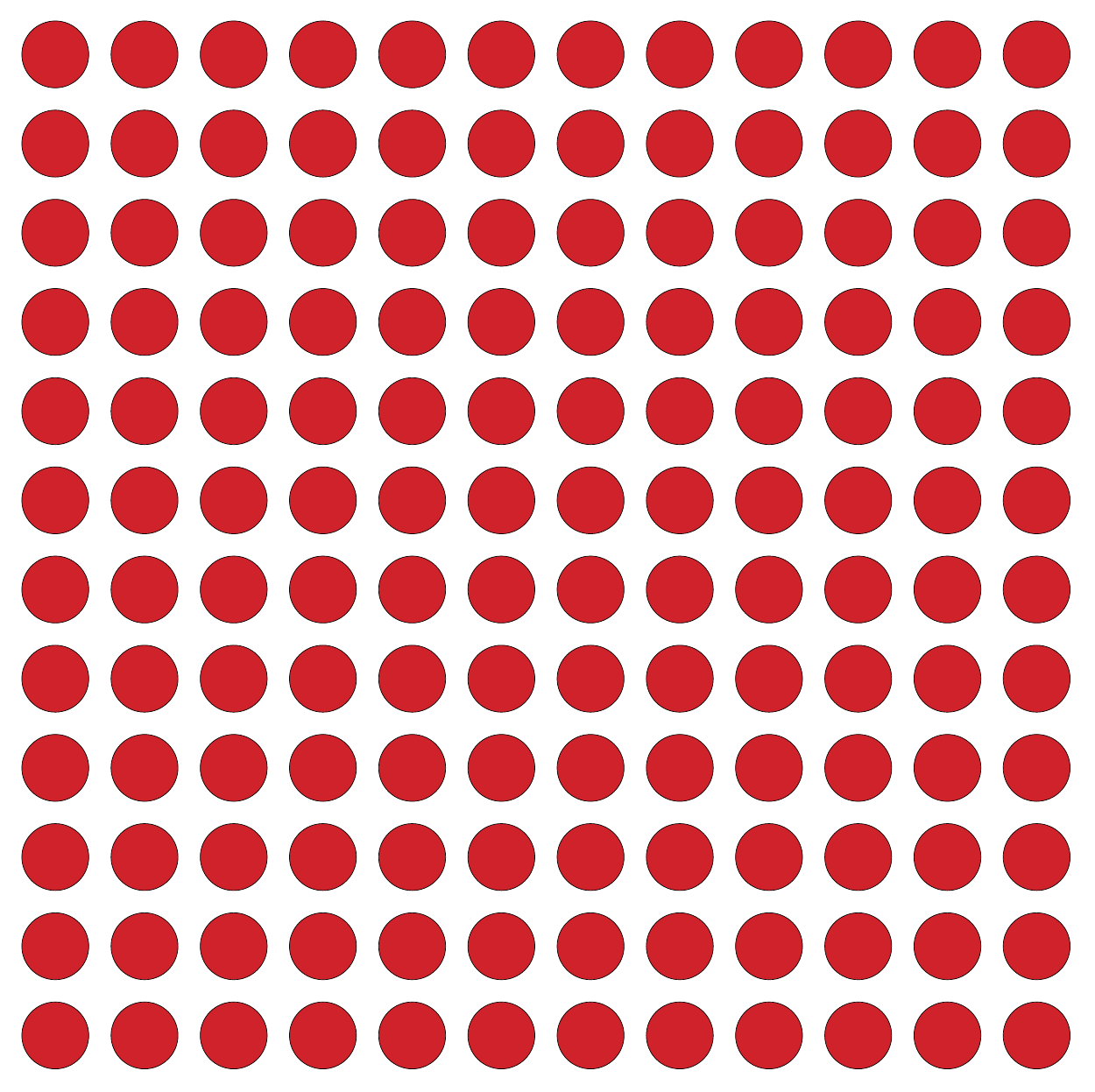}}\hfill
    \subfloat{\includegraphics[width=0.3\textwidth]{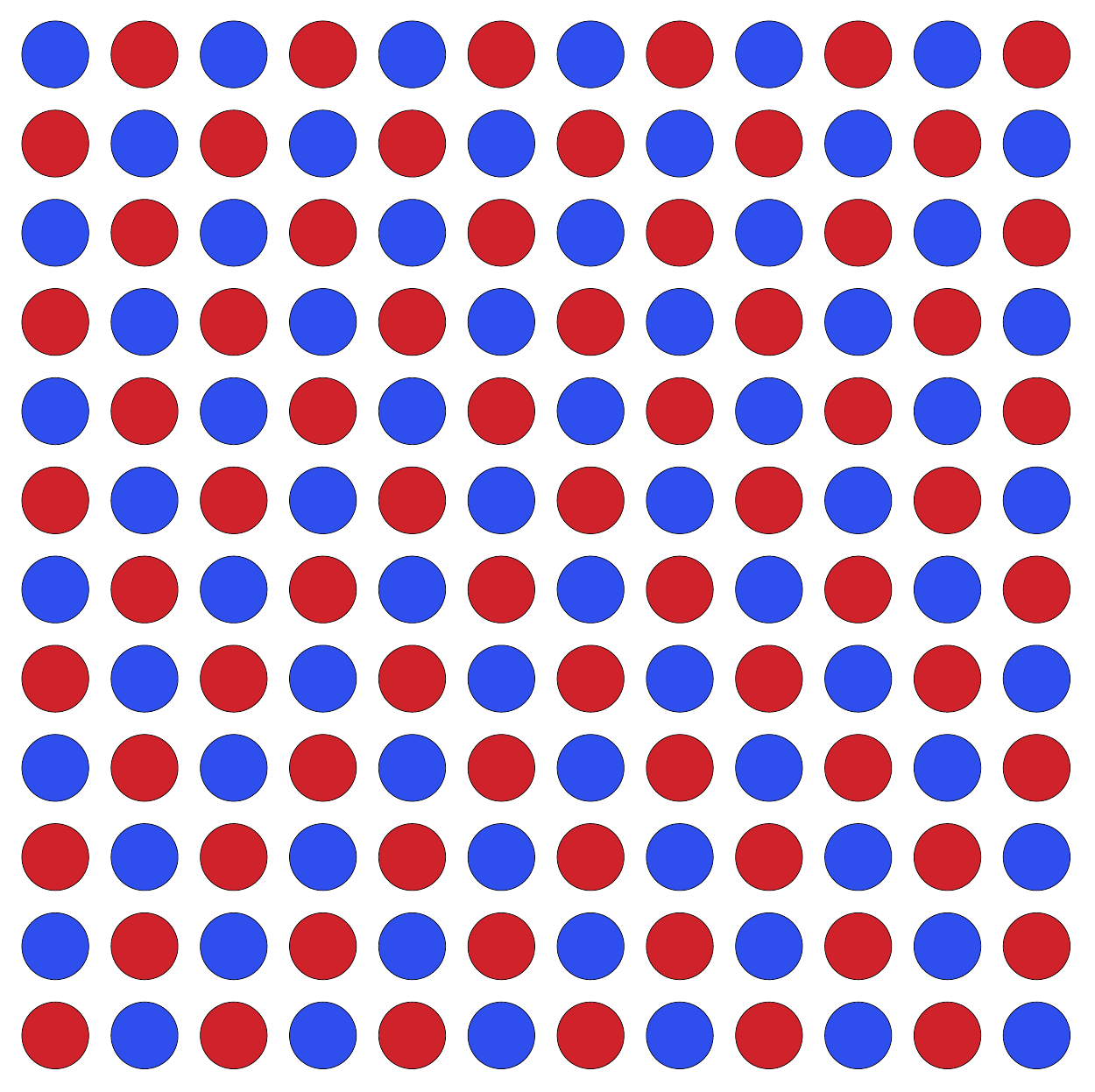}}\hfill
    \subfloat{\includegraphics[width=0.3\textwidth]{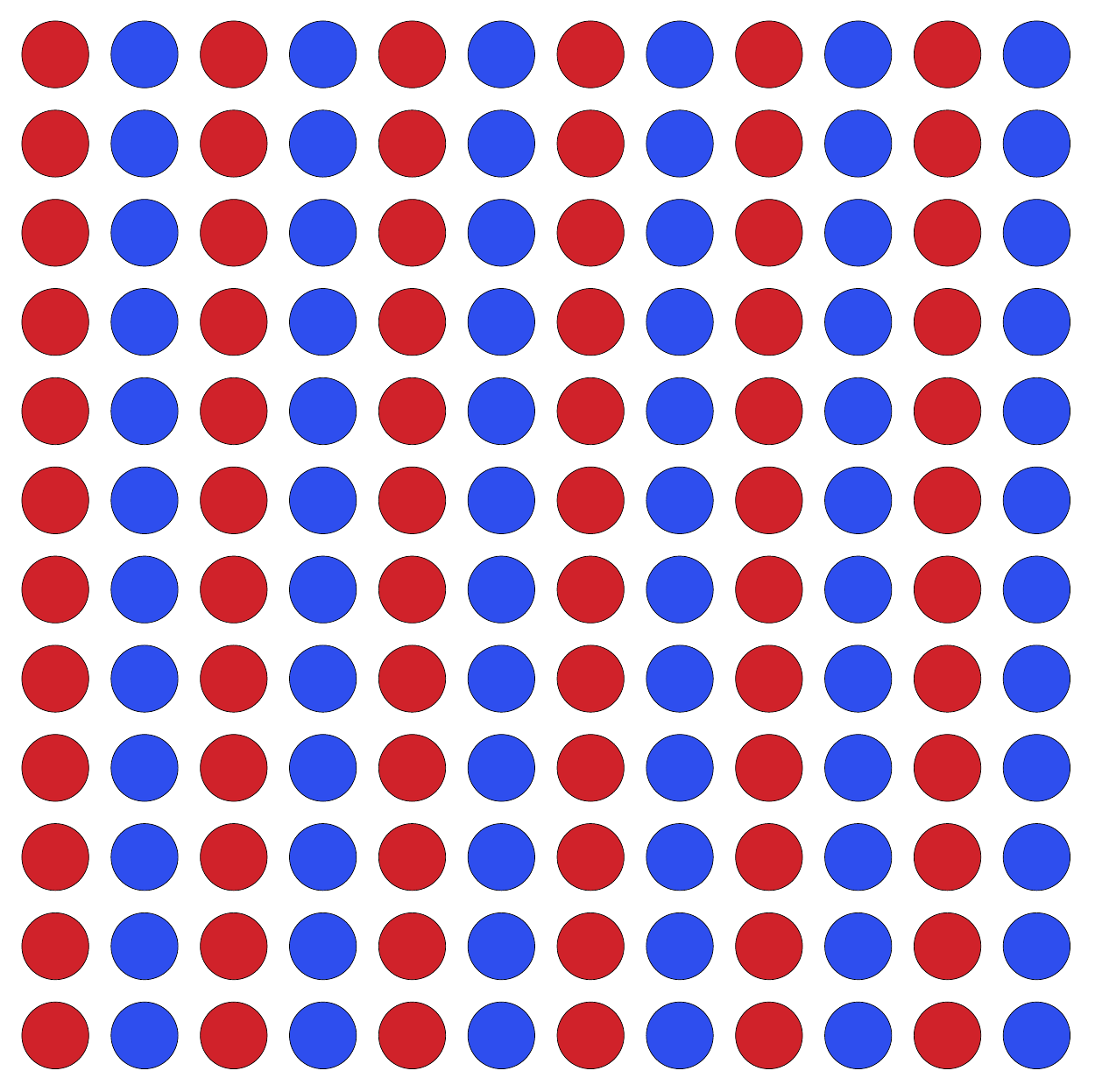}}
    \caption{The ordering pattern corresponding to the major modes with from left to right $\mathfrak{q}^{F}=(0,0)$, $\mathfrak{q}^{AF}=(\pi,\pi)$ and $\mathfrak{q}^{AS}=(\pi,0)$.}
    \label{fig:pat-major}
\end{figure}

\begin{figure}[htbp]
    \centering
    \subfloat{\includegraphics[width=0.3\textwidth]{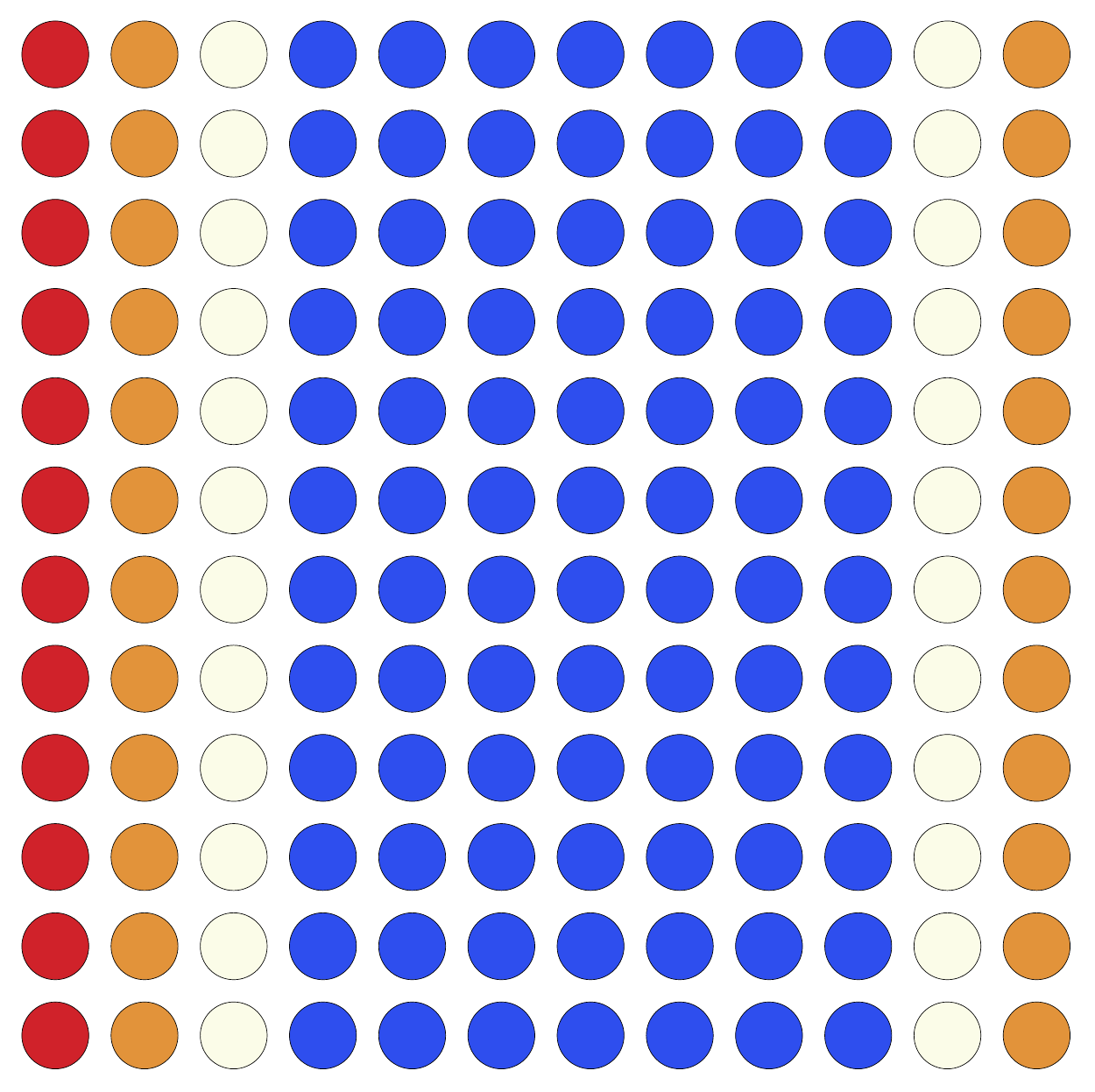}}\hfill
    \subfloat{\includegraphics[width=0.3\textwidth]{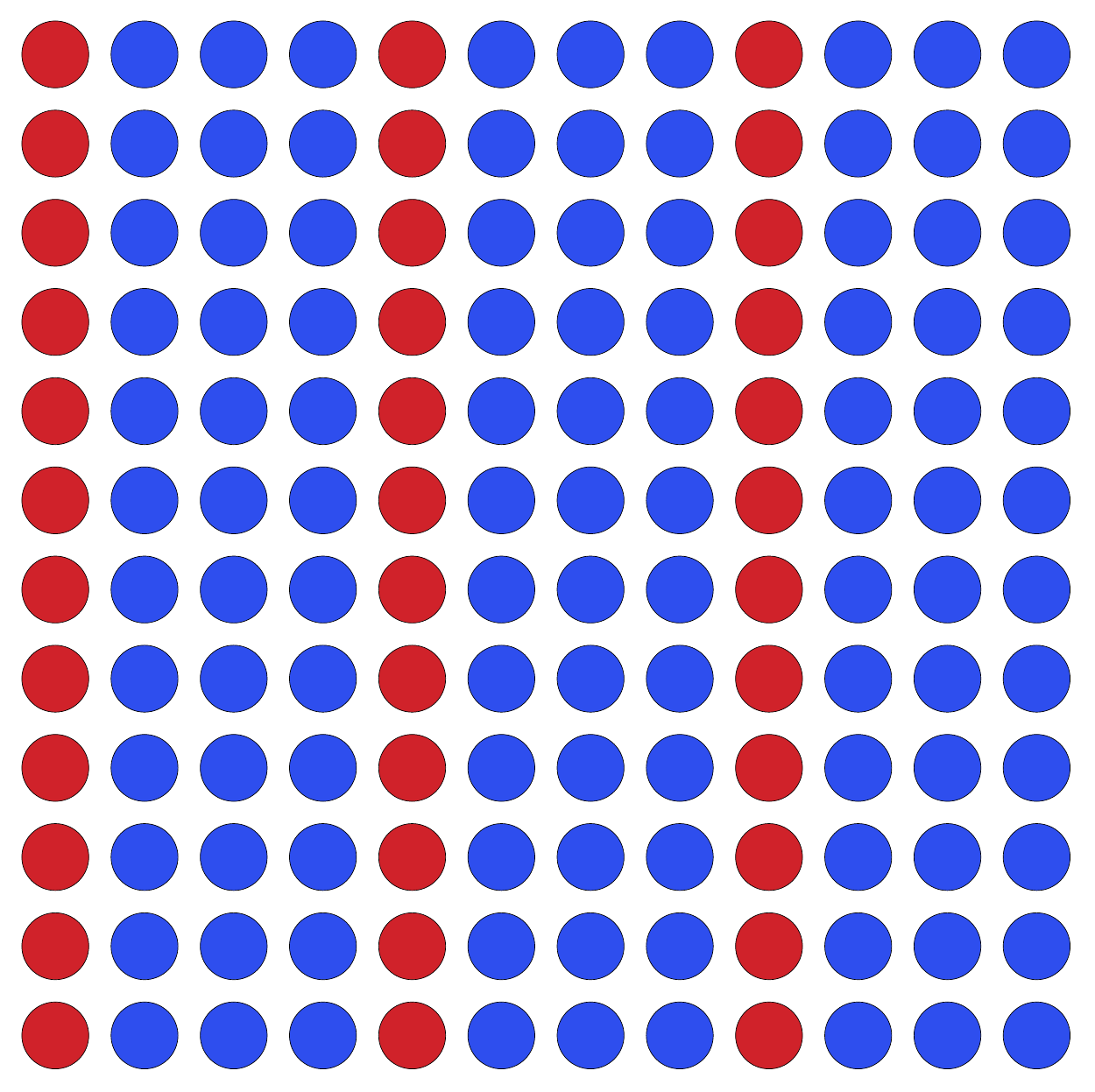}}\hfill
    \subfloat{\includegraphics[width=0.3\textwidth]{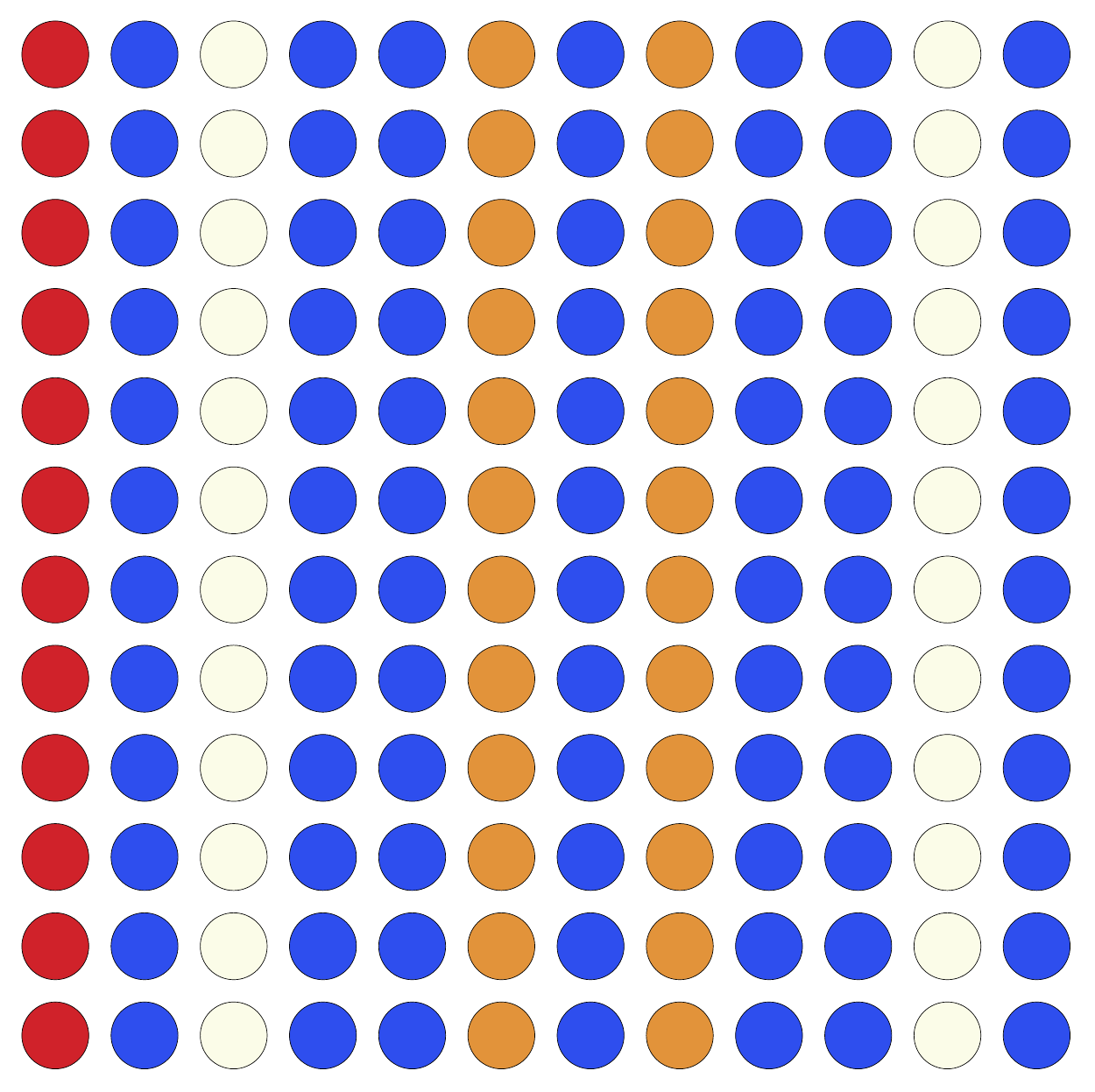}}\\
    \subfloat{\includegraphics[width=0.3\textwidth]{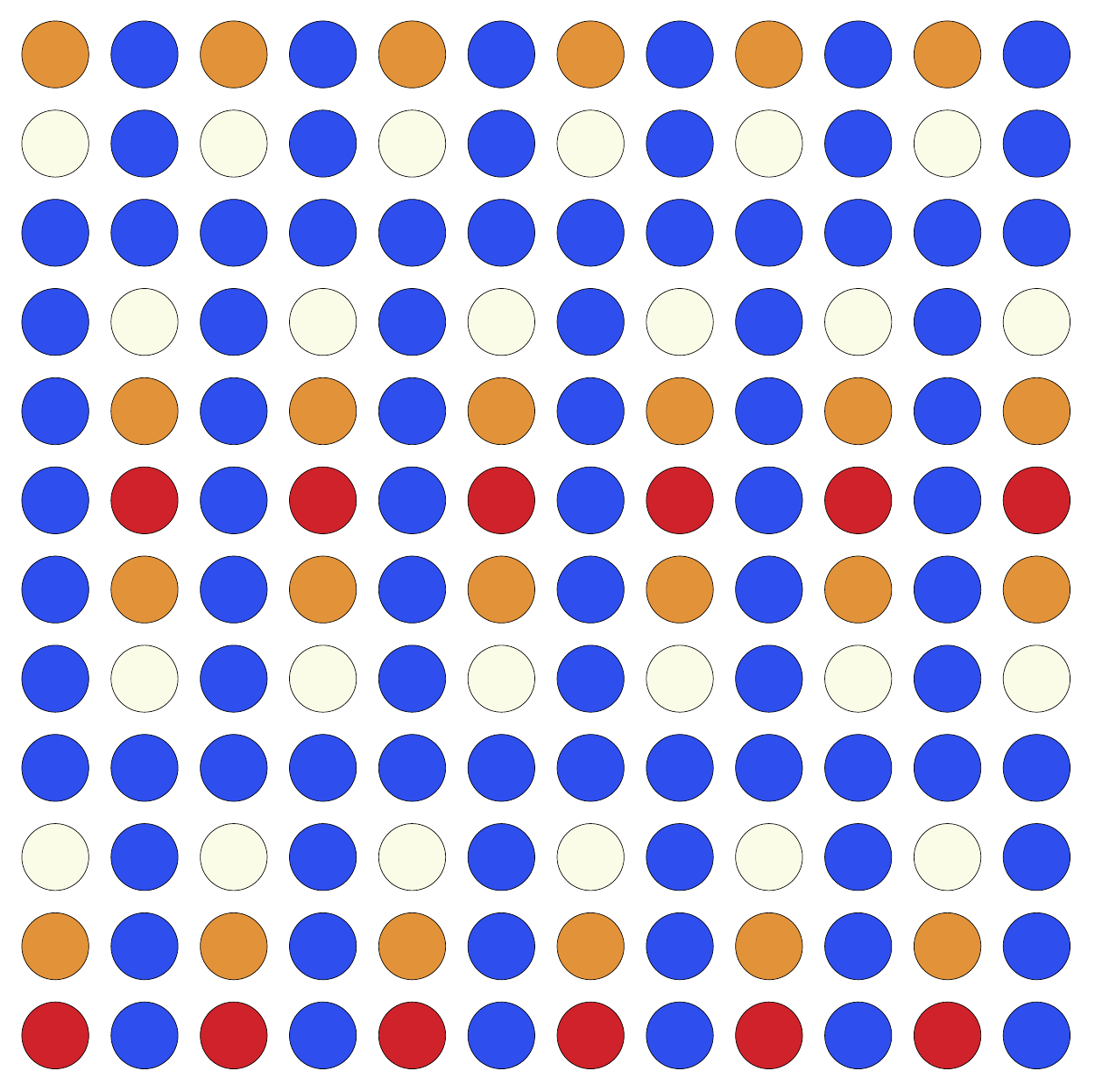}}\hfill
    \subfloat{\includegraphics[width=0.3\textwidth]{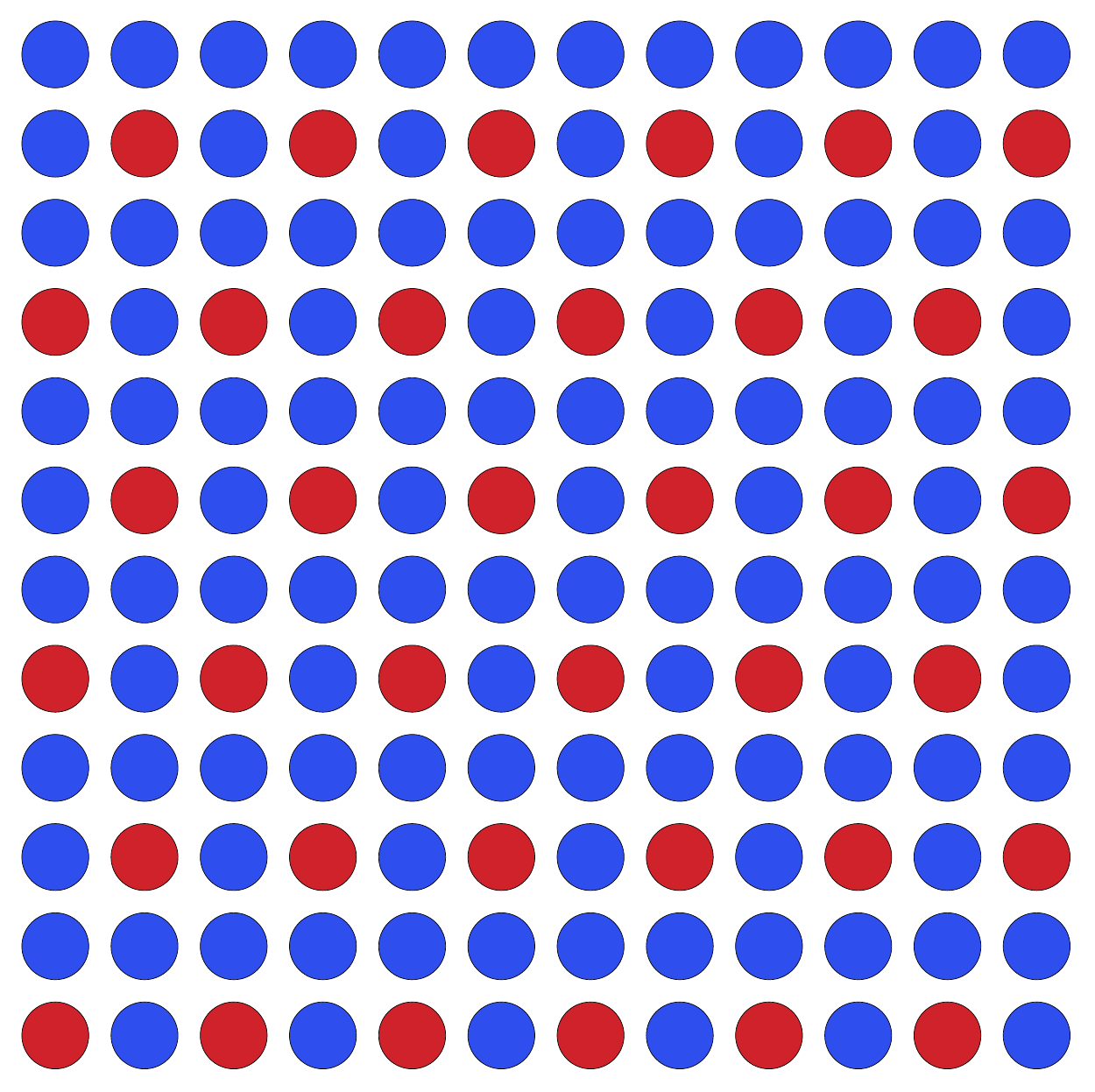}}\hfill
    \subfloat{\includegraphics[width=0.3\textwidth]{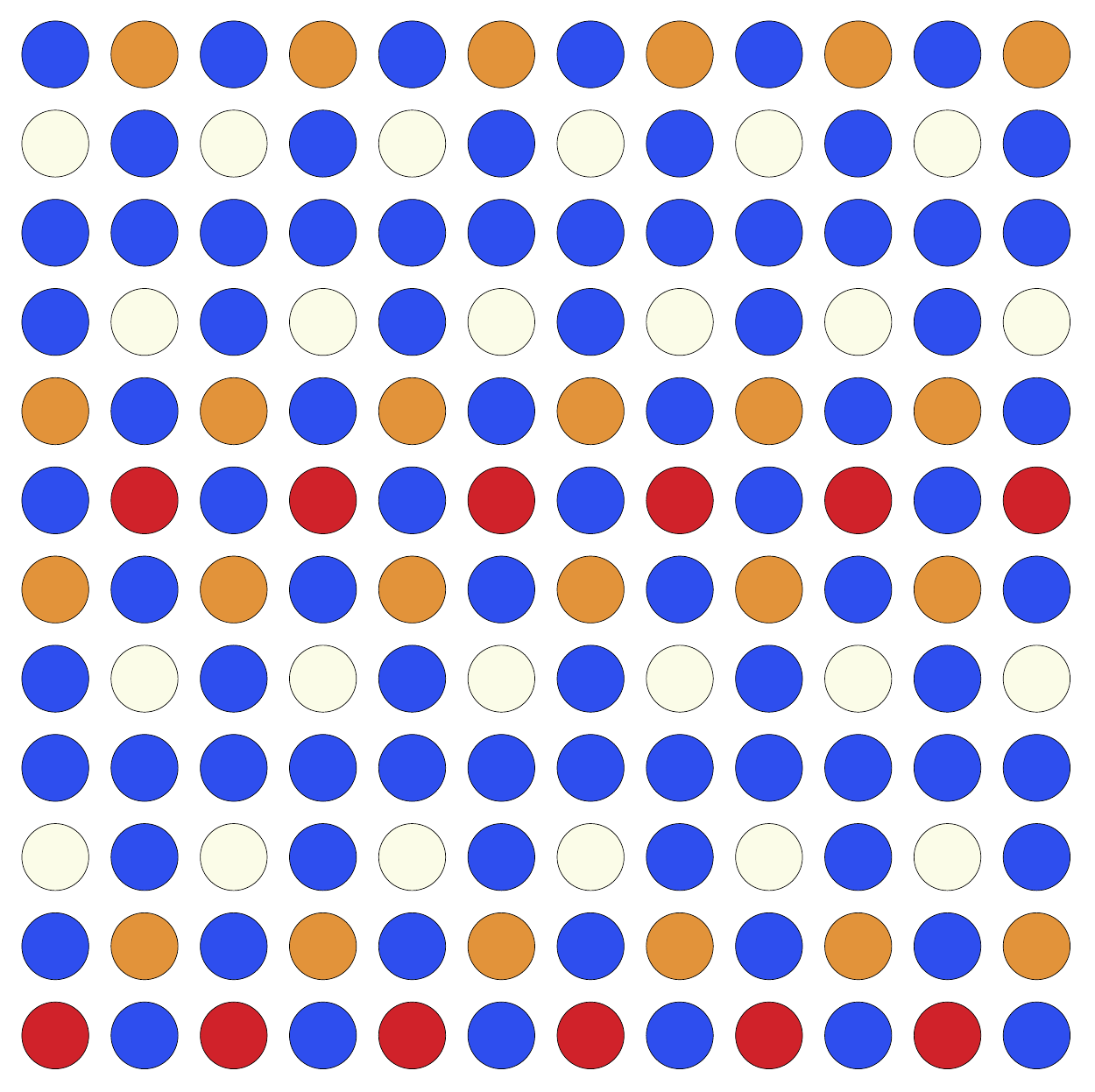}}\\
    \subfloat{\includegraphics[width=0.3\textwidth]{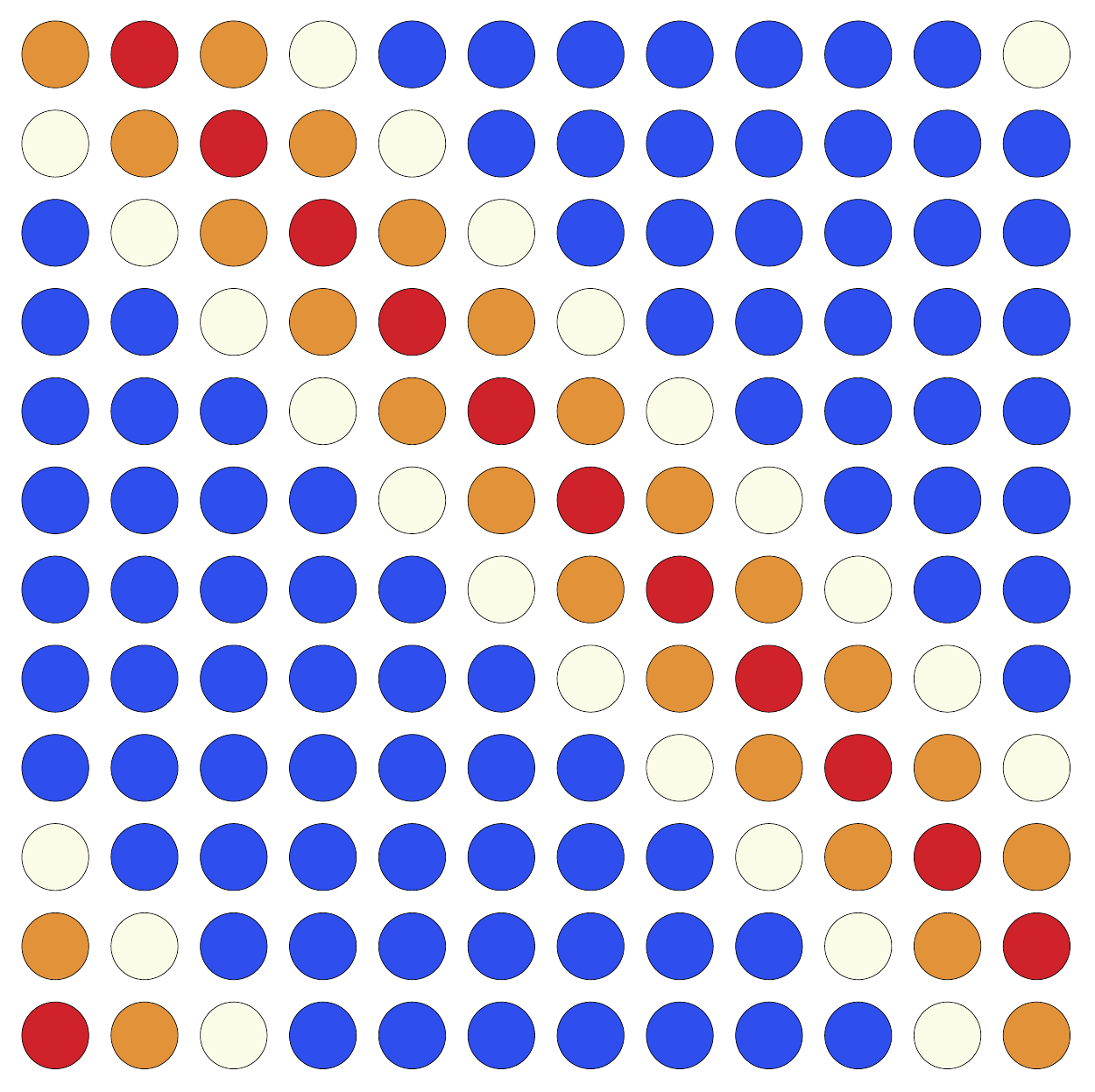}}\hfill
    \subfloat{\includegraphics[width=0.3\textwidth]{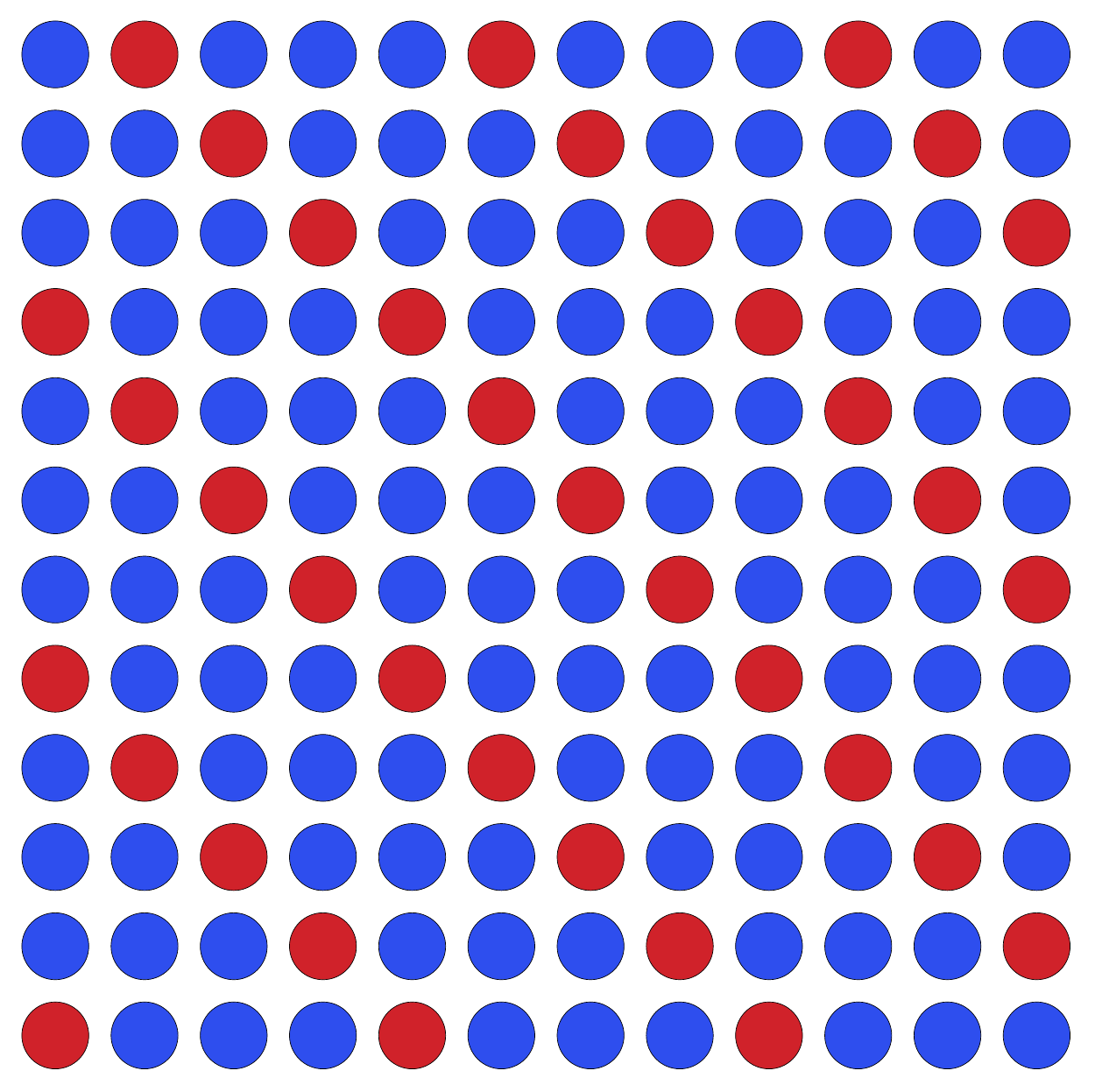}}\hfill
    \subfloat{\includegraphics[width=0.3\textwidth]{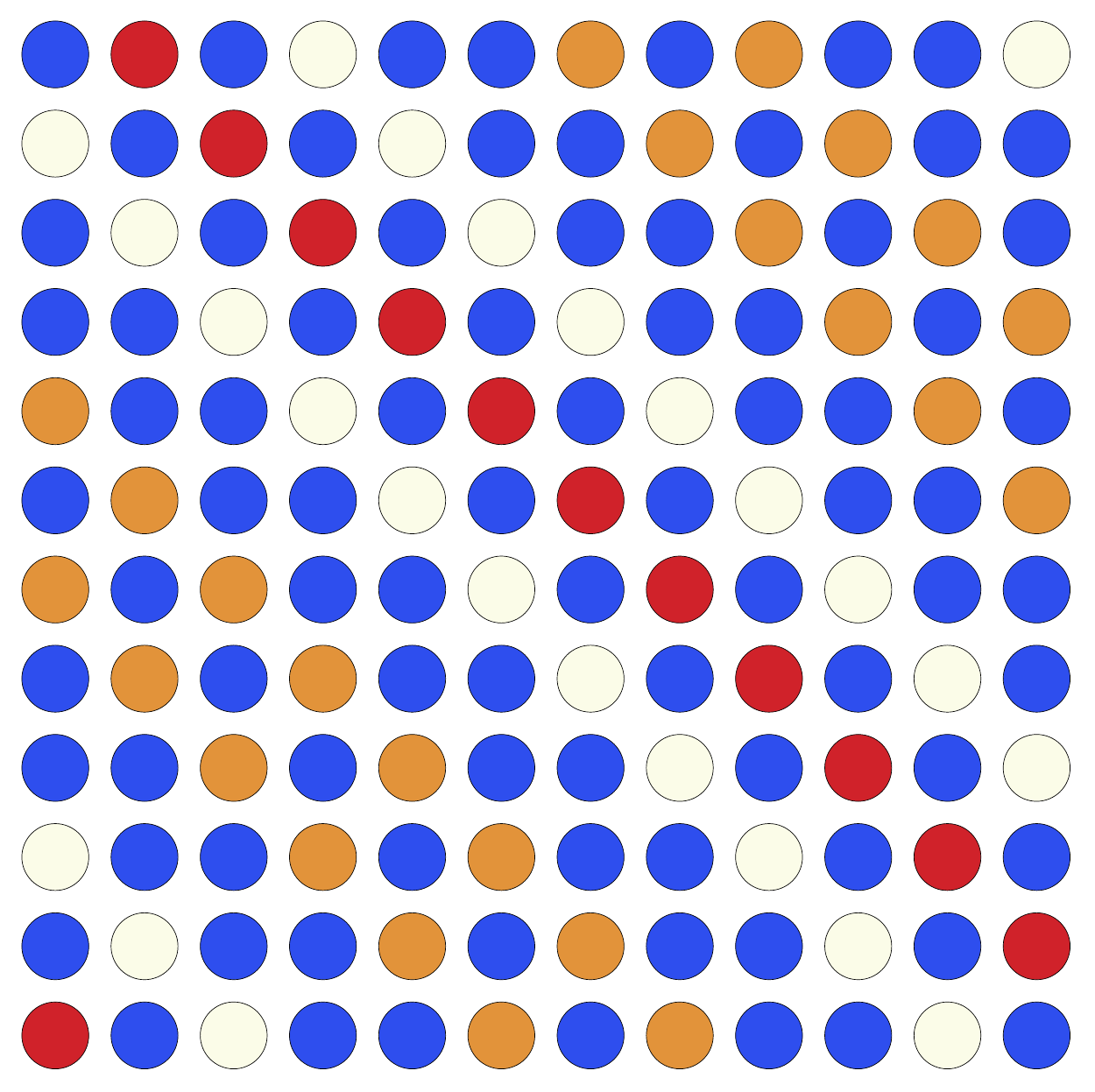}}
    \caption{The ordering pattern corresponding to fan modes. Top row: Striped modes with from left to right  $\mathfrak{q}=\pi/6 (1,0)$, $\mathfrak{q}=\pi/6 (3,0)$ and $\mathfrak{q}=\pi/6 (5,0)$. Middle row: Modulated stripe modes with from left to right $\mathfrak{q}=\pi/6 (6,1)$, $\mathfrak{q}=\pi/6 (6,3)$ and $\mathfrak{q}=\pi/6 (6,5)$. Bottom row: Diagonal stripe modes with from left to right $\mathfrak{q}=\pi/6 (1,1)$, $\mathfrak{q}=\pi/6 (3,3)$ and $\mathfrak{q}=\pi/6 (5,5)$. }
    \label{fig:pat-fans}
\end{figure}

\begin{figure}[htbp]
    \centering
    \subfloat{\includegraphics[width=0.3\textwidth]{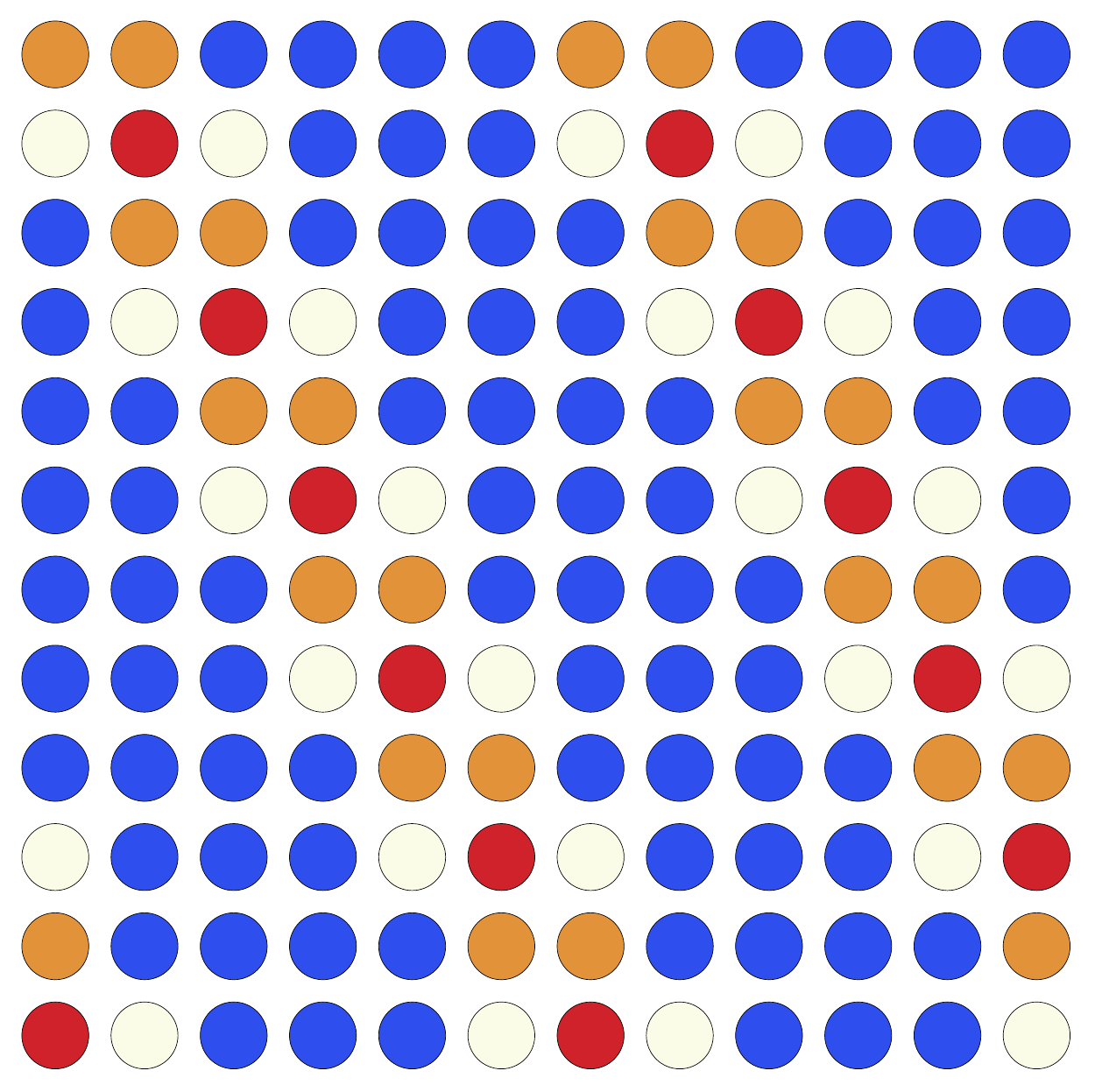}}\hfill
    \subfloat{\includegraphics[width=0.3\textwidth]{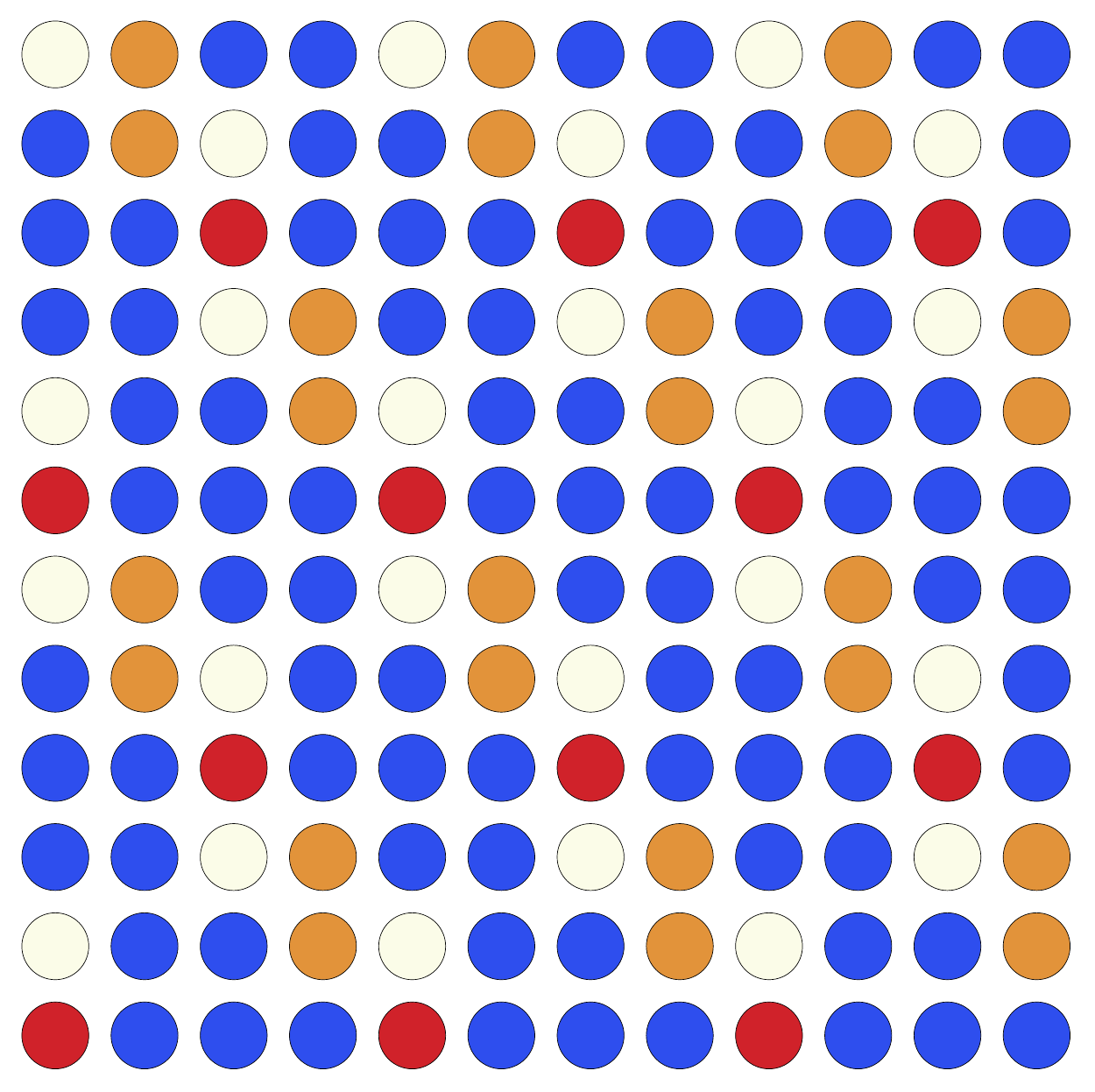}}\hfill
    \subfloat{\includegraphics[width=0.3\textwidth]{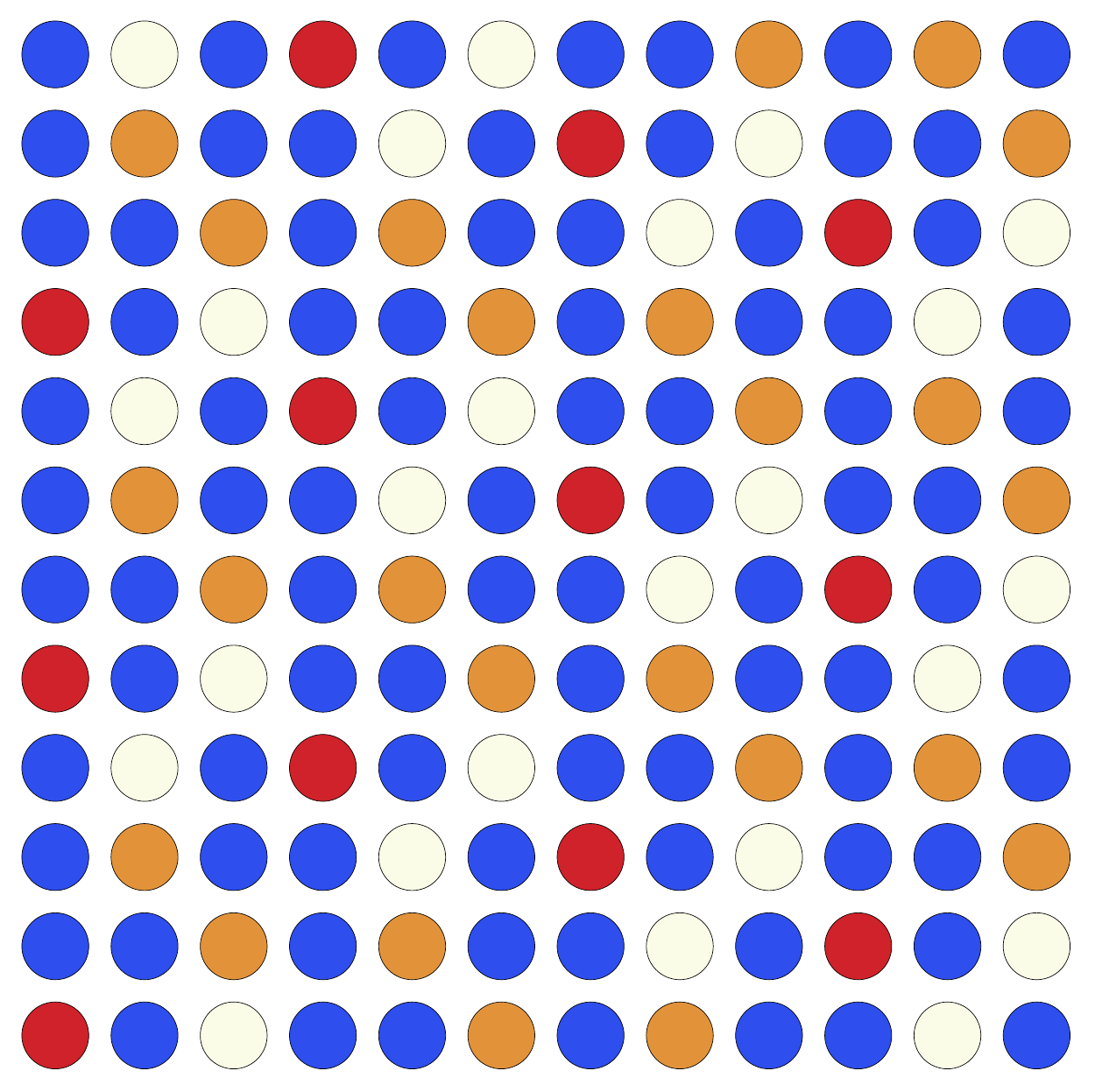}}
    \caption{The ordering pattern corresponding to multiplicity $M=8$ modes with from left to right $\mathfrak{q}=\pi/6 (2,1)$, $\mathfrak{q}=\pi/6 (3,2)$ and $\mathfrak{q}=\pi/6 (5,3)$.}
    \label{fig:pat-M8}
\end{figure}

We start off with the major modes for $K_{3}>0$, which are shown in Figure \ref{fig:pat-major}.
Next, in Figure \ref{fig:pat-fans} we show a few of the fan modes 
Finally, we present a few samples of $M=8$ modes from the interior of the IBZ in Figure \ref{fig:pat-M8}
\section{The natural coordinate system}
\label{app:natural}
\begin{figure}[ptb]
\centering
\includegraphics{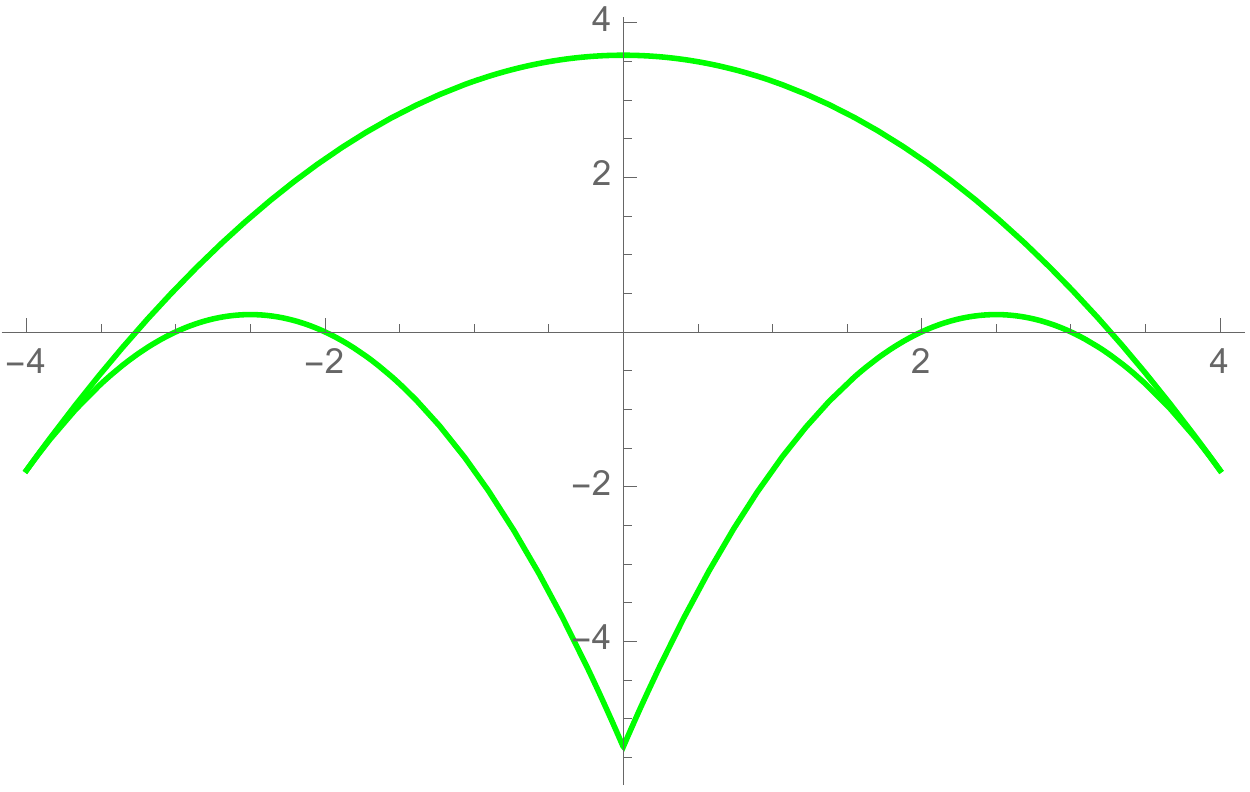} 
\caption{The shape of $\widehat{\Phi}_{\infty}$, the image of the IBZ $\widehat{\mathfrak{U}}_{\infty}$ expressed in the natural coordinates $\varphi_j$.}%
\label{fig:phi_domain}%
\end{figure}

Slightly rewriting Eq.~(\ref{eq:Fdepend}), we have $F_{1}\left(\mathfrak{q}\right)^{2}-4=2F_{2}\left(\mathfrak{q}\right)+F_{3}\left(\mathfrak{q}\right)$, which suggests a new basis vector along $(0,2,1)$. Trivially, a vector along $(0,1,-2)$ is then orthogonal to both the latter and the invariant axis along $(1,0,0)$. Normalizing these vectors yields the frame $\hat{\mathbf{n}}_j$ given in the main text. The explicit form of the two independent coordinates is%
\begin{align}
\varphi_{1} &  =2\cos q_{1}+2\cos q_{2} \equiv 2\xi_{1}+2\xi_{2}, \\
\varphi_{2} &  =\frac{2}{\sqrt{5}}\left(  \cos\left(  q_{1}-q_{2}\right)
+\cos\left(  q_{1}+q_{2}\right)  -2\cos2q_{1}-2\cos2q_{2}\right)  \nonumber \\
&  =\frac{4}{\sqrt{5}}\left(  2+\cos q_{1}\cos q_{2}-2\cos^{2}q_{1}-2\cos
^{2}q_{2}\right) \equiv \frac{4}{\sqrt{5}}\left(  2+\xi_{1}\xi_{2}-2\xi_{1}^{2}%
-2\xi_{2}^{2}\right),
\end{align}
where we have introduced $\xi_{1}=\cos q_{1}$ and $\xi_{2}=\cos q_{2}$. 
The mapping $\left(q_{1},q_{2}\right)\rightarrow\left(  \xi_{1},\xi_{2}\right)$ maps $\widehat{\mathfrak{U}}_{\infty}$ to $\widehat{\Xi}_{\infty}=\left\{\left(  \xi_{1},\xi_{2}\right)  |-1\leq\xi_{1}\leq\xi_{2}\leq1\right\}$.  Clearly, on this domain $\varphi_{1}\in\left[-4,4\right]$. Eliminating $\xi_{2}$ then yields%
\begin{equation}
\varphi_{2}=-2\frac{10\text{$\xi_{1}^{2}$}-5\text{$\xi_{1}$}\varphi_{1}+\varphi_{1}^{2}-4}{\sqrt{5}}.
\end{equation}
For a given $\varphi_{1}$ we need to ensure that $\left(\xi_{1},\xi
_{2}\right)  \in\Xi_{\infty},$ which yields the constraint $\xi_{1}\in\left[
\max\left(-1,\frac{1}{2}\varphi_{1}-1\right)  ,\frac{1}{4}\varphi
_{1}\right]$. This yields the two limiting curves
\begin{align}
\varphi_{2}^{\max}(\varphi_{1}) &  =\frac{2}{\sqrt{5}}\left(  4-\frac{3}{8}\varphi_{1}%
^{2}\right)  \label{eq:phi2max}\\
\varphi_{2}^{\min}(\varphi_{1}) &  =-\frac{2}{\sqrt{5}}\left(  2-\left\vert \varphi
_{1}\right\vert \right)  \left(  3-\left\vert \varphi_{1}\right\vert \right) \label{eq:phi2min},
\end{align}
where the first curve corresponds to the diagonal edge of $\widehat{\mathfrak{U}}_{\infty}$, and the lower curve to the horizontal leg when $\varphi_1>0$ and the vertical leg when $\varphi_1<0$. The three extreme points $\mathfrak{q}^{F}$, $\mathfrak{q}^{AF}$ and $\mathfrak{q}^{AS}$ are mapped to $\mathbf{\varphi}^{F}=(4,-4/\sqrt{5})$, $\mathbf{\varphi}^{AF}=(-4,-4/\sqrt{5})$ and $\mathbf{\varphi}^{AS}=(0,-12/\sqrt{5})$ respectively. The domain of values $\widehat{\Phi}_{\infty}$ in the $\varphi$ parametrization that corresponds to $\widehat{\mathfrak{U}}_{\infty}$ is shown in Figure \ref{fig:phi_domain}.

\section{Details of the simulations}\label{app:simulations}

The Monte Carlo (MC) simulations were implemented in C++ and performed on a local computing cluster. Following the discussion in Section \ref{sec:simulations}, we consider rays in phase space through the centroids of the predicted faces of $D_N$ on $N\times N$ square lattices with $N=4,6,8,12$. The specific state points were chosen by sampling the inverse temperature $T=1/\beta$ on $40$ equally spaced points in the range $T\in[1,3]$ and $45$ points in the lower temperature range. Finally, on the lattice of size $N=12$, we refined the temperature resolution even further to $\Delta T= 0.01$ for $T<0.5$. 

To compute the average of observables and their standard deviations we employed block averaging, using $n_B=100$ blocks. In order to choose the appropriate block length to ensure independence of the block averages, we estimated the autocorrelation `time' $\tau$ in MC sweeps (one attempted flip per spin in the lattice) for a number of observables. For $N=4$ and $N=6$ we were able to establish that the correlation time of the order parameter of the dominant mode was systematically larger, yet of similar order of magnitude than that of any of the other modes, and moreover did not depend strongly on the specific face of $D_N$ considered. For the larger lattice sizes, we therefore limited ourselves to measuring the correlation time for the standard ferromagnetic order parameter on the ray through the face associated with $\mathfrak{q}^{F}=(0,0)$. For $T>0.4$ we systematically measured $\tau<100$ for all lattice sizes, allowing $n_b=10^4$. For $T\le 0.4$ the correlation time increases rapidly and we employed block sizes of $n_b=10^5\text{--}10^6$, with the exact value optimized for the specific temperature and system size. Finally, we note that the computational bottleneck of our simulations is actually the calculation of the order parameters. Due to their extreme small size these systems are intrinsically noisy and subject to e.g.\ drift. These means that local magnetisations quickly average out. We therefore needed to resort to calculating the order parameter on the basis of instantaneous configurations, and subsequently average these, which requires a costly Fourier transform at every sweep.

\bibliography{ising}

\end{document}